\documentclass[12pt]{report}

\usepackage{epsf,amsfonts,hyperref}
\renewcommand{\appendix}[1]{
    \addtocounter{section}{1}
    \setcounter{equation}{0}
    \renewcommand{\thesection}{\Alph{section}}
    \section*{Appendice \thesection\protect\indent #1}
    \addcontentsline{toc}{section}{Appendice \thesection\ \ \ #1}
}

\newcommand\1 {{\mathbf 1}}

\renewcommand\l{\lambda}
\renewcommand\L{\Lambda}

\newcommand{\Tr}{{\,\rm Tr}\:}
\newcommand{\tr}{{\,\rm tr}\:}

\newcommand{\diag}{{\,\rm diag}}
\newcommand{\ovl}{\overline}

\newcommand{\Res}{{\rm Res}}
\newcommand{\Gammax}{{\mathop\Gamma^{x}}}
\newcommand{\Gammay}{{\mathop\Gamma^{y}}}
\newcommand{\ds}{\displaystyle}

\newcommand{\td}[1]{{\tilde{#1}}}

\newcommand{\om}{\omega}

\newcommand{\ee}[1]{{{\rm e}^{#1}}}

\renewcommand{\d}{{{\partial}}}
\newcommand{\D}{{{\hbox{d}}}}

\newcommand{\Pint}{{\int\kern -1.em -\kern-.25em}}

\renewcommand{\Re}{{\mathrm{Re}}}
\renewcommand{\Im}{{\mathrm{Im}}}

\renewcommand{\and}{{\qquad {\rm and} \qquad}}

\newcommand{\virg}{{\,\, , \qquad}}
\newcommand{\point}{{\,\, .}}

\newcommand{\beq}{\begin{equation}}
\newcommand{\eeq}{\end{equation}}
\newcommand{\bea}{\begin{eqnarray}}
\newcommand{\eea}{\end{eqnarray}}
\newcommand{\bacc}{\left\{ \begin{array}{l}}
\newcommand{\eacc}{\end{array}\right.}
\renewcommand{\thesection}{\arabic{section}}

\makeatletter
\@addtoreset{equation}{section}
\makeatother
\newtheorem{theorem}{Th\'eor\`eme}[section]
\newtheorem{remark}{Remarque}[section]
\newtheorem{proposition}{Proposition}[section]
\newtheorem{lemma}{Lemme}[section]
\newtheorem{conjecture}{Conjecture}[section]
\newtheorem{corollary}{Corollaire}[section]
\newtheorem{definition}{D\'efinition}[section]

\def\br{\begin{remark}\rm\small}
\def\er{\end{remark}}
\def\bt{\begin{theorem}}
\def\et{\end{theorem}}
\def\bd{\begin{definition}}
\def\ed{\end{definition}}
\def\bp{\begin{proposition}}
\def\ep{\end{proposition}}
\def\bl{\begin{lemma}}
\def\el{\end{lemma}}
\def\bc{\begin{corollary}}
\def\ec{\end{corollary}}
\def\bconj{\begin{conjecture}}
\def\econj{\end{conjecture}}
\def\beaq{\begin{eqnarray}}
\def\eeaq{\end{eqnarray}}

\newcommand{\R}{{\bf\rm  R}}
\newcommand{\C}{{\bf\rm  C}}

\newcommand{\Remark}{\bigskip \noindent{\bf Remarque: }}
\newcommand{\refeq}[1]{eq.(\ref{#1})}

\newcommand\encadremath[1]{\vbox{\hrule\hbox{\vrule\kern8pt
\vbox{\kern8pt \hbox{$\displaystyle #1$}\kern8pt}
\kern8pt\vrule}\hrule}}
\def\enca#1{\vbox{\hrule\hbox{
\vrule\kern8pt\vbox{\kern8pt \hbox{$\displaystyle #1$}
\kern8pt} \kern8pt\vrule}\hrule}}

\begin{document}

\sloppy


\pagestyle{empty}
\hfill SPT-05/047
\addtolength{\baselineskip}{0.20\baselineskip}
\vspace{56pt}
\begin{center}
{\large \bf UNIVERSIT\'E PARIS 7 DENIS DIDEROT}
\vspace{46pt}

{\Large \bf HABILITATION \`A DIRIGER DES RECHERCHES}
\vspace{46pt}

{pr\'esent\'ee par}
\vspace{16pt}

{\large \bf Bertrand EYNARD}
\vspace{56pt}

{Sujet :}
\vspace{16pt}

{\Huge \bf {Le mod\`ele \`a deux matrices,}}
\vspace{16pt}

{\Large \bf Polyn\^omes biorthogonaux, probl\`eme de Riemann--Hilbert et g\'eom\'etrie alg\'ebrique}


\end{center}

\vspace{70pt}


\vspace{20pt}
%





\newpage
\pagestyle{plain}
\setcounter{page}{1}


\newcommand{\citeeynmatsg}{{\bf [P1]}}
\newcommand{\bibeynmatsg}{ {\citeeynmatsg}
 {Genus one contribution to free energy in hermitian two-matrix model,  (B.E., D. Korotkin, A. Kokotov), 25 pages,
Nucl.Phys. B694 (2004) 443-472, SPHT T04/020.
xxx, hep-th/0403072.}}

\newcommand{\citeBEformulaD}{{\bf [P2]}}
\newcommand{\bibBEformulaD}{{\citeBEformulaD}
{The PDEs of biorthogonal polynomials arising in the 2-matrix model,  (M. Bertola, B.E.), 20 pages,
Preprint SPHT T03/139, nlin.SI/0311033.}}

\newcommand{\citeeynmultimat}{{\bf [P3]}}
\newcommand{\bibeynmultimat}{{\citeeynmultimat}
 {Master loop equations, free energy and correlations for the chain of matrices,  (B.E.) , 43 pages,
SPHT T03/125. JHEP11(2003)018. xxx, hep-th/0309036, ccsd-00000572.}}

\newcommand{\citeBEmixed}{{\bf [P4]}}
\newcommand{\bibBEmixed}{{\citeBEmixed}
 Mixed Correlation Functions of the Two-Matrix Model  (M. Bertola, B.E.) , 16 pages,
SPHT T03/028, CRM-2961 (2003). J. Phys. A36 (2003) 7733-7750.
xxx, hep-th/0303161.}

\newcommand{\citeeynmatgzero}{{\bf [P5]}}
\newcommand{\bibeynmatgzero}{{\citeeynmatgzero}
 Large N expansion of the 2-matrix model (B.E.) , 41 pages,
      SPHT T02/128, CRM-2868. JHEP 01 (2003) 051.
      xxx, hep-th/0210047.}

\newcommand{\citeBEHRH}{{\bf [P6]}}
\newcommand{\bibBEHRH}{{\citeBEHRH}
 Differential systems for bi-orthogonal polynomials appearing in two-matrix models and the associated Riemann-Hilbert problem. (M. Bertola, B.E., J. Harnad) , 60 pages,
      SPHT 02/097, CRM-2852. Comm. Math. Phys. 243 no.2 (2003) 193-240,
      xxx, nlin.SI/0208002.}

\newcommand{\citeBEHtauiso}{{\bf [P7]}}
\newcommand{\bibBEHtauiso}{{\citeBEHtauiso}
 Partition functions for Matrix Models and Isomonodromic Tau Functions. (M. Bertola, B.E., J. Harnad) , 17 pages,
      SPHT 02/050, CRM 2841 , J. Phys. A Math. Gen. 36 No 12 (28 March 2003) 3067-3083.
      xxx, nlin.SI/0204054.}

\newcommand{\citeBEHduality}{{\bf [P8]}}
\newcommand{\bibBEHduality}{{\citeBEHduality}
 Duality, Bi-orthogonal Polynomials and Multi-Matrix Models. (M. Bertola, B.E., J. Harnad) , 35 pages
      SPHT 01/047, CRM-2749 , Commun.Math.Phys. 229 (2002) 73-120.
      xxx, nlin.SI/0108049.}

\newcommand{\citeeynchaint}{{\bf [P9]}}
\newcommand{\bibeynchaint}{{\citeeynchaint}
 Correlation functions of eigenvalues of multi-matrix models, and the limit of a time dependent matrix. (B.E.) , 27 pages.
      SPHT 98/001 DTP 97-59 , Journal of Physics A 40 (1998) 8081,
      xxx, cond-mat/9801075.}

\newcommand{\citeEMchain}{{\bf [P10]}}
\newcommand{\bibEMchain}{{\citeEMchain}
 Matrices coupled in a chain: eigenvalue correlation. (M.L. Mehta, B.E.) , 8 pages.
      SPHT 97/112. Journal of Physics A 19 (1998) 4449,
      xxx, cond-mat/9710230.}

\newcommand{\citeeynchain}{{\bf [P11]}}
\newcommand{\bibeynchain}{{\citeeynchain}
 Eigenvalue distribution of large random matrices, from one matrix to several coupled matrices. (B.E.) , 31 pages.
      SPHT 97031. Nuc. Phys. B506,3 633-664 (1997).
      xxx, cond-mat/9707005.}

\begin{tableofcontents}

\end{tableofcontents}


\chapter{Introduction: les matrices al\'eatoires en physique et en math\'ematiques}

Je vais pr\'esenter dans cette habilitation, des r\'esultats  obtenus ces derni\`eres ann\'ees sur le mod\`ele \`a deux matrices et les familles de polyn\^omes biorthogonales.

\subsection{Bref historique du sujet}

Les mod\`eles de matrices al\'eatoires ont \'et\'e introduits en 1951 par E. Wigner \cite{Wigner}, en physique nucl\'eaire,
comme mod\`ele pour la r\'epartition des niveaux d'\'energie des noyeaux lourds.
Cette th\'eorie a eu un succ\`es consid\'erable, et s'est impos\'ee comme outil tr\`es puissant pour \'etudier
de nombreux autres ph\'enom\`enes physiques \cite{Guhr,DGZ, courseynard, gross:1991}.
En vrac: chaos quantique, conducteurs m\'esoscopiques, croissance des cristaux, probl\`emes de fronti\`eres libres (probl\`eme de Saffman--Taylor),
gravitation quantique, th\'eorie des cordes, repliement des prot\'eines, etc...
La raison de ce succ\`es tient dans les propri\'et\'es d'int\'egrabilit\'e des mod\`eles de matrices.
Les matrices al\'eatoires ont jou\'e un r\^ole important aussi en math\'ematiques \cite{Mehta, BI, Moerbeke:2000}.
Elles sont \'etroitement reli\'ees aux polyn\^omes orthogonaux, aux probl\`emes isomonodromiques, probl\`eme de Riemann--Hilbert.
Les int\'egrales de matrices sont le prototype des fonctions $\tau$.
De plus, r\'ecement, on a compris que toutes ces notions avaient \'egalement beaucoup \`a voir avec la g\'eom\'etrie alg\'ebrique.

\medskip

Le mod\`ele \`a une matrice \cite{DGZ, Mehta}, qui est reli\'e aux polyn\^omes orthogonaux habituels, a \'et\'e beaucoup \'etudi\'e.
Toutefois, il ne donne acc\`es qu'aux courbes hyperelliptiques, et aux syst\`emes d'\'equations diff\'erentielles d'ordre 2.
Il aparaissait int\'eressant et important de g\'en\'eraliser les notions qui aparaissaient dans le mod\`ele \`a une matrice, \`a un cas un peu plus g\'en\'eral.
C'est pourquoi le mod\`ele \`a deux matrices a alors attir\'e l'attention des physiciens et des math\'ematiciens.

Le {\bf mod\`ele \`a deux matrices}, est reli\'e \`a deux familles de polyn\^omes biorthogonales.
Ces deux familles satisfont deux syst\`emes diff\'erentiels d'ordre diff\'erents, dont on peut montrer qu'ils sont duaux
l'un de l'autre.
La courbe spectrale de ces syst\`emes, est une courbe alg\'ebrique, pas n\'ec\'essairement hyperelliptique,
en particulier, elle peut avoir des singularit\'es rationelles $y\sim x^{p/q}$ avec $(p,q)$ entiers quelconques.

Du point de vue des physiciens, le mod\`ele \`a deux matrices a \'et\'e introduit pour mod\'eliser le mod\`ele d'Ising \cite{Kazakov}
sur un r\'eseau al\'eatoire, i.e. en termes de th\'eories de champs conformes, un mod\`ele minimal $(3,4)$.
Il a rapidement \'et\'e compris \cite{KazakoVDK} que le mod\`ele \`a deux matrices donnait acc\`es \`a tous les mod\`eles minimaux
rationels $(p,q)$.
C'est dans ce cadre qu'ont \'et\'e invent\'ees et \'etudi\'ees les ''\'equations de boucles'' \cite{staudacher, DGZ}, et le d\'eveloppement topologique \cite{thooft:1974}.
Plus r\'ecement, les mod\`eles de matrices ont jou\'e un r\^ole tr\`es important en th\'eorie des cordes \cite{Dijgrafvafa}.

\subsection{Principaux r\'esultats obtenus}

Parmis les r\'esultats les plus importants que j'ai obtenu, avec mes collaborateurs, sur ce sujet, et qui sont pr\'esent\'es dans cette th\`ese:

\noindent $\bullet$ L'expression des fonctions de corr\'elations de valeurs propres en terme de d\'eterminants de noyaux de polyn\^omes bi-orthogonaux (\underline{th\'eor\`eme III.\ref{theoremeynMehta}}, \citeEMchain).

\noindent $\bullet$ Un th\'eor\`eme de Christoffel--Darboux pour ces noyaux (\underline{th\'eor\`eme III.\ref{TheoremCD}}, \citeBEHduality).

\noindent $\bullet$ L'obtention de deux syst\`emes diff\'erentiels pour les deux familles bi-orthogonales de polyn\^omes, leur expression explicite et le calcul de leur trace (\underline{th\'eor\`eme III.\ref{Theoremfolding}}, \underline{th\'eor\`eme III.\ref{TheoremDFn}}, \underline{th\'eor\`eme III.\ref{TheoremDexplicit}}, \citeBEformulaD), et un premier pas vers le calcul de la fonction $\tau$ isomonodromique \citeBEHtauiso.

\noindent $\bullet$ La dualit\'e spectrale entre les deux syst\`emes diff\'erentiels (\underline{th\'eor\`eme III.\ref{Theoremdualite}}, \citeBEHduality).

\noindent $\bullet$ L'expression exacte de la solution fondamentale de ces syst\`emes diff\'erentiels (\underline{th\'eor\`eme III.\ref{Theoremsolfondtdphi}},  \citeBEHRH).

\noindent $\bullet$ Les asymptotiques \`a $x$ grand des solutions fondamentales (\underline{th\'eor\`eme III.\ref{Theoremasymptdphi}},  \citeBEHRH), et les matrices de Stokes associ\'ees.

\noindent $\bullet$ La propri\'et\'e de dualit\'e spectrale a permis d'\'ecrire un probl\`eme de Riemann-Hilbert associ\'e (\underline{paragraphe III.\ref{sectionpbRH}},  \citeBEHRH).

\noindent $\bullet$ L'expression d'une fonction de corr\'elation mixte (i.e. qui ne peut pas s'\'ecrire directement en termes de valeurs propres des deux matrices) (\underline{th\'eor\`eme III.\ref{Theoremtracemixt}}, \citeBEmixed).

\noindent $\bullet$ La limite $N$ grand des int\'egrales de matrices formelles, en particulier l'obtention d'une \'equation alg\'ebrique pour la r\'esolvante (\underline{th\'eor\`eme IV.\ref{Cormasterloopeq}}, \underline{th\'eor\`eme IV.\ref{TheoremmasterloopNlarge}}, \citeeynchain, \citeeynchaint, \citeeynmultimat),
et le calcul de certaines fonctions de corr\'elations mixtes dans la limite $N$ grand (\underline{th\'eor\`eme IV.\ref{Theoremtracemixteloop}}, \citeeynchain, \citeeynchaint, \citeeynmultimat).

\noindent $\bullet$ Le d\'eveloppement \`a $N$ grand des int\'egrales de matrices formelles (\citeeynchain, \citeeynchaint, \citeeynmultimat, \citeeynmatgzero, \citeeynmatsg)
en particulier, le fait que le terme en$1/N^2$ du d\'eveloppement de l'\'energie libre est reli\'e au d\'eterminant d'un Laplacien sur une courbe alg\'ebrique (\underline{th\'eor\`eme IV.\ref{Theoremresolvantesubleading}}, \underline{th\'eor\`eme IV.\ref{TheoremFsubleading}}, \citeeynmatgzero, \citeeynmatsg).

\noindent $\bullet$ Une conjecture, (tout de m\^eme bas\'ee sur de s\'erieux arguments heuristiques) pour les asymptotiques \`a $N$ grands des polyn\^omes biorthogonaux (\underline{conjecture V.\ref{conjasymp}}, \citeeynchain, \citeeynchaint), du type Deift \&\ co \cite{dkmvz}.

\subsection{Plan}

La th\`ese est organis\'ee comme suit:

- Le chapitre 2 donne les d\'efinitions du mod\`ele \`a deux matrices,
et explique les quantit\'es que l'on cherche \`a calculer.

- le chapitre 3 pr\'esente la m\'ethode des polyn\^omes biorthogonaux,
et les syst\`emes diff\'erentiels associ\'es. Un certains nombres de r\'esultats
comme la  dualit\'e, les asymptotiques, y sont expos\'es.

- le chapitre 4 pr\'esente la m\'ethode des \'equations de boucles,
et la courbe alg\'ebrique associ\'ee. Un certains nombres de mes contributions
comme le calcul du d\'eveloppement en $1/N^2$ de l'\'energie libre y sont expos\'es.

- le chapitre 5 montre, comment en combinant les deux m\'ethodes pr\'ec\'edentes,
on peut formuler une conjecture pour les asymptotiques $N$-grand des polyn\^omes biorthogonaux, en termes de g\'eom\'etrie  alg\'ebrique.

- le chapitre 6 est une conclusion qui propose les suites possibles envisageables de ces travaux.

\chapter{Trois d\'efinitions du mod\`ele \`a deux matrices}
\label{chapterdefs}

Il existe plusieurs d\'efinitions du mod\`ele \`a deux matrices, {\bf non \'equivalentes}.
Suivant le contexte, c'est \`a dire \`a quelle application des matrices al\'eatoires on
s'int\'eresse, il faut choisir l'une des d\'efinitions.
Dans la limite de grande taille des matrices, \`a l'ordre dominant, les trois d\'efinitions conduisent
souvent aux m\^emes r\'esultats, mais ce n'est pas vrai au del\`a de l'ordre dominant.
De plus un certains nombre de m\'ethodes de calcul sont communes aux trois mod\`eles, et les trois mod\`eles
font appels aux m\^emes concepts de g\'eom\'etrie et int\'egrabilit\'e.
De l\`a, il sort que les trois mod\`eles ne sont pas toujours bien distingu\'es dans la litt\'erature, et l'on rencontre
de nombreuses confusions.

Nous allons introduire ces trois mod\`eles, et expliquer comment ils sont reli\'es entre eux (le mod\`ele 1 et le mod\`ele 2 sont reli\'es par
une transform\'ee de Fourrier, et le mod\`ele 2 et le mod\`ele 3, par leur d\'eveloppement en s\'erie asymptotique).

Avant de les  introduire, il faut rappeler quelques notions de base.

\section{Quelques d\'efinitions utiles}

\subsection{Mesure de Lebesgue pour les matrices hermitiennes}

On d\'efinit la mesure sur $H_N$ ($=$espace des matrices hermitiennes de taille $N$):
\beq\label{defLebesgue}
M\in H_N \qquad \to \quad
\D M:= {1\over U_N}\,\prod_{i=1}^{N} dM_{i,i}
\,\prod_{i<j} d\Re M_{i,j}
\,\prod_{i<j} d\Im M_{i,j}
\eeq
o\`u $U_N$ est le "volume du groupe unitaire":
\beq\label{defUN}
U_N:= {\rm Vol}(U(N)/U(1)^N\times S_N) = {\pi^{N(N-1)/2}\over \prod_{k=1}^N k! }
\eeq

\subsection{Changement de variable: matrice $\leftrightarrow$ valeurs propres $+$ unitaire}

Toute matrice hermitienne $M\in H_N$ peut s'\'ecrire \cite{fulton} (pas de fa\c con unique, i.e. permutation des valeurs propres $S_N$ et phases diagonales $U(1)^N$):
\beq\label{hermdiagU}
M=U\L U^\dagger
\eeq
o\`u $\L=\diag(\l_1,\dots,\l_N)$ est diagonale, et $U\in U(N)$ est unitaire.
On a:
\beq\label{chgtmatricesvp}
\D M= \Delta(\l)^2\, \D U\, \D \L
\eeq
o\`u $\D\L=\prod_i \D\l_i$, $\D U$ est la mesure de Haar normalis\'ee sur $U(N)$,
$\Delta(\l)$ est le d\'eterminant de Vandermonde:
\beq\label{defVandermonde}
\Delta(\l) := \prod_{i>j} (\l_i-\l_j)
\eeq
Pour $N=1$, on d\'efinit $\Delta:=1$.

\subsection{Int\'egrales sur le groupe unitaire:
formules de Itzykson-Zuber-Harish Chandra, Morozov et Eynard Prats Ferrer}

Soient $A=\diag(a_1,\dots,a_N)$ et $B=\diag(b_1,\dots,b_N)$ deux matrices diagonales
non d\'eg\'en\'er\'ees, et $dU$ la mesure de Haar sur le groupe $U(N)$,
on a:

\bt\label{TheoremEynPrats}
Th\'eor\`eme  (Eynard, Prats-Ferrer  \cite{eynprats}):

Pour toute fonction polyn\^omiale invariante (i.e. produit de traces de produits des deux matrices) $F$, on a:
\bea\label{Eynprats}
&& \int_{U(N)}\, F(A,UBU^\dagger)\,\ee{-\Tr A U B U^\dagger}\,dU \cr
&=& {c_N\over N!\,\Delta(a)\Delta(b)}\,\sum_{\sigma,\tau\in S_N}\,(-1)^{\sigma\tau}\,\ee{-\Tr A_\tau B_\sigma}\,\int_{T(N)} \ee{-\Tr T T^\dagger}\,F(A_\tau+T,B_\sigma+T^\dagger)\, dT
\eea
o\`u $T(N)$ est l'ensemble des matrices complexes triangulaires sup\'erieures strictes, muni de la mesure produit des mesures de Lebesgues des parties r\'eelles et imaginaires
de tous les \'el\'ements de matrice.
Si $\sigma$ est une permutation, $B_\sigma$ est la matrice $\diag(b_{\sigma(i)})$, et la constante $c_N$ vaut:
\beq
c_N = {\prod_{k=1}^{N-1} k!\over (-2\pi)^{N(N-1)\over 2}}
\eeq
\et

En particulier avec $F=1$ on a:

\bt Formule Itzykson-Zuber -Harish Chandra \cite{IZ,HC, fulton}:
\beq\label{IZ}
\int_{U(N)} \D U\, \ee{\Tr A U B U^\dagger}
= {\det{E}\over \Delta(a)\Delta(b)}\,\, c_N\,(-\pi)^{N(N-1)\over 2}
\virg
\eeq
o\`u la matrice $E$ est donn\'ee par:
\beq\label{defIn}
E_{ij}:=\ee{a_i b_j}
\point
\eeq
\et

Et, avec $F(A,B)=\Tr {1\over x-A}\,{1\over y-B}$ vue comme s\'erie formelle en puissances de $1/x$ et $1/y$:

\bt\label{TheoremMorozov}
Th\'eor\`eme de Morozov simplifi\'e (eynard \cite{eynmorozov}):

La fonction g\'en\'eratrice d\'efinie par le d\'eveloppement formel en puissances de $1/x$ et $1/y$:
\beq
\left<\tr {1\over x-A}U{1\over y-B}U^\dagger\right> :=\sum_{i,j}{1\over x-a_i}\,{1\over y-b_j}\,\left<|U_{ij}|^2\right>
\eeq
o\`u
\beq\label{defMorozov}
\left<|U_{ij}|^2 \right> := {\int_{U(N)} \D U\,  |U_{ij}|^2\,\ee{\Tr A U B U^\dagger}\over \int_{U(N)} \D U\, \ee{\Tr A U B U^\dagger}} \virg
\eeq
est donn\'ee par:
\bea\label{Morozov}
\left<\tr {1\over x-A}U{1\over y-B}U^\dagger\right>
&=& 1-\det{\left(\1_N - {1\over x-A}E{1\over y-B}E^{-1}\right)}
\eea
\et
Les corr\'elations $\left<|U_{ij}|^2 \right>$ ont \'et\'e calcul\'ees par A. Morozov \cite{Morozov}, puis simplifi\'ee par moi m\^eme dans \cite{eynmorozov},
et utilis\'ee pour le r\'esultat obtenu par Bertola--Eynard dans \citeBEmixed\ et pr\'esent\'e au paragraphe \ref{trmixtepol} dans le cadre de cette habilitation.

\subsection{Ensemble des matrices normales sur un chemin}

\bd
Pour tout contour $\gamma$ du plan complexe,
on d\'efinit l'ensemble de matrices $H_n(\gamma)$:
\beq
H_n(\gamma):= \{ M\in GL_n(\C)\, / \,
\exists \L=\diag(\l_1,\dots,\l_n) \, ,\, \l_i\in \gamma
\, , \, \exists U\in U(n)
\, , \, M= U\L U^\dagger
\}
\eeq
On le munit d'une mesure analogue \`a \refeq{chgtmatricesvp}:
\beq\label{mesurenormale}
\D M= \Delta(\l)^2\, \D U\, \D \L
\eeq
\ed

L'ensemble des matrices hermitiennes est $H_n=H_n(\R)$.

Notons que si $M\in H_n(\gamma)$, on a:
\beq
[M,M^\dagger]=0
\eeq
d'o\`u le nom de matrices normales \cite{WZ}.

\section{Le mod\`ele hermitien et le mod\`ele normal}

Le mod\`ele hermitien est reli\'e \`a deux familles de polyn\^omes biorthogonaux (que l'on \'etudiera dans le chapitre \ref{chapterpol}).
Une g\'en\'eralisation naturelle des polyn\^omes biorthogonaux (dits polyn\^omes semi-classiques, cf \cite{marcopath}),
nous conduit \`a introduire une g\'en\'eralisation du mod\`ele hermitien, que j'app\`elerai le mod\`ele normal sur un contour.

\subsection{Les donn\'ees}

Donnons nous deux polyn\^omes $V_1$ et $V_2$ de degr\'es respectifs $d_1+1$
et $d_2+1$, \`a coefficients complexes, appel\'es {\bf potentiels}:
\beq
V_1(x) = g_{0} + \sum_{k=1}^{d_1+1} {g_k\over k} x^k
\virg
V_2(y) = \td{g}_{0} + \sum_{k=1}^{d_2+1} {\td{g}_k\over k} y^k
\virg
\eeq
un entier $N>{\rm min}\,(d_1,d_2)$,
et un nombre complexe non nul $T$, appel\'e {\bf temp\'erature}.

Il existe $d_1$ chemins homologiquement ind\'ependants allant de $\infty$ \`a $\infty$,
$\Gammax_{k}$, $k=1,\dots,d_1$, tels que l'int\'egrale:
\beq
\ds \int_{\Gammax_{k}} \ee{-{N\over T}V_1(x)}\,dx
\eeq
soit absolument convergente. Il suffit que chaque chemin parte de $\infty$ dans un secteur ou $\Re {V_1(x)\over T}>0$, et revienne
\`a $\infty$ dans un autre secteur o\`u $\Re {V_1(x)\over T}>0$.

De m\^eme, il existe $d_2$ chemins homologiquement ind\'ependants allant de $\infty$ \`a $\infty$,
$\Gammay_{k}$, $k=1,\dots,d_2$, tels que l'int\'egrale:
\beq
\int_{\Gammay_{k}} \ee{-{N\over T}V_2(y)} \,dy
\eeq
converge. Il suffit que chaque chemin parte de $\infty$ dans un secteur ou $\Re {V_2(y)\over T}>0$, et revienne
\`a $\infty$ dans un autre secteur o\`u $\Re {V_2(y)\over T}>0$.

Deux telles bases de chemins \'etant choisies, donnons nous une matrice $\kappa_{i,j}$ \`a coefficients complexes, non identiquement nulle, et d\'efinissons:
\beq\label{defGamma}
\Gamma:=\sum_{i,j} \kappa_{i,j}\,\, \Gammax_{i}\times\Gammay_{j}
\eeq

\br
nous aurions pu consid\'erer un mod\`ele plus g\'en\'eral, o\`u $V_1$ et $V_2$ ne sont pas des polyn\^omes, mais tels que
$V'_1$ et $V'_2$ soient des fractions rationelles.
Dans ce cas, il faudrait introduire des bases d'homologies de tous les contours sur lesquels $\int dx\,\ee{-{N\over T}V_1(x)}$ a un sens.
Cel\`a inclut les chemins allant d'un p\^ole de $V_1$ \`a un autre p\^ole, et cel\`a inclut aussi des segments finis,...
C'est le cas semi-classique introduit par \cite{marcopath}.
Nous  ne le discuterons pas ici par souci de clart\'e et simplicit\'e.
Le cas g\'en\'eral est pr\'esent\'e dans \cite{BEHsemiclas}, il n'est pas tr\`es diff\'erent du cas polynomial pr\'esent\'e ici.
\er

\subsection{Mod\`ele hermitien}

C'est le cas o\`u il est possible de choisir les bases d'homologies
telles que:
\beq\label{GammaRR}
\Gamma=\R\times \R
\eeq
c'est \`a dire si et seulement si les conditions suivantes sont
satisfaites:\par
$d_1$ et $d_2$ sont impairs, et $g_{d_1+1}/T$ et $\td{g}_{d_2+1}/T$ ont une
partie r\'eelle strictement positive.
Et dans le cas o\`u $d_1=d_2=1$, la forme
quadratique ${1\over T}\pmatrix{g_{2}&1\cr 1&\td{g}_{2}}$
a des valeurs propres de parties r\'eelles strictement positives.
Autrement dit si et seulement si
$\Re\left( {1\over T}(V_1(x)+V_2(y)-xy)\right)$ est born\'ee
inf\'erieurement sur $\R\times \R$.

\bigskip
On introduit alors

\bd Mesure du mod\`ele hermitien sur $H_N\times H_N$:
\beq\label{defmodelehermitien}
d\mu(M_1,M_2) :=  {1\over \td{Z}_{\rm Herm}}\,
\ee{-{N\over T}\tr \left[ V_1(M_1)+V_2(M_2)-M_1 M_2 \right]}
\,dM_1\, dM_2
\eeq
o\`u la normalisation $\td{Z}_{\rm Herm}$
\beq\label{defZHerm}
\td{Z}_{\rm Herm} := \ee{-{N^2\over T^2} \td{F}_{\rm Herm}}:=
\int_{H_N\times H_N} \ee{-{N\over T}\tr
\left[ V_1(M_1)+V_2(M_2)-M_1 M_2 \right]}
\,dM_1\, dM_2
\eeq
est appel\'ee fonction de partition,
et $\td{F}_{\rm Herm}$ est appel\'ee \'energie libre.
\ed

En \'ecrivant comme \refeq{hermdiagU}:
\beq
M_1=U\, X\, U^\dagger
\virg
M_2=(U V)\, Y\, (UV)^\dagger
\eeq
avec $X=\diag(x_1,\dots,x_N)$ et $Y=\diag(y_1,\dots,y_N)$, et $U$ et $V$ unitaires,
on a:
\beq
d\mu(M_1,M_2) = {1\over \td{Z}}\, \Delta(x)^2 \Delta(y)^2 \,
\ee{-{N\over T}\tr \left[ V_1(X)+V_2(Y)\right]}\,
\ee{{N\over T}\tr X V Y V^\dagger }\,
\D X \D Y \D U \D V
\eeq
En utilisant \ref{IZ}, l'int\'egrale sur $U$ et $V$ donne une mesure induite pour les valeurs propres $x_i$
de $M_1$ et $y_i$ de $M_2$:
\bd Mesure sur $\R^N\times \R^N$ pour les valeurs propres du mod\`ele hermitien:
\beq\label{mesurevphermitien}
d\nu(x_1,\dots,x_N;y_1,\dots,y_N)
:=  {\det{(\ee{{N\over T}x_i y_j})}\,
\Delta(x)\, \Delta(y)\over N!\,Z_{\rm Herm}}\,
\prod_{i=1}^N \ee{-{N\over T}V_1(x_i)} dx_i\,\,
\prod_{i=1}^N \ee{-{N\over T}V_2(y_i)} dy_i\,\,
\eeq
La normalisation
\bea\label{ZHermNfact}
Z_{\rm Herm}
&:=& {1\over N!}\,\int_{\R^N\times \R^N} \det{(\ee{{N\over T}x_i y_j})}\,
\Delta(x)\, \Delta(y)\,
\prod_{i=1}^N \ee{-{N\over T}[V_1(x_i)+V_2(y_i)]} dx_i\,dy_i\, \cr
&=& \int_{\R^N\times \R^N}
\Delta(x)\, \Delta(y)\, \prod_{i=1}^N \ee{-{N\over T}[ V_1(x_i)+V_2(y_i)-x_i y_i]} dx_i\, dy_i\, \cr
&:=& \ee{-{N^2\over T^2}F_{\rm Herm}}
\eea
est appel\'ee fonction de partition, et $F_{\rm Herm}$ est appel\'ee \'energie libre.
\ed

Notons que:
\beq\label{relZmatZvp}
Z_{\rm Herm} = \td{Z}_{\rm Herm} \,\left( {N\over \pi\, T}\right)^{N(N-1)/2}
\eeq

\br
L'espace des param\`etres du mod\`ele hermitien est de dimension $d_1+d_2+4$:
les $d_1+2$ coefficients de $V_1$, les $d_2+2$ coefficients de $V_2$,
la temperature $T$,
et il faut retrancher la redondance des coefficients constant de $V_1$ et $V_2$.
\er

\br
Le mod\`ele de matrices hermitiennes, est celui qui peut s'appliquer \`a des probl\`emes de physique de la mati\`ere condens\'ee \cite{Guhr, Mehta, courseynard}.
En effet les matrices qui apparaissent dans ces probl\`emes repr\'esentent des Hamiltoniens, et sont toujours hermitiennes
(elles peuvent avoir aussi des sym\'etries supl\'ementaires).
\er

\subsection{Le mod\`ele Normal sur  une classe d'homologie de contours $\Gamma$}
\label{sectiondefmodelenormal}

Il appara\^\i t naturel d'\'etendre la d\'efinition de la fonction de partition $Z_{\rm Herm}$, donn\'ee par
la seconde partie de l'\'equation \refeq{ZHermNfact} \`a n'importe quel contour $\Gamma$
comme d\'efini par \refeq{defGamma}.

On d\'efinit donc, comme g\'en\'eralisation de \refeq{mesurevphermitien},
la mesure pour les valeurs propres:
\bd Mesure sur $\prod_{i=1}^N \Gammax_{k_i}\times  \Gammay_{l_i}$ pour les valeurs propres du mod\`ele normal sur un chemin $\Gamma$:
\bea\label{mesurevpnormal}
&& d\nu_{{\rm Norm},\,k_1,\dots,k_N;l_1,\dots,l_N}(x_1,\dots,x_N;y_1,\dots,y_N) \cr
&&:={\Delta(x)\,\Delta(y)\det{\left(\kappa_{k_i,l_j}\ee{{N\over T}x_i y_j}\right)}\over N! Z_{\rm Norm}(\kappa)}\,\prod_{i=1}^N \ee{-{N\over T}[V_1(x_i)+V_2(y_i)]}\, dx_i\, dy_i\,
\eea
o\`u $Z_{\rm Norm}(\kappa)=\ee{-{N^2\over T^2}F_{\rm Norm}(\Gamma)}$ est la fonction de partition, et $F_{\rm Norm}(\kappa)$ est l'\'energie libre.
\ed
On a:
\bea\label{defZNormal}
Z_{\rm Norm}(\kappa)
&:=&{1\over N!}\,\ee{-{N^2\over T^2}F_{\rm Norm}(\kappa)}\cr
&:=& {1\over N!}\,\sum_{k_1,\dots,k_N,l_1,\dots,l_N} \prod_{i=1}^N\int_{x_i\in\Gammax_{k_i}}\prod_{j=1}^N  \int_{y_j\in\Gammay_{l_j}}\,\,\, \cr
&& \qquad \Delta(x)\,\Delta(y)\det{\left(\kappa_{k_i,l_j}\ee{{N\over T}x_i y_j}\right)}\,\prod_i \ee{-{N\over T}[V_1(x_i)+V_2(y_i)]}\, dx_i\, dy_i\,
\cr
&=& {1\over N!}\,\sum_{\sigma\in S_N} (-1)^\sigma \sum_{k_1,\dots,k_N,l_1,\dots,l_N}  \prod_{i=1}^N\int_{x_i\in\Gammax_{k_i},y_{\sigma_i}\in\Gammay_{l_{\sigma_i}}}\,\,\, \cr
&& \qquad \Delta(x)\,\Delta(y)
\prod_i \kappa_{k_i,l_{\sigma_i}} \ee{{N\over T}x_i y_{\sigma_i}}
\,\prod_i \ee{-{N\over T}[V_1(x_i)+V_2(y_i)]}\, dx_i\, dy_i\,
\cr
\eea
o\`u l'on a d\'ecompos\'e le d\'eterminant comme une somme sur les permutations de $S_N$.
L'int\'egrale \`a effectuer est la m\^eme pour chaque permutation (il suffit de  renommer $y_{\sigma_i}\to y_i$ et $l_{\sigma_i}\to l_i$), et donc:
\bea
Z_{\rm Norm}(\kappa)
&=& \sum_{k_1,\dots,k_N,l_1,\dots,l_N} \prod_i \kappa_{k_i,l_i} \prod_{i=1}^N\int_{x_i\in\Gammax_{k_i},y_i\in\Gammay_{l_i}}\,\,\, \cr
&& \qquad \Delta(x_i)\,\Delta(y_i)
\,\prod_i \ee{-{N\over T}[V_1(x_i)+V_2(y_i)-x_i y_i]}\, dx_i\, dy_i\,
\cr
\eea

\br
Notons que d'autres g\'en\'eralisations du mod\`ele hermitien sont possibles, en particulier, on pourra\^\i t choisir
de mettre des valeurs absolues aux Vandermonde carr\'es.
La d\'efinition ci-dessus est celle qui s'interpr\`ete en termes de polyn\^omes biorthogonaux pr\'esent\'es au chapitre \ref{chapterpol},
et surlaquelle ont port\'e une grande partie de mes travaux.
\er

\br
En rassemblant les paires $(x_i,y_i)$ qui sont sur le m\^eme produit de chemins $(\Gammax_{k}\times \Gammay_{l})$, on peut \'ecrire:
\bea\label{defintxyGammaN}
Z_{\rm Norm}(\kappa)
&=& \sum_{\sum_{{k,l}} n_{{k,l}}=N}\, {N!\over \prod_{{k,l}} n_{{k,l}}!} \prod_{{k,l}} \kappa_{{k,l}}^{n_{{k,l}}}\,
\prod_{k,l} \prod_{i=1}^{n_{k,l}} \int_{x_{{k,l},i}\in \Gammax_{k}} \int_{y_{{k,l},i}\in \Gammay_{l}} \cr
&& \qquad \quad \Delta(x)\, \Delta(y)\,
\prod_{k,l} \prod_{i=1}^{n_{k,l}} \ee{-{N\over T}[V_1(x_{{k,l},i})+V_2(y_{{k,l},i})-x_{{k,l},i} y_{{k,l},i}]} dx_{{k,l},i}\,dy_{{k,l},i} \cr
&=& \sum_{\sum_{{k,l}} n_{{k,l}}=N}\, {N!\over \prod_{{k,l}} n_{{k,l}}!} \prod_{{k,l}} \kappa_{{k,l}}^{n_{{k,l}}}\,
\prod_{k,l} \prod_{i=1}^{n_{k,l}} \int_{x_{{k,l},i}\in \Gammax_{k}} \int_{y_{{k,l},i}\in \Gammay_{l}} \cr
&& \qquad \quad \Delta(x)\, \Delta(y)\,
{\det{\ee{{N\over T} x_{{k,l},i}y_{{k,l},j}}}\over n_{k,l}!}
\,\prod_{k,l} \prod_{i=1}^{n_{k,l}} \ee{-{N\over T}[V_1(x_{{k,l},i})+V_2(y_{{k,l},i})]} dx_{{k,l},i}\,dy_{{k,l},i} \cr
\eea
et l'on reconna\^\i t des int\'egrales sur $U_{n_{k,l}}$:
\bea\label{Znormmultimat}
Z_{\rm Norm}(\kappa)
&=& N!\,\sum_{\sum_{{k,l}} n_{{k,l}}=N}\,  \prod_{{k,l}} {\kappa_{{k,l}}^{n_{{k,l}}}\over  n_{{k,l}}! \,\pi^{n_{k,l}(n_{k,l}-1)\over 2}}\,
  \int_{M_{x,{k,l}}\in H_{n_{k,l}}(\Gammax_{k})} \int_{M_{y,{k,l}}\in H_{n_{k,l}}(\Gammay_{l})} \cr
&& \qquad \quad
\,\prod_{k,l}  \ee{-{N\over T}\tr [V_1(M_{x,{k,l}})+V_2(M_{y,{k,l}})-M_{x,{k,l}} M_{y,{k,l}}]} dM_{x,{k,l}}\,dM_{y,{k,l}} \cr
&& \prod_{(k,l)>(k',l')} \det{(M_{x,k,l}\otimes \1_{n_{k',l'}}-\1_{n_{k,l}}\otimes M_{x,k',l'})} \cr
&& \prod_{(k,l)>(k',l')} \,\det{(M_{y,k,l}\otimes \1_{n_{k',l'}}-\1_{n_{k,l}}\otimes M_{y,k',l'})} \cr
\eea
o\`u l'on a ordonn\'e les paires $(k,l)$ par ordre lexicographique.
\er

\Remark
L'espace des modules du mod\`ele normal est de dimension $d_1+d_2+3+d_1 d_2$.
Les modules sont les $d_1+2$ coefficients de $V_1$, les $d_2+2$ coefficients de $V_2$,
la temperature $T$, et les $d_1 d_2$ coefficients $\kappa_{k,l}$,
et il faut retrancher la redondance des coefficients constant de $V_1$ et $V_2$ et un facteur global pour $\kappa$.

\section{Le mod\`ele Normal \`a sym\'etrie bris\'ee}

Nous allons consid\'erer individuellement chaque terme de la somme \refeq{Znormmultimat}, autrement
dit on se place \`a nombre de valeurs propres fix\'e sur chaque contour.
Cela revient \`a briser le groupe de symm\'etrie $U(N)$ en $\prod_{k,l} U(n_{k,l})$.
Comme nous le verrons au chapitre \ref{chapterasymp}, ce mod\`ele est tr\`es utile pour faire le lien entre polyn\^omes biorthogonaux et d\'eveloppement \`a $N$ grand.

\subsection{Les donn\'ees}

On se donne deux entiers $d_1$ et $d_2$ sup\'erieurs ou \'egaux \`a $1$, et un entier $N>{\rm min}\,(d_1,d_2)$.

On se donne $d_1\times d_2$ entiers $n_{k,l}$, $1\leq k\leq d_1,\,  1\leq l\leq d_2$
tels que
\beq
\sum_{k,l} n_{k,l}=N
\eeq
et, pour des raisons qui deviendront claires au chapitre \ref{chapterloop}, on note:
\beq
\epsilon_{k,l} := {n_{k,l}\over N}
\eeq
Les $\epsilon_{k,l}$ sont appel\'es {\bf fractions de remplissage}.
On note
\beq
n:=\{n_{k,l}\} \qquad {\rm et} \qquad \epsilon:=\{\epsilon_{k,l}\}
\eeq

Donnons nous aussi deux polyn\^omes $V_1$ et $V_2$ de degr\'es respectifs $d_1+1$
et $d_2+1$, \`a coefficients complexes, appel\'es {\bf potentiels}:
\beq
V_1(x) = g_{0} + \sum_{k=1}^{d_1+1} {g_k\over k} x^k
\virg
V_2(y) = \td{g}_{0} + \sum_{k=1}^{d_2+1} {\td{g}_k\over k} y^k
\virg
\eeq
et un nombre complexe non nul $T$ appel\'e {\bf temp\'erature}.

Et on se donne des bases de classes d'homologies de chemins dans $\C$ d\'efinies comme au paragraphe pr\'ec\'edent:
\beq
\Gammax_{k}\,\, k=1,\dots, d_1
\virg
\Gammay_{l}\,\, l=1,\dots, d_2
\eeq

\subsection{Le mod\`ele}

\bd Mesure du mod\`ele normal \`a sym\'etrie bris\'ee sur $\prod_{k,l} H_{n_{k,l}}(\Gammax_{k})\times H_{n_{k,l}}(\Gammay_{l})$:
\bea\label{defmsurematnormb}
d\mu(M_{x,k,l};M_{y,k,l})
&:=&
{1\over \td{Z}_{\rm Normb}(n)}\prod_{k,l}  \ee{-{N\over T}\tr [V_1(M_{x,{k,l}})+V_2(M_{y,{k,l}})-M_{x,{k,l}} M_{y,{k,l}}]} dM_{x,{k,l}}\,dM_{y,{k,l}} \cr
&& \prod_{(k,l)<(k',l')} \det{(M_{x,k,l}\otimes \1_{n_{k',l'}}-\1_{n_{k,l}}\otimes M_{x,k',l'})} \cr
&& \prod_{(k,l)<(k',l')} \,\det{(M_{y,k,l}\otimes \1_{n_{k',l'}}-\1_{n_{k,l}}\otimes M_{y,k',l'})} \cr
\eea
o\`u les paires $(k,l)$ sont ordonn\'ees par ordre lexicographique,
et o\`u la fonction de partition et l'\'energie libre sont donn\'ees par:
\bea\label{defZmatNormb}
\td{Z}_{\rm Normb}(n) & := & \ee{-{N^2\over T^2}F_{\rm Normb}(n)}  \cr
& := & \prod_{k,l} \int_{M_{x,k,l}\in H_{n_{k,l}}(\Gammax_{k})} \int_{M_{y,k,l}\in H_{n_{k,l}}(\Gammay_{l})} \cr
&& \prod_{k,l}\,  \ee{-{N\over T}\tr [V_1(M_{x,{k,l}})+V_2(M_{y,{k,l}})-M_{x,{k,l}} M_{y,{k,l}}]} dM_{x,{k,l}}\,dM_{y,{k,l}} \cr
&& \prod_{(k,l)<(k',l')} \det{(M_{x,k,l}\otimes \1_{n_{k',l'}}-\1_{n_{k,l}}\otimes M_{x,k',l'})} \cr
&& \prod_{(k,l)<(k',l')} \,\det{(M_{y,k,l}\otimes \1_{n_{k',l'}}-\1_{n_{k,l}}\otimes M_{y,k',l'})} \cr
\eea
\ed
En utilisant la d\'ecomposition \refeq{hermdiagU}, et en int\'egrant sur les groupes unitaires avec la formule \refeq{IZ},
on obtient la mesure induite pour les valeurs propres
\bd Mesure pour les valeurs propres du mod\`ele normal \`a sym\'etrie bris\'ee,
sur $\prod_{k,l} \left(\Gammax_{k}\times\Gammay_{l}\right)^{n_{k,l}}$
(i.e. $\forall k=1,\dots, d_1\,\, , \, \forall l=1,\dots, d_2 \,\, , \, \forall i=1,\dots,n_{k,l} \,\, , \, x_{k,l,i}\in\Gammax_{k} \,\, , \, y_{k,l,i}\in\Gammay_{l}$):
\bea\label{defdnunormb}
&& d\nu_{\rm Normb}(x_{k,l,i};y_{k,l,i}) \cr
& := &
{\Delta(x)\, \Delta(y)\,
\over Z_{\rm Normb}(n)}
\,\prod_{k,l} \left( \det{\left(\ee{{N\over T}x_{k,l,i}y_{k,l,j}}\right)} \, \prod_{i=1}^{n_{k,l}} \ee{-{N\over T}[V_1(x_{{k,l},i})+V_2(y_{{k,l},i})]} \,dx_{k,l,i}\,dy_{k,l,i}\right) \cr
\eea
o\`u la fonction de partition et l'\'energie libre sont donn\'ees par:
\bea\label{defZNormb}
Z_{\rm Normb}(n) & := & \ee{-{N^2\over T^2}F_{\rm Normb}(n)}  \cr
& := &
\prod_{k,l} {1\over n_{{k,l}}!} \prod_{i=1}^{n_{k,l}} \int_{x_{{k,l},i}\in \Gammax_{k}} \int_{y_{{k,l},i}\in \Gammay_{l}} \cr
&& \qquad \quad \Delta(x)\, \Delta(y)\,
\,\prod_{k,l} \prod_{i=1}^{n_{k,l}} \ee{-{N\over T}[V_1(x_{{k,l},i})+V_2(y_{{k,l},i})-x_{{k,l},i} y_{{k,l},i}]} dx_{{k,l},i}\,dy_{{k,l},i} \cr
\eea
\ed

\subsection{Relation avec le mod\`ele Normal}

Il est clair que le mod\`ele normal est tel que:

\beq\label{ZNormNormb}
Z_{\rm Norm}(\kappa)=N!\, \sum_{n}
\left(\prod_{k,l} \kappa_{k,l}^{n_{k,l}} \right)\,\,
Z_{\rm Normb}(n)
\eeq
Autrement dit, les $\kappa_{k,l}$ sont les variables duales des $n_{k,l}$,
et les deux mod\`eles sont reli\'es par une transform\'ee de Fourrier (avec la contrainte $\sum n_{k,l}=N$).

\Remark
L'espace des modules du mod\`ele normal bris\'e est de dimension $d_1+d_2+3+d_1 d_2$,
comme le mod\`ele normal.
Les $d_1 d_2-1$ coefficients $\kappa_{k,l}$ (moins la normalisation redondante)
ont \'et\'es remplac\'es par les $d_1 d_2-1$ fractions de remplissages ind\'ependantes $\epsilon_{k,l}$.
D'apr\`es \refeq{ZNormNormb}, les $\kappa_{k,l}$ apparaissent comme des variables duales des
$\epsilon_{k,l}$.

\section{Le mod\`ele formel}
\label{pardefmodeleformel}

Le mod\`ele formel pourra\^it aussi \^etre appel\'e mod\`ele combinatoire.
En effet, il n'est pas d\'efini par une int\'egrale de matrices (ni de valeurs propres), mais par une s\'erie formelle en puissances de la temp\'erature $T$,
qui sert de s\'erie g\'en\'eratrice pour le d\'enombrement de certains types de graphes \cite{DGZ, BIPZ, Matrixsurf, EML, gross:1991}.
Pour ce mod\`ele, les questions de convergences et de bases d'homologies ne se posent pas.

\subsection{Les donn\'ees}

On se donne deux entiers $d_1$ et $d_2$ sup\'erieurs ou \'egaux \`a $1$.
Donnons nous aussi deux polyn\^omes $V_1$ et $V_2$ de degr\'es respectifs $d_1+1$
et $d_2+1$, \`a coefficients complexes appel\'es {\bf potentiels}:
\beq
V_1(x) = g_{0} + \sum_{k=1}^{d_1+1} {g_k\over k} x^k
\virg
V_2(y) = \td{g}_{0} + \sum_{k=1}^{d_2+1} {\td{g}_k\over k} y^k
\virg
\eeq
et deux nombres complexes non nuls $T$ et $N$. $T$ est appel\'e {\bf temp\'erature}.

Le potentiel $V_1(x)+V_2(y)-xy$ poss\`ede $d_1 d_2$ extrema dans $\C\times\C$, que l'on note:
$(x_{I},y_{I})$, $I=1,\dots,d_1 d_2$, tels que:
\beq
\bacc
V'_1(x_{I})=y_{I}\cr
V'_2(y_{I})=x_{I}
\eacc
\eeq
et on se donne $d_1 d_2$ nombres complexes arbitraires $\epsilon_I \,\, , \, I=1,\dots,d_1 d_2$, que l'on appelle {\bf fractions de remplissages}, tels que:
\beq\label{epsilondataformel}
\sum_{I=1}^{d_1 d_2} \epsilon_I=1
\eeq

\bd on d\'efinit les {\bf constantes de couplages}:
\beq
\begin{array}{llrcl}
\forall I=1,\dots,d_1 d_2 & \forall k=2,\dots,d_1+1 & g_{k,I}&:=&{V_1^{(k)}(x_I)\over k-1!}  \cr
\forall I=1,\dots,d_1 d_2 & \forall k=2,\dots,d_2+1 & \td{g}_{k,I}&:=&{V_2^{(k)}(y_I)\over k-1!}  \cr
\forall i\neq j & \forall k=1,\dots,\infty,  & h_{k,i}&:=& \sum_{J\neq I} {\epsilon_j\over (x_J-x_I)^k} \cr
\forall i\neq j & \forall k=1,\dots,\infty,  & \td{h}_{k,i}&:=& \sum_{J\neq I} {\epsilon_j\over (y_J-y_I)^k} \cr
\forall i\neq j & \forall k,l=1,\dots,\infty,  & h_{k,I;l,J}&:=& {k+l-1!\over k-1!\, l-1!}\,{1\over (x_J-x_I)^k\, (x_I-x_J)^l} \cr
\forall i\neq j & \forall k,l=1,\dots,\infty,  & \td{h}_{k,I;l,J}&:=& {k+l-1!\over k-1!\, l-1!}\,{1\over (y_J-y_I)^k\, (y_I-y_J)^l} \cr
\end{array}
\eeq
\ed
Notons que:
\beq
g_{2,I}\td{g}_{2,I}-1
= \td{g}_{d_2+1} g_{d_1+1}^{d_2} \prod_{J\neq I} (x_I-x_J)
= g_{d_1+1} {\td{g}}_{d_2+1}^{d_1} \prod_{J\neq I} (y_I-y_J)
\eeq

\br
Notons, que contrairement au mod\`ele pr\'ec\'edent, $N$ et $n_{I}=N\epsilon_{I}$ ne sont pas n\'ec\'essairement des entiers.
\er

\subsection{Combinatoire de surfaces discr\'etis\'ees}

\bd
Soit ${\cal G}$ l'ensemble des graphes ferm\'es (sans bords),
pas n\'ec\'essairement connexes,
form\'es de polygones orient\'es (une face sup\'erieure et une face inf\'erieure),
coll\'es le long de leur bord ou par leurs centres, avec les r\`egles suivantes:
Chaque polygone \`a $k$ cot\'es ($k\geq 1$ on peut avoir des $1$-gones et des $2$-gones) porte un ''signe'' ($+$ ou $-$) et une ''couleur'' ($I=1,\dots,d_1 d_2$).
Deux polygones peuvent \^etre coll\'es par leur bord seulement si ils ont la m\^eme couleur, (et en respectant l'orientation des faces).
Deux polygones peuvent etre coll\'es par leurs centres seulement si ils sont de couleur diff\'erente et de m\^eme  signe.
\ed

\bd
Pour un graphe $G\in {\cal G}$, on note:\\
$\bullet$ $n_{k,I}(G)=$ nombre de $k-$gones de signe $+$ et de couleur $I$.\\
$\bullet$ $\td{n}_{k,I}(G)=$ nombre de $k-$gones de signe $-$ et de couleur $I$.\\
$\bullet$ $n_{p,I}(G)=$ nombre d'arr\^etes communes \`a deux polygones de couleur $I$ et de signes quelconques.\\
$\bullet$ $n_{++,I}(G)=$ nombre de paire de polygones de signe $+$ et $+$, de couleur $I$ coll\'es par un bord.\\
$\bullet$ $n_{--,I}(G)=$ nombre de paire de polygones de signe $-$ et $-$, de couleur $I$ coll\'es par un bord.\\
$\bullet$ $n_{k,I;l,J}(G)=$ nombre de paire de polygones form\'ees d'un $k-$gone de signe $+$ et de couleur $I$, et d'un $l-$gone de signe $+$ et de couleur $J$, coll\'es par leurs centres.\\
$\bullet$ $\td{n}_{k,I;l,J}(G)=$ nombre de paire de polygones form\'ees d'un $k-$gone de signe $-$ et de couleur $I$, et d'un $l-$gone de signe $-$ et de couleur $J$, coll\'es par leurs centres.\\
$\bullet$ $l_{I}(G)=$ nombre de sommets de couleur $I$.\\
$\bullet$ $\#{\rm Aut}(G)=$ cardinal du groupe des automorphismes de $G$.\\
$\bullet$ $\chi(G)=$ Caract\'eristique d'Euler-Poincar\'e de $G$, en comptant les liens entre centres de polygones comme des cylindres. $\chi(G)$ est toujours pair.
\beq
\chi(G)=\sum_I l_I(G) - n_p(G) + \sum_k\sum_I n_{k,I}(G)+\td{n}_{k,I}(G)
\eeq
$\bullet$ et $n_T(G)$ est donn\'e par:
\bea\label{defnTG}
n_T(G)
&:=&\sum_I \left(\sum_{k=1}^{\infty} {k\over 2} (n_{k,I} + \td{n}_{k,I})-\sum_{k=3}^{d_1+1} n_{k,I} -\sum_{k=3}^{d_2+1}  \td{n}_{k,I} \right) \cr
&& +\sum_{I<J}\sum_{k\geq 1}\sum_{l\geq 1} {k+l\over 2}(n_{k,I;l,J}+\td{n}_{k,I;l,J})
\eea
\ed

\br
On a l'in\'egalit\'e:
\bea\label{nTGinequ}
n_T(G)
&\geq &{1\over 2}\left(\sum_I\sum_{k=1}^{\infty} (n_{k,I} + \td{n}_{k,I})+\sum_{I<J}\sum_{k\geq 1}\sum_{l\geq 1} (n_{k,I;l,J}+\td{n}_{k,I;l,J})
 \right)
\eea
\er

\bd Le poids de Feynmann d'un graphe $G$ est:
\bea
{\cal W}(G)
&:=& {N^{\chi(G)}\over \#{\rm Aut}(G)} T^{n_T(G)}
\prod_{I=1}^{d_1 d_2}  \prod_{k=3}^{d_1+1}(g_{k,I}+Th_{k,I})^{n_{k,I}(G)}\prod_{k=3}^{d_2+1} ({\td{g}}_{k,I}+T\td{h}_{k,I})^{\td{n}_{k,I}(G)} \cr
&& \prod_{I=1}^{d_1 d_2} h_{1,I}^{n_{1,I}(G)}  h_{2,I}^{n_{2,I}(G)} \prod_{k>d_1+1} h_{k,I}^{n_{k,I}(G)}\,\,\,
 \prod_{I=1}^{d_1 d_2} \td{h}_{1,I}^{\td{n}_{1,I}(G)}  \td{h}_{2,I}^{\td{n}_{2,I}(G)} \prod_{k>d_2+1} \td{h}_{k,I}^{\td{n}_{k,I}(G)} \cr
&& \prod_{I<J}  \prod_{k=1}^{\infty}\prod_{l=1}^{\infty}  h_{k,I;l,J}^{n_{k,I;l,J}(G)} \td{h}_{k,I;l,J}^{\td{n}_{k,I;l,J}(G)}    \prod_I \epsilon_I^{l_I(G)}
 \prod_I  {\td{g}_2}^{n_{++I}}\, {g_2}^{n_{--I}} \, (g_{2,I}\td{g}_{2,I}-1)^{-n_{p,I}(G)} \cr
\eea
\ed

\subsection{Le mod\`ele}

On d\'efinit la fonction de partition du mod\`ele formel comme une s\'erie g\'en\'eratrice (au sens combinatoire), i.e. une s\'erie formelle en puissances de $T$:
\bd\label{definitionZformel}
 Fonction de partition du mod\`ele formel
\bea\label{defZformdevFeyn}
Z_{\rm Form}(\epsilon)
&:=& \prod_I \ee{-{N^2\over T}\epsilon_I \left(V_1(x_I)+V_2(y_I)-x_I y_I\right)}\,
\prod_{I<J} (x_I-x_J)^{{N^2}\epsilon_I \epsilon_J}\prod_{I<J} (y_I-y_J)^{{N^2}\epsilon_I \epsilon_J}  \cr
&&
\sum_{G\in {\cal G}} {\cal W}(G)
\eea
\ed
Notons que l'in\'egalit\'e \ref{nTGinequ} implique que pour chaque puissance de $T$, seul un nombre fini de graphes contribuent,
et donc \ref{defZformdevFeyn} d\'efinit bien une s\'erie formelle en puissances de T:
\beq
Z_{\rm Form}
= \sum_{n=0}^\infty A_n T^n
\eeq
On d\'efinit aussi l'\'energie libre formelle:
\bea
F_{\rm Form}
&:= & -{T^2\over N^2}\ln{Z_{\rm Form}} \cr
&=& \sum_I T\epsilon_I \left(V_1(x_I)+V_2(y_I)-x_I y_I\right)\, -T^2\sum_{I<J} {\epsilon_I \epsilon_J}\ln{(x_I-x_J)}\, -T^2\sum_{I<J} {\epsilon_I \epsilon_J}\ln{(y_I-y_J)}  \cr
&&
-{T^2\over N^2}\sum_{G\in {\cal G}_{\rm conn}} {\cal W}(G)
\eea
qui est aussi une s\'erie formelle en puissances de $T$, qui est la m\^eme somme que pour $Z_{\rm Form}$, mais r\'eduite aux diagrammes connexes seulement:
\beq\label{devFTn}
F_{\rm Form} = \sum_{n=0}^\infty B_n T^n
\eeq

\subsection{Relation avec le mod\`ele normal \`a sym\'etrie bris\'ee}

Ces d\'efinitions formelles viennent de l'interpr\'etation combinatoire de \ref{defZmatNormb} avec $n_k = N\epsilon_k$, par la m\'ethode des diagrammes de Feynman,
comme s\'erie g\'en\'eratrice de diagrammes (ces graphes sont introduits dans \cite{BDE}).
En effet, on r\'eecrit \refeq{defZmatNormb} sous la forme:
\bea\label{Zformbyblocks}
Z_{\rm Normb}(n)
& = &
\int\, \prod_{I=1}^{d_1 d_2} \,  \D{M_{x,I}} \D{M_{y,I}}\, \ee{-{N\over T}\Tr[V_1(M_{x,I})+V_2(M_{y,I})-M_{x,I}M_{y,I}]} \cr
&& \prod_{I<J}\, \det{(M_{x,I}\otimes \1_{n_{J}}-\1_{n_{I}}\otimes M_{x,J})}\,
\, \det{(M_{y,I}\otimes \1_{n_{J}}-\1_{n_{I}}\otimes M_{y,J})} \cr
& = &
\int\, \prod_{I=1}^{d_1 d_2} \,  \D{M_{x,I}} \D{M_{y,I}}\, \ee{-S} \cr
\eea
puis on pose $M_{x,I}=x_I \1_{n_I} + X_I$, $M_{y,I}=y_I \1_{n_I} + Y_I$, et l'action $S$, \'ecrite en termes des matrices $X_I$ et $Y_I$ est:
\bea
S&:=&{N\over T}\sum_I \left(  {g_{2,I} \over 2}\tr X_I^2+ {\td{g}_{2,I} \over 2}\tr Y_I^2 -  \tr X_I Y_I\right) \cr
&& +{N\over T}\sum_I \left( \sum_{k\geq 3} {g_{k,I} \over k}\tr X_I^k+ \sum_{k\geq 3} {\td{g}_{k,I} \over k}\tr Y_I^k \right) \cr
&& +{N\over T}\sum_I \left( \sum_{k\geq 1} {h_{k,I} \over k}\tr X_I^k+ \sum_{k\geq 1} {\td{h}_{k,I} \over k}\tr Y_I^k \right) \cr
&& + \sum_{I<J}\sum_{k\geq 1}\sum_{l\geq 1} {h_{k,I;l,J}\over k l} \tr X_I^k \tr X_J^l \cr
&& + \sum_{I<J}\sum_{k\geq 1}\sum_{l\geq 1} {\td{h}_{k,I;l,J}\over k l} \tr Y_I^k \tr Y_J^l \cr
\eea
La m\'ethode de Feynmann consiste \`a d\'evelopper en s\'erie de Taylor l'exponentielle de tous les termes sauf la partie quadratique de la premi\`ere ligne,
puis \`a calculer les int\'egrales gaussiennes restantes par le th\'eor\`eme de Wick \cite{thooft:1974, BIPZ, Matrixsurf, DGZ, courseynard}.
Les chemins d'int\'egration ne jouent aucun r\^ole ici. Seuls les voisinages des points $x_I$, $y_I$  jouent un r\^ole.
La m\'ethode de Feynman associe \`a chaque terme du d\'eveloppement, un diagramme $G\in {\cal G}$, avec un poids qui est celui \'ecrit plus  haut.
Ceci est pr\'esent\'e en appendice de \cite{BDE}.

Toutefois, il faut savoir que, m\^eme lorsque $N$ et les $n_I=N\epsilon_I$ sont des entiers, les d\'efinitions \ref{defZformdevFeyn} et \ref{defZmatNormb} ne coincident pas toujours.

Elles ne coincident que dans certains cas (pour des ensembles ouverts de potentiels $V_1$, $V_2$, pour certaines valeurs de fractions de remplissage, et pour des choix appropri\'es de contours $\Gammax_{k}$ et $\Gammay_{l}$).
Il faut en particulier que l'on puisse trouver une base de  chemins $\Gammax_{k},\Gammay_{l}$ telle que chaque paire de chemins passe par un seul
extremum $(x_I,y_I)$ et telle que le potentiel effectif pour une paire de valeur propre de $(M_1,M_2)$ ait un unique minimum.

Lorsqu'elles coincident, \ref{defZformdevFeyn} est le d\'eveloppement limit\'e de \ref{defZmatNormb} en puissances de $T$.
En g\'en\'eral, ce d\'eveloppement est un d\'eveloppement asymptotique, qui ne converge pas vers une fonction unique \cite{DGZ, EML}.

\subsection{D\'eveloppement topologique}

Pour chaque puissance de $T$, la s\'erie \ref{devFTn} compte un nombre fini de diagrammes connexes, et donc chaque
$B_n$ est un polyn\^ome en $1/N^2$:
\beq
B_n(N) = \sum_{k=0}^{\deg B_n} B_{n,h} N^{-2h}
\eeq
Ceci autorise \`a d\'efinir les s\'eries \`a puissances de $N$ fix\'ee (i.e. \`a topologie fix\'ee puisque la puissance de $N$ est la caract\'eristique d'Euler).
\beq
F_{\rm Form}^{(h)}:=\sum_{n=0}^\infty B_{n,h} T^n
\eeq
Chaque $F^{(h)}$ est une s\'erie formelle en puissances de $T$, qui compte les diagrammes de ${\cal G}$ avec le m\^eme poids que pour $Z_{\rm Form}$,
restreints aux diagrammes connexes de genre $h$ \cite{thooft:1974, DGZ, courseynard}.
Ainsi, par d\'efinition du mod\`ele, on a le d\'eveloppement de l'\'energie libre en puissances de $N^{-2}$:
\bd D\'eveloppement topologique de l'\'energie libre du mod\`ele formel:
\beq\label{devtopo}
F_{\rm Form} = \sum_{h=0}^\infty F_{\rm Form}^{(h)}\, N^{-2h}
\eeq
\ed
Il s'av\`ere que chaque $F_{\rm Form}^{(h)}$ est une fonction alg\'ebrique de tous ses param\`etres (les coefficients de $V_1$ et $V_2$, les $\epsilon_k$, $T$), et donc
poss\`ede un rayon de covergence non nul comme s\'erie en puissances de $T$.
Une partie de mon travail pr\'esent\'e pour cette habilitation,
a consist\'e \`a calculer $F_{\rm Form}^{(1)}$, et \`a l'identifier avec le d\'eterminant du Laplacien sur une courbe alg\'ebrique \citeeynmatgzero, \citeeynmatsg.
Au chapitre \ref{chapterloop} nous  verrons l'expression de $F_{\rm Form}^{(0)}$ et $F_{\rm Form}^{(1)}$.

Par contre, la somme \refeq{devtopo} est une s\'erie asymptotique en puissances de $1/N^2$, et, en g\'en\'eral, ne se resomme pas en une fonction analytique des param\`etres.

\Remark
L'espace des modules du mod\`ele formel est de dimension $d_1+d_2+3+d_1 d_2$,
comme le mod\`ele normal et le mod\`ele normal bris\'e.

\Remark
Le mod\`ele formel est celui qui est utilis\'e pour les applications des matrices al\'eatoires au d\'enombrement des
surfaces discr\'etis\'ees, et subs\'equement, pour les applications aux th\'eories des champs conformes,
\`a la gravitation quantique et \`a la th\'eorie des cordes en physique \cite{DGZ, BIPZ, Matrixsurf, gross:1991, courseynard, Dijgrafvafa}.

\section{Observables}
\label{chapterdefobservables}

Nous allons maintenant d\'ecrire les quantit\'es que l'on cherche \`a calculer dans ces mod\`eles.
Lorsque cel\`a est possible, on cherche bien s\^ur \`a calculer ces quantit\'es pour tout $N$,
mais l'objectif principal en physique est de les calculer pour $N$ grand, dans divers r\'egimes.

\subsection{Fonction de partition, \'energie libre et moments}

La fonction de partition et l'\'energie libre sont des observables fondamentales.
Elles ont \'et\'e d\'efinies pr\'ec\'edement pour les trois types de mod\`eles de matrices.

Il est conjectur\'e\footnote{Cel\`a a \'et\'e prouv\'e dans le cas d'une seule matrice \citeBEHtauiso.} que la fonction de partition du mod\`ele normal est une fonction $\tau$ isomonodromique
au sens de Jimbo-Miwa-Ueno \cite{MiwaJimbo, JM, sato, uenotak}, et il est connu que la fonction de partition du mod\`ele formel est
une fonction tau KP \cite{DGZ}.
Dans les trois mod\`eles, l'\'energie libre a une limite $N$ grand, reli\'ee \`a la hi\'erarchie de Toda ''sans dispersion'' \cite{DGZ, ZinnZuber}.

Les d\'eriv\'ees de la fonction de partition par rapport aux coefficients des potentiels donnent acc\`es aux moments:
\bea\label{defobstrsimples}
T_{k_1;\dots;k_r;\ovl{l_1};\dots;\ovl{l_s}}
& := &
N^{-r-s} \left<
\tr M_1^{k_1}\dots \tr M_1^{k_r} \tr M_2^{l_1}\dots \tr M_2^{l_s} \right> \cr
& = & \left(-{T\over N^2}\right)^{r+s}\,{\prod_{i=1^r} k_i \prod_{i=1}^s l_i\over Z}\,
\d_{g_{k_1}}\,\dots \d_{g_{k_r}}\,\d_{\td{g}_{l_1}}\,\dots \d_{\td{g}_{l_s}}\, Z \cr
\eea
Souvent, il est pr\'ef\'erable de calculer les cumulants:
\bea\label{defobstrsimplesconn}
T_{{\rm conn.}\,k_1;\dots;k_r;\ovl{l_1};\dots;\ovl{l_s}}
& := &
N^{r+s-2} \left<
\tr M_1^{k_1}\dots \tr M_1^{k_r} \tr M_2^{l_1}\dots \tr M_2^{l_s} \right>_{\rm conn.} \cr
& = & -\left(-T\right)^{r+s-2}\,{\prod_{i=1^r} k_i \prod_{i=1}^s l_i}\,
\d_{g_{k_1}}\,\dots \d_{g_{k_r}}\,\d_{\td{g}_{l_1}}\,\dots \d_{\td{g}_{l_s}}\, F \cr
\eea
Dans le mod\`ele Normal et Normal-bris\'e, les $T_{{\rm conn.}\,k_1;\dots;k_r;\ovl{l_1};\dots;\ovl{l_s}}$ sont des int\'egrales convergentes.
Dans le mod\`ele formel, les $T_{{\rm conn.}\,k_1;\dots;k_r;\ovl{l_1};\dots;\ovl{l_s}}$ sont des fonctions g\'en\'eratrices
de graphes ouverts \cite{DGZ, courseynard}, avec $r$ bords de signe $+$, et $s$ bords de signe $-$.

\subsection{Valeurs moyennes de traces mixtes}
\label{obstrmixtes}

Notons que les d\'eriv\'ees de la fonction de partition et de l'\'energie libre donnent acc\`es seulement aux moments non mixtes, i.e.
aux valeurs moyennes de produits de traces, telles que chaque trace ne contienne qu'un seul type de matrice ($M_1$ ou $M_2$).

Pour les trois mod\`eles de matrices, on peut souhaiter calculer la valeur moyenne de toute traces de la forme:
\bea\label{defcormix}
&& T_{k_{1,1},\ovl{l_{1,1}},\dots,k_{1,r_1},\ovl{l_{1,r_1}};\dots;
k_{r,1},\ovl{l_{r,1}},\dots,k_{r,r_r},\ovl{l_{r,r_r}}} \cr
&& := N^{-\sum_i r_i}\,
\left<
\prod_{s=1}^r
\Tr \left(M_1^{k_{s,1}}M_2^{l_{s,1}}\dots M_1^{k_{s,r_s}}M_2^{l_{s,r_s}}\right)
\right>
\eea
dont l'exemple le plus simple est:
\beq\label{defcormix2}
T_{k,\ovl{l}} := {1\over N}\,
\left<  \Tr M_1^{k}M_2^{l} \right>
\eeq
Ces observables ne peuvent pas \^etre obtenues en d\'erivant l'\'energie libre par rapport
aux coefficients des potentiels.
Elles ne peuvent pas non plus \^etre obtenues \`a partir des densit\'es de valeurs propres
(en effet, on ne peut pas exprimer $\tr M_1^k M_2^l$ \`a partir des valeurs propres de $M_1$ et $M_2$).
Leur calcul est plus difficile, et repr\'esente l'un des d\'efis de ce mod\`ele.

Pour le mod\`ele normal, l'observable \refeq{defcormix2} a \'et\'e calcul\'ee par (Bertola,Eynard) dans \citeBEmixed,
et une g\'en\'eralisation \`a toutes les autres observables mixtes a \'et\'e obtenue par (Eynard, Prats-Ferrer)
dans \cite{eynprats}, en utilisant le th\'eor\`eme II.\ref{TheoremEynPrats}.

\bigskip

Dans le mod\`ele formel,
les observables mixtes du type  \refeq{defcormix} sont les fonctions g\'en\'eratrices
de surfaces ouvertes, dont les bords peuvent porter plusieurs signes, autrement dit,
avec des conditions de bords non-triviales \cite{DGZ}.
L'\'etude des op\'erateurs de bords est un sujet actuel et actif en physique des th\'eories
conformes et en th\'eorie des cordes, et o\`u beaucoup reste \`a comprendre \cite{Kostov}.
Les mod\`eles de matrices pourraient apporter des r\'esultats \`a ce domaine.

Une large classe de ces observables (incluant \refeq{defcormix2}) a \'et\'e calcul\'ee dans la limite $N\to\infty$ par (Eynard) dans
\citeeynmultimat\ et dans \cite{eynFsubg0}, et le calcul de toutes les autres observables mixtes est en voie d'\^etre achev\'e \`a ce jour.

\subsection{Densit\'es et corr\'elations de valeurs propres}

Consid\'erons ici uniquement le mod\`ele normal. Nous allons suivre les traces de Mehta \cite{Mehta, MehtaShukla}.

\bigskip

Donnons nous $r$ et $s$ deux entiers compris entre $0$ et $N$ (et l'un au moins non nul),
et des entiers $k_1,\dots, k_r$ compris entre 1 et $d_1$ et $l_1,\dots, l_s$ compris entre 1 et $d_2$.
\bd\label{Defrhors} Densit\'es de probabilit\'es int\'egr\'ees de \refeq{mesurevpnormal},
sur $\left(\prod_{i=1}^r \Gammax_{k_i}\,\,\times \prod_{j=1}^s\Gammay_{l_j}\right)$:
\bea\label{defrhors}
&& \rho_{r;s}^{k_1,\dots,k_r;l_1,\dots,l_s}(x_1,\dots,x_r;y_1,\dots,y_s)\,
dx_1\dots dx_r dy_1\dots dy_s \cr
&& :=
\sum_{k_{r+1},\dots,k_N}\sum_{l_{s+1},\dots,l_N}\,\,
\prod_{i=1}^N
\int_{\forall i>r,\, x_i\in \Gammax_{k_i}} \int_{\forall i>s,\, y_i\in \Gammay_{l_i}} \cr
&& \qquad\quad\,\, d\nu_{{\rm Norm},\,k_1,\dots,k_N;l_1,\dots,l_N}(x_1,\dots,x_N;y_1,\dots,y_N) \cr
\eea
\ed
C'est la probabilit\'e que $x_1\in\Gammax_{k_1},\dots,x_r\in\Gammax_{k_r}$ soient valeurs propres de $M_1$
et $y_1\in\Gammay_{l_1},\dots,y_s\in\Gammay_{l_s}$ soient valeurs propres de $M_2$, simultan\'ement.

\bd Fonctions de corr\'elations de valeurs propres
\bea\label{defRrs}
&& R_{r;s}^{k_1,\dots,k_r;l_1,\dots,l_s}(X_1,\dots,X_r;Y_1,\dots,Y_s) \cr
&& := N^{-r-s}\,
\left< \prod_{i=1}^r \Tr \delta^{(x,k_i)}(X_i-M_1)\,\, \prod_{i=1}^s \Tr \delta^{(y,l_i)}(Y_i-M_2) \right> \cr
&& := N^{-r-s}\,
\sum_{k_{r+1},\dots,k_N}\sum_{l_{s+1},\dots,l_N}\,\,
\prod_{i=1}^N
\int_{x_i\in \Gammax_{k_i}} \int_{y_i\in \Gammay_{l_i}} \cr
&& \prod_{i=1}^r \left(\sum_{j=1}^N \delta(X_i-x_j)\right)\,\,
\prod_{i=1}^s \left(\sum_{j=1}^N \delta(Y_i-y_j)\right) \cr
&& \qquad \qquad d\nu_{{\rm Norm},\,k_1,\dots,k_N;l_1,\dots,l_N}(x_1,\dots,x_N;y_1,\dots,y_N) \cr
\eea
o\`u $\delta^{(x,k)}(x)$ (resp. $\delta^{(y,k)}(y)$) est la distribution $\delta$ de Dirac sur le contour $\Gammax_{k}$
(resp. $\Gammay_{k}$).
\ed
On peut aussi d\'efinir la partie connexe (les cumulants), on la note:
\beq
 R_{{\rm conn.}\,r;s}^{k_1,\dots,k_r;l_1,\dots,l_s}(X_1,\dots,X_r;Y_1,\dots,Y_s)
\eeq

\br
Les deux fonctions $R_{r;s}$ et $\rho_{r;s}$ sont presque \'egales. Lorsque tous les $x_i$ sont distincts et tous les $y_i$ sont
distincts on a:
\beq
R_{r;s}^{k_1,\dots,k_r;l_1,\dots,l_s}(x_1,\dots,x_r;y_1,\dots,y_s)
=
{N!^2 N^{-r-s}\over N-r! N-s!} \rho_{r;s}^{k_1,\dots,k_r;l_1,\dots,l_s}(x_1,\dots,x_r;y_1,\dots,y_s)
\eeq
\er

\br
On peut retrouver les quantit\'es d\'efinies en \refeq{defobstrsimples}:
\bea
&& T_{p_1;\dots;p_r;\ovl{q_1};\dots;\ovl{q_s}} \cr
&&=  \sum_{k_1,\dots,k_r}\sum_{l_1,\dots,l_s} \int_{x_i\in \Gammax_{k_i}}\int_{y_i\in \Gammay_{l_i}}
x_1^{p_1}\dots x_r^{p_r}y_1^{q_1}\dots y_s^{q_s} \cr
&& \qquad \qquad R_{r;s}^{k_1,\dots,k_r;l_1,\dots,l_s}(x_1,\dots,x_r;y_1,\dots,y_s)\, dx_1\dots dx_r dy_1\dots dy_s
\eea
\er

\br
On a la relation de r\'ecurrence \'evidente:
\bea\label{recrhors}
&& \rho_{r-1;s}^{k_1,\dots,k_{r-1};l_1,\dots,l_s}(x_1,\dots,x_{r-1};y_1,\dots,y_s)\,\,
dx_1\dots dx_{r-1} dy_1\dots dy_s \cr
&& =
\sum_{k_r} \int_{x_r\in\Gammax_{k_r}} \rho_{r;s}^{k_1,\dots,k_r;l_1,\dots,l_s}(x_1,\dots,x_r;y_1,\dots,y_s)\,\,
dx_1\dots dx_r dy_1\dots dy_s  \cr
\eea
\er

\subsection{Fonctions g\'en\'eratrices}
\label{sectiondefobservablesseries}

Il est plus commode d'introduire des fonctions g\'en\'eratrices pour les moments \cite{DGZ}.
\bd
La {\bf r\'esolvante} de la matrice $M_1$ est:
\beq\label{defWx}
W_{1}(x):=\sum_{k=0}^\infty {T_{k}\over x^{k+1}}
= {1\over N}\left<\tr{1\over x-M_1}\right>
= \sum_{k} \int_{\ds\Gammax_{k}} {1\over x-x'}\rho_{1,0}^{k}(x')\,\D{x'}
\eeq
\ed
Notons que $W_{1}(x)$ n'est qu'une s\'erie formelle en $1/x$ (pour les trois mod\`eles),
et il faut faire attention si on veut l'interpr\'eter comme une s\'erie convergente, et comme une fonction analytique de $x$.
En fait $W_{1}(x)$ est une s\'erie asymptotique au voisinage de $x\to\infty$ qui d\'efinit plusieurs fonctions analytiques diff\'erentes.

De m\^eme,
\bd
La {\bf r\'esolvante} de la matrice $M_2$ est:
\beq\label{defWy}
W_{2}(y):=\sum_{l=0}^\infty {T_{\ovl{l}}\over y^{l+1}}
= {1\over N}\left<\tr{1\over y-M_2}\right>
=\sum_{l} \int_{\Gammay_{l}} {1\over y-y'}\rho_{0,1}^{l}(y')\,\D{y'}
\eeq
\ed
Comme pr\'ec\'edement, c'est une s\'erie formelle,
qui correspond \`a plusieurs fonctions analytiques diff\'erentes:

\medskip

On introduit aussi des fonctions \`a deux points (voir notations de \refeq{defobstrsimplesconn}):
\bd Fonctions \`a deux points
\beq
W_{1;2}(x;y):=\sum_{k=0}^\infty \sum_{l=0}^\infty {T_{{\rm conn.}\,\,k;\ovl{l}}\over x^{k+1}y^{l+1}}
= \left<\tr{1\over x-M_1}\tr{1\over y-M_2}\right>- N^2 W_1(x)W_2(y)
\eeq
\beq
W_{1;1}(x;x'):=\sum_{k=0}^\infty \sum_{l=0}^\infty {T_{{\rm conn.}\,\,k;l}\over x^{k+1}{x'}^{l+1}}
= \left<\tr{1\over x-M_1}\tr{1\over x'-M_1}\right>- N^2 W_1(x)W_1(x')
\eeq
\beq
W_{2;2}(y;y'):=\sum_{k=0}^\infty \sum_{l=0}^\infty {T_{{\rm conn.}\,\,\ovl{k};\ovl{l}}\over y^{k+1}{y'}^{l+1}}
= \left<\tr{1\over y-M_2}\tr{1\over y'-M_2}\right> - N^2 W_2(y)W_2(y')
\eeq
\ed
Notons ce sont des s\'eries formelles en $x$ et $y$, i.e. des s\'eries asymptotiques qui d\'efinisent plusieurs fonctions analytiques diff\'erentes.

\medskip
On introduit aussi des fonctions \`a deux points contenant les traces mixtes du type \refeq{defcormix2}:
\bd Fonction de corr\'elation \`a deux points mixte
\beq
W_{1,2}(x,y):=\sum_{k=0}^\infty \sum_{l=0}^\infty {T_{k,\ovl{l}}\over x^{k+1}y^{l+1}}
= {1\over N}\left<\tr{1\over x-M_1}{1\over y-M_2}\right>
\eeq
\ed

\bigskip

\section*{Conclusion}

L'objet de l'\'etude des mod\`eles de matrices est essentiellement de calculer toutes ces fonctions, en particulier dans la limite $N\to\infty$.



\chapter{M\'ethode des polyn\^omes biorthogonaux et int\'egrabilit\'e}
\label{chapterpol}

Cette m\'ethode ne s'applique qu'au mod\`ele normal, on peut aussi lui donner un sens pour le mod\`ele formel, mais nous ne le ferons pas ici.
Nous nous placerons d\'esormais dans le cadre du mod\`ele normal d\'efini dans le paragraphe II.\ref{sectiondefmodelenormal}.
Il y a de nombreuses r\'ef\'erences sur les polyn\^omes orthogonaux: \cite{Mehta, Szego, DeiftBook},  voir aussi \cite{AvM, EMLbiortho, DGZ}.

\section{Polyn\^omes biorthogonaux}

\bd\label{defpolbiorth}
Consid\'erons deux familles de polyn\^omes moniques $\pi_n(x)=x^n+\dots$
et $\sigma_n(y)=y^n+\dots$, telles que:
\beq\label{deforthorel}
\int_\Gamma \D{x}\D{y}\, \pi_n(x) \sigma_m(y)\,\, \ee{-{N\over T}[V_1(x)+V_2(y)-xy]} = h_n \delta_{nm}
\eeq
\ed

\subsection{G\'en\'eralisation de la formule de Heine}

\bt Formule de Heine g\'en\'eralis\'ee pour les polyn\^omes biorthogonaux
\beq\label{Heinepsi}
\pi_n(x) =
{\int_{H_n(\Gamma)} \D{M_1}\D{M_2}\, \det(x-M_1)\,\ee{-{N\over T}[V_1(M_1)+V_2(M_2)-M_1 M_2]}
\over \int_{H_n(\Gamma)} \D{M_1}\D{M_2} \,\ee{-{N\over T}[V_1(M_1)+V_2(M_2)-M_1 M_2]}}
\eeq
autrement dit, $\pi_n$ est la valeur moyenne du polyn\^ome caract\'eristique de la matrice $M_1$.
De m\^eme:
\beq\label{Heinephi}
\sigma_n(y) =
{\int_{H_n(\Gamma)} \D{M_1}\D{M_2}\, \det(y-M_2)\,\ee{-{N\over T}[V_1(M_1)+V_2(M_2)-M_1 M_2]}
\over \int_{H_n(\Gamma)} \D{M_1}\D{M_2} \,\ee{-{N\over T}[V_1(M_1)+V_2(M_2)-M_1 M_2]}}
\eeq
\et
Ce th\'eor\`eme semble \^etre d\^u \`a Jean Zinn-justin \cite{zinnprivate}. Il est expos\'e dans \citeeynchain.
Ecrit avec la mesure induite sur les valeurs propres, il donne une formule semblable \`a celle de Heine \cite{Szego} pour les polyn\^omes orthogonaux.

Il est clair que l'existence des polyn\^omes biorthogonaux est \'equivalente \`a la non-annulation
du d\'enominateur.
Pour le mod\`ele hermitien, avec des potentiels $V_1$ et $V_2$ r\'eels, l'int\'egrand du d\'enominateur
est strictement positif, et le d\'enominateur ne s'annule jamais  (voir aussi \cite{EMLbiortho}).
Pour le cas g\'en\'eral, il n'a pas encore \'et\'e trouv\'e de crit\`ere simple sur $V_1$ et $V_2$
pour garantir l'existence des polyn\^omes orthogonaux.

Remarquons d'apr\`es \citeBEHRH, que le d\'enominateur est un polyn\^ome de degr\'e $\leq n$ dans
les $\kappa$, et donc la non-existence des polyn\^omes biorthogonaux (i.e.
$\exists n$ tel que le d\'enominateur s'annule) est un sous-ensemble de mesure nulle (d\'enombrable) de $\C^{d_1\times d_2}$.
Autrement dit, pour $V_1$ et $V_2$ donn\'es, les polyn\^omes biorthogonaux existent pour presque
tout $\kappa$. Cet argument est d\^u \`a M. Bertola.

\subsection{Notation, fonctions d'ondes}

Nous allons suivre les notations de \citeBEHduality, \citeBEHRH:
\bd\label{defwavefunctions} {\bf Fonctions d'ondes}
\beq\label{defpsiphi}
\psi_n(x):={1\over \sqrt{h_n}}\,\pi_n(x)\,\ee{-{N\over T} V_1(x)}
\virg
\phi_n(y):={1\over \sqrt{h_n}}\,\sigma_n(y)\,\ee{-{N\over T} V_2(y)}
\eeq
\ed
\bd Transform\'ees de Fourier-Laplace des fonctions d'ondes
\bea\label{deftrfourrierpsiphi}
k=1,\dots, d_1,\,\,\td\psi^{(k)}_n(y):= \int_{\Gammax_{k}} \psi_n(x) \ee{{N\over T} xy} \D{x}
\cr
k=1,\dots, d_2,\,\,\td\phi^{(k)}_n(x):= \int_{\Gammay_{k}} \phi_n(y) \ee{{N\over T} xy} \D{y}
\eea
\ed
\bd Transform\'ees de Hilbert des fonctions  d'ondes
\beq\label{deftrhilbertphi}
\td\psi^{(0)}_n(y):= \ee{{N\over T}V_2(y)}
\int_{\Gamma} {\psi_n(x) \over y-y'} \ee{-{N\over T}V_2(y')} \ee{{N\over T} xy'} \D{x}\D{y'}
\eeq
\beq\label{deftrhilbertpsi}
\td\phi^{(0)}_n(x):= \ee{{N\over T}V_1(x)}
\int_{\Gamma} {\phi_n(y) \over x-x'} \ee{-{N\over T}V_1(x')} \ee{{N\over T} x'y} \D{x'}\D{y}
\eeq
\ed

Notons que les $\psi_n$, $\phi_n$, ainsi que les $\td\psi^{(k)}$ et $\td\phi^{(k)}$ pour $k>0$ sont
des fonctions enti\`eres.
Par contre, les transform\'ees de Hilbert sont analytiques par morceaux dans $\C/\cup_j \Gammax_{j}$
(resp. $\C/\cup_j \Gammay_{j}$).
La fonction $\td\phi^{(0)}_n(x)$ est discontinue chaque fois que $x$ croise un contour $\Gammax_{j}$,
et sa discontinuit\'e vaut ($x_+$ et $x_-$ d\'esignent respectivement des points \`a gauche et
\`a droite du contour):
\beq\label{disctdpsi0}
\td\phi^{(0)}_n(x_+) = \td\phi^{(0)}_n(x_-) + 2i\pi \sum_{l=1}^{d_2} \kappa_{j,l}\, \td\phi^{(l)}_n(x)
\eeq
de m\^eme, si $y\in \Gammay_{j}$:
\beq\label{disctdphi0}
\td\psi^{(0)}_n(y_+) = \td\psi^{(0)}_n(y_-) + 2i\pi \sum_{l=1}^{d_1} \kappa_{l,j}\, \td\psi^{(l)}_n(y)
\eeq
On peut aussi d\'efinir des fonctions enti\`eres \`a partir des transform\'ees de Hilbert,
par prolongement analytique en d\'epla\c cant les contours.
pour chaque composante connexe $A$ de $\C/\cup_j \Gammax_{j}$,
on obtient une fonction enti\`ere sur $\C$, qui coincide avec $\td\psi^{(0)}_n(x)$ sur $A$.
Cette proc\'edure permet d'obtenir $d_1$ (resp. $d_2$) fonctions enti\`eres ind\'ependantes, not\'ees:
\beq
\td\phi^{(0,k)}_n(x)\, ,\, k=1,\dots,d_1
\virg
\td\psi^{(0,k)}_n(y)\, ,\, k=1,\dots,d_2
\eeq

\section{Relations de r\'ecurrences et matrices \`a bandes finies}

Les polyn\^omes $\pi_n$ forment une base, et l'on peut d\'ecomposer le polyn\^ome $x\pi_n(x)$ sur cette
base, et donc pour les $\psi_n$:
\beq
x\psi_n(x) = \sum_{m=0}^{n+1} Q_{nm}\, \psi_m(x)
\eeq
de m\^eme:
\beq
y\phi_n(y) = \sum_{m=0}^{n+1} P_{nm}\, \phi_m(y)
\eeq
o\`u $Q$ et $P$ sont des matrices semi--infinies.

De m\^eme, on peut d\'ecomposer le polyn\^ome $\pi'_n(x)$ sur la base des $\pi_n$,
le terme de plus haut degr\'e \'etant $n\pi_{n-1}$.
La d\'eriv\'ee de $\psi_n$ peut donc se d\'ecomposer sur la base des $\psi_m$, avec
$m\leq n+d_1$ (\`a cause de la d\'eriv\'ee de $\ee{-{N\over T}V_1}$).
Les coefficients de ce d\'eveloppement sont obtenus par l'identit\'e
\beq
\forall n,m,\qquad
\int_\Gamma \D{x}\D{y}\, \d_x\left(\psi_n(x) \phi_m(y)\,\, \ee{{N\over T}xy}\right) = 0
\eeq
qui implique:
\beq
-{T\over N}\d_x \psi_n(x) = \sum_{m=n-1}^{n+d_1} P_{mn} \psi_m(x)
\eeq
On a vu plus haut que cette somme est en fait limit\'ee \`a $m\leq n+d_1$,
i.e.
\beq
P_{nm} = 0 \quad {\rm si} \quad m>n+1 \,\, {\rm ou} \,\, m<n-\deg V'_1
\eeq
de m\^eme:
\beq
Q_{nm} = 0 \quad {\rm si} \quad m>n+1 \,\, {\rm ou} \,\, m<n-\deg V'_2
\eeq
i.e. les matrices $Q$ et $P$ sont \`a bande finie.

On pose:
\beq
\alpha_k(n):= Q_{n,n-k}
\virg
\beta_k(n):= P_{n,n-k}
\eeq
Le fait que les polyn\^omes sont moniques implique:
\beq\label{defgammanalphabeta}
\alpha_{-1}(n)=\beta_{-1}(n)=\sqrt{h_{n+1}\over h_{n}}:=\gamma_n
\eeq
Autrement dit:
\beq
Q=\pmatrix{
\alpha_0(0)       & \gamma_0            & 0        & \dots   &              &         \cr
\alpha_1(1)       & \alpha_0(1)         & \gamma_1 & 0       &              &         \cr
\vdots            &                     & \ddots   & \ddots  & 0            &         \cr
\alpha_{d_2}(d_2) & \ddots              &          &         & \gamma_{d_2} &         \cr
0                 & \alpha_{d_2}(d_2+1) & \ddots   &         &              & \ddots  \cr
\vdots            & 0                   & \ddots   &         &              &         \cr
0                 &                     & 0        & \ddots  &              &         \cr
}
\eeq
\beq
P=\pmatrix{
\beta_0(0)       & \gamma_0           & 0        & \dots   &              &         \cr
\beta_1(1)       & \beta_0(1)         & \gamma_1 & 0       &              &         \cr
\vdots           &                    & \ddots   & \ddots  & 0            &         \cr
\beta_{d_1}(d_1) & \ddots             &          &         & \gamma_{d_1} &         \cr
0                & \beta_{d_1}(d_1+1) & \ddots   &         &              & \ddots  \cr
\vdots           & 0                  & \ddots   &         &              &         \cr
0                &                    & 0        & \ddots  &              &         \cr
}
\eeq

\subsection{Notations pour les matrices semi--infinies}
\label{notationmatricesinfty}

Si on d\'efinit la matrice semi--infinie:
\beq\label{defLambdashift}
\L:=\pmatrix{
0      & 1      & 0      &  \dots    \cr
\vdots & 0      & 1      & \ddots    \cr
0      & \dots  & \ddots & \ddots    \cr
}
\qquad {\rm i.e.}\quad
\L_{ij}=\delta_{i+1,j}
\eeq
et les matrices:
\beq
\alpha_k := \diag(\alpha_k(n))_{n=0,\dots,\infty}
\virg
\beta_k := \diag(\beta_k(n))_{n=0,\dots,\infty}
\virg
\gamma := \diag(\gamma_n)_{n=0,\dots,\infty}
\eeq
on peut \'ecrire:
\beq
Q = \gamma\L + \sum_{k=0}^{d_2} \alpha_k {\L^t}^{k}
\virg
P = \gamma\L + \sum_{k=0}^{d_1} \beta_k {\L^t}^{k}
\eeq
On introduit aussi les projecteurs sur le sous espace des polynomes de degr\'e $\leq n$:
\beq
\Pi_n:= \diag(1,\dots,1 (n+1\,{\rm termes}),\dots,1,0,\dots)
\eeq
On a les relations:
\beq
\L\Pi_n = \Pi_{n-1}\L
\virg
\L\L^t = 1
\virg
\L^t\L = 1 - \Pi_0
\eeq

\br\label{remassociativite}
Notons que l'alg\`ebre des matrices infinies consid\'er\'ee ici, n'est pas associative ($ A(BC)\neq (AB)C$), car on ne peut
pas \'echanger l'ordre de sommation pour les sommes infinies.
Il n'y a associativit\'e que si toutes les sommes impliqu\'ees sont finies, c'est le cas pour les matrices \`a bandes finies et les projecteurs $\Pi_n$.
\er

\subsection{Relations entre $P$ et $Q$}

Le fait que la d\'ecomposition de $\pi'_n$ sur la base des $\pi$ commence par:
\beq
\pi'_n(x) = n \pi_{n-1}(x)+\dots
\eeq
implique pour $P$ et $Q$ les relations:
\beq\label{PVQ}
(P^t)_{n,m} = (V'_1(Q))_{n,m} \quad {\rm si}\,\, m\geq n
\virg
(P^t)_{n,n-1} = (V'_1(Q))_{n,n-1} - {Tn\over N  \gamma_{n-1}}
\eeq
De m\^eme:
\beq
(Q^t)_{n,m} = (V'_2(P))_{n,m} \quad {\rm si}\,\, m\geq n
\virg
(Q^t)_{n,n-1} = (V'_2(P))_{n,n-1} - {Tn\over N \gamma_{n-1}}
\eeq
Ces relations sont souvent appel\'ees ''\'equations du mouvement'' \cite{DGZ}.

Cela implique en particulier:
\beq
 \L^{d_2}\alpha_{d_2} = \td{g}_{d_2+1} (\gamma\L)^{d_2}
\virg
 \L^{d_1} \beta_{d_1} =  {g}_{d_1+1} (\gamma\L)^{d_1}
\eeq

On a aussi la relation de Heisenberg aussi appel\'ee ''\'equation des cordes'' \cite{DGZ}:
\beq
[P^t,Q] = {T\over N} 1
\eeq

\subsection{Relations de r\'ecurrences, r\'esum\'e}

On a l'ensemble de relations:
\beq\label{xydxdypsiphi}
\begin{array}{rclcrcl}
x\psi_n & = & \sum_{k=-1}^{d_2} \alpha_k(n) \psi_{n-k}
& , &
y\phi_n & = & \sum_{k=-1}^{d_1} \beta_k(n) \phi_{n-k} \cr
-{T\over N}\d_x \psi_n & = & \sum_{k=-1}^{d_1} \beta_k(n+k) \psi_{n+k}
& , &
-{T\over N}\d_y \phi_n & = & \sum_{k=-1}^{d_2} \alpha_k(n+k) \phi_{n+k} \cr
\end{array}
\eeq
De plus, il est facile de montrer \citeBEHRH\ que pour tout $j\in[0,d_2]$:
\beq\label{eqrecphij}
\begin{array}{rcl}
x\td\phi^{(j)}_n & = & \sum_{k=-1}^{d_1} \alpha_k(n+k) \td\phi^{(j)}_{n+k} + \delta_{j,0}\sqrt{h_0}\ee{{N\over T}V_1(x)} \cr
{T\over N}\d_x \td\phi^{(j)}_n & = & \sum_{k=-1}^{d_1} \beta_k(n) \phi^{(j)}_{n-k}
+\delta_{j,0} \sqrt{h_0}\ee{{N\over T}V_1(x)} \left({V'_1(x)-V'_1(Q)\over x-Q}\right)_{0,n}
\end{array}
\eeq
et pour tout $j\in[0,d_1]$:
\beq
\begin{array}{rcl}
y\td\psi^{(j)}_n & = & \sum_{k=-1}^{d_1} \beta_k(n+k) \td\psi^{(j)}_{n+k} + \delta_{j0}\sqrt{h_0}\ee{{N\over T}V_2(y)} \cr
{T\over N}\d_y \td\psi^{(j)}_n & = & \sum_{k=-1}^{d_2} \alpha_k(n) \td\psi^{(j)}_{n-k}
+\delta_{j,0} \sqrt{h_0}\ee{{N\over T}V_2(y)} \left({V'_2(y)-V'_2(P)\over y-P}\right)_{0,n}
\end{array}
\eeq

\section{Noyaux et densit\'es}

On introduit \citeEMchain:
\bd les noyaux
\beq
K_{12}(x,y) := \sum_{j=0}^{N-1} \psi_j(x)\phi_j(y)
\eeq
\beq
K^{(k)}_{11}(x,x') := \sum_{j=0}^{N-1} \psi_j(x)\td\phi^{(k)}_j(x')
\eeq
\beq
K^{(k)}_{22}(y',y) := \sum_{j=0}^{N-1} \td\psi^{(k)}_j(y')\phi_j(y)
\eeq
\bea
K_{21}(y',x') & := & \sum_{k=1}^{d_1}\sum_{l=1}^{d_2} \kappa_{k,l} \sum_{j=0}^{N-1} \td\psi^{(k)}_j(y')\td\phi^{(l)}_j(x')
- \ee{-{N\over T}[V_1(x)+V_2(y)-xy]} \cr
& = & - \sum_{k=1}^{d_1}\sum_{l=1}^{d_2} \kappa_{k,l} \sum_{j=N}^{\infty} \td\psi^{(k)}_j(y')\td\phi^{(l)}_j(x')
\eea
\ed
Ces noyaux permettent de calculer les densit\'es d\'efinies en \refeq{defrhors}.

\bt\label{theoremeynMehta} (Eynard, Mehta \citeEMchain) les densit\'es de valeurs propres d\'efinies en def.II.\ref{Defrhors} sont donn\'ees par:
\beq\label{theynmehta}
\rho_{r;s}^{k_1,\dots,k_r;l_1,\dots,l_s}(x_1,\dots,x_r;y_1,\dots,y_s)
= \det\pmatrix{
\quad K^{(k_j)}_{11}(x_i,x_j) \qquad & K_{12}(x_i,y_j) \cr
\quad K_{21}(y_i,x_j) \qquad & K^{(l_i)}_{22}(y_i,y_j)
}
\eeq
\et
La  preuve est expos\'ee dans  \citeEMchain,
c'est une g\'en\'eralisation du th\'eor\`eme de Dyson pour les polyn\^omes orthogonaux \cite{thDyson}.
Elle utilise la relation de r\'ecurence \refeq{recrhors} et les propri\'et\'es auto-reproduisantes des noyaux:
\beq
\sum_{k,k'} \kappa_{k,k'} \int_{\ds\Gammax_{k}} \D{u}\,\, K^{(k')}_{11}(x,u)K^{(l)}_{11}(u,x') =
K^{(l)}_{11}(x,x')
\eeq
\beq
\sum_{k,k'} \kappa_{k,k'} \int_{\Gammay_{k}} \D{u}\,\, K^{(k')}_{22}(y,u)K^{(l)}_{22}(u,y') =
K^{(l)}_{22}(y,y')
\eeq
\beq
\sum_{k,l} \kappa_{k,l}
\int_{\ds\Gammax_{k}} \D{u}\,\, K_{21}(y,u)K^{(l)}_{11}(u,x) =
K_{21}(y,x)
\eeq
\beq
\sum_{k,l} \kappa_{k,l}
\int_{\Gammay_{l}} \D{u}\,\, K^{(k)}_{22}(y,u)K_{21}(u,x) =
K_{21}(y,x)
\eeq

\bigskip
Grace \`a ce th\'eor\`eme, le calcul de ces quatre noyaux (notons que ceux ci s'obtiennent par transform\'ee de Fourrier
\`a partir de $K_{12}$), est suffisant pour calculer toutes les fonctions de corr\'elations de valeurs propres du type
\refeq{defrhors}.

\section{Matrices de Christoffel--Darboux}

\bt\label{ChrisDarb}
La matrice:
\beq
A_n:=[Q,\Pi_{n-1}]
\eeq
poss\`ede seulement un sous bloc de taille $d_2+1\times d_2+1$ non identiquement nul:
\beq
(A_n)_{k,l}=\left\{
\begin{array}{l}
\alpha_{k-l}(k)\,\, {\rm si}\,\,\, n\leq k\leq n+d_2-1\, ,\, n-d_2\leq l\leq n-1 \, ,\, k-l\leq d_2 \cr
-\gamma_{n-1}\,\, {\rm si}\,\,\, k=n-1\,{\rm et}\, l=n \cr
0\,\,\, {\rm autrement}\cr
\end{array}\right.
\eeq
De m\^eme, $B_n:=[P,\Pi_{n-1}]$, poss\`ede seulement un bloc de taille $d_1+1$ non nul:
\beq
(B_n)_{k,l}=\left\{
\begin{array}{l}
\beta_{k-l}(k)\,\, {\rm si}\,\,\, n\leq k\leq n+d_1-1\, ,\, n-d_1\leq l\leq n-1 \, ,\, k-l\leq d_1 \cr
-\gamma_{n-1}\,\, {\rm si}\,\,\, k=n-1\,{\rm et}\, l=n \cr
0\,\,\, {\rm autrement}\cr
\end{array}\right.
\eeq
\et
\bd
Les matrices $A_n$ et $B_n$ seront appel\'ees matrices de {\bf Christoffel--Darboux}.
\ed

\medskip

On a imm\'ediatement la g\'en\'eralisation du th\'eor\`eme de Christoffel--Darboux \citeBEHduality:
\bt\label{TheoremCD} th\'eor\`eme de Christoffel--Darboux g\'en\'eralis\'e
\beq\label{thCD}
(x'-x)K^{(k)}_{11,n}(x,x') = \sum_{i=n-1}^{n+d_2-1}\sum_{j=n-d_2}^{n} {A_n}_{i,j} \, \psi_j(x)\, \td\phi^{(k)}_i(x')
\eeq
\beq
(y'-y)K^{(k)}_{22,n}(y',y) = \sum_{i=n-1}^{n+d_1-1}\sum_{j=n-d_1}^{n} {B_n}_{i,j}\, \td\psi^{(k)}_i(y') \, \phi_j(y)
\eeq
\et
Gr\^ace \`a ce th\'or\`eme, le calcul des noyaux se r\'eduit au calcul de seulement $d_2+1$ polynomes
de type $\psi_n$, et $d_1+1$ polynomes de type $\phi_n$.
Ceci est particuli\`erement utile dans la limite $n\to\infty$.
Il n'est pas clair qui a d\'ecouvert ce th\'eor\`eme pour la premi\`ere fois.

\bigskip

\bc
On a des relations additionnelles (Bertola, Eynard, Harnad \citeBEHduality), par exemple:
\beq
(x+{T\over N}\d_y)K_{12,n}(x,y) = -\sum_{i=n-1}^{n+d_2-1}\sum_{j=n-d_2}^{n} {A_n}_{i,j} \, \psi_{j}(x)\, \phi_i(y)
\eeq
\beq
(y+{T\over N}\d_x)K_{12,n}(x,y) = -\sum_{i=n-1}^{n+d_1-1}\sum_{j=n-d_1}^{n} {B_n}_{i,j} \, \psi_{i}(x)\, \phi_j(y)
\eeq
\beq\label{thCDadd}
{T\over N}(\d_x+\d_{x'})K^{(k)}_{11,n}(x,x')
= \sum_{i=n-1}^{n+d_1-1}\sum_{j=n-d_1}^{n} {B_n}_{i,j} \, \psi_{i}(x)\, \td\phi^{(k)}_j(x')
\eeq
\ec

\section{Repliement sur une fen\^etre}

On a vu que les noyaux peuvent s'exprimer seulement \`a l'aide des polynomes
de rang $\in[n-d_2,n]$ et des transform\'ees de Fourrier de rang $\in[n-1,n+d_2-1]$.

\bd Les vecteurs suivants, form\'es de $d_2+1$ fonctions d'ondes $\psi$ (resp. transform\'ees de Fourrier $\td\phi$) cons\'ecutifs sont appell\'es fen\^etre et fen\^etre duale:
\beq\label{deffenetrepsi}
\Psi_n(x):=(\psi_{n-d_2}(x),\dots,\psi_n(x))^t
\virg
\td\Phi_n^{(k)}(x):=(\td\phi^{(k)}_{n-1}(x),\dots,\td\phi^{(k)}_{n+d_2-1}(x))^t
\eeq
\ed
et de m\^eme:
\bd Les vecteurs suivants, form\'es de $d_1+1$ fonctions d'ondes $\phi$ (resp. transform\'ees de Fourrier $\td\psi$) cons\'ecutifs sont appell\'es fen\^etre et fen\^etre duale:
\beq\label{deffenetrephi}
\Phi_n(y):=(\phi_{n-d_1}(y),\dots,\phi_n(y))^t
\virg
\td\Psi_n^{(k)}(y):=(\td\psi^{(k)}_{n-1}(y),\dots,\td\psi^{(k)}_{n+d_1-1}(y))^t
\eeq
\ed
Il est clair qu'on peut utiliser la premi\`ere relation \refeq{xydxdypsiphi} (multiplication par $x$)
 pour exprimer  n'importequel polyn\^ome $\psi_m$ avec $m\in[0,\infty[$, comme une
combinaison lin\'eaire \`a coefficients polynomiaux en $x$
de polyn\^omes de la fen\^etre.
\bd La matrice $F_n(x)$ de taille $\infty\times(d_2+1)$ telle que
\beq\label{defpsifolding}
\forall m=0,\dots,\infty \virg \psi_{m}(x) = \sum_{j=n-d_2}^{n} {\left(F_n(x)\right)}_{m,j}\,\psi_j(x)
\eeq
est appel\'ee matrice de repliement sur la fen\^etre $\Psi_n$.
\ed
On a $\deg {\left(F_n\right)}_{m,j}\leq n-d_2-m$ si $m<n-d_2$ et $\deg {\left(F_n\right)}_{m,j}\leq m-n$ si $m>n$.
La matrice $F_n(x)$ a \'et\'e calcul\'ee par Bertola-Eynard dans \citeBEformulaD, et vaut:
\bt\label{Theoremfolding} (Bertola, Eynard \citeBEformulaD)
La matrice $F_n(x)$ est donn\'ee par
\beq\label{foldinfmatrix}
F_n(x) = ((Q-x)^{-1}_L - (Q-x)^{-1}_R)\, A_n
\eeq
\et
o\`u $(Q-x)^{-1}_R$ et $(Q-x)^{-1}_L$ sont respectivement les inverses \`a droite (triangulaire inf\'erieure) et \`a gauche (triangulaire sup\'erieure)
de la matrice $Q-x$. On les calcule en introduisant les deux matrices suivantes
respectivement strictement triangulaires sup\'erieures et inf\'erieures:
\beq\label{QLRcal}
{\cal Q}_L:= 1 - {1\over \td{g}_{d_2+1}}\,(\gamma^{-1}\L) ^{d_2}\,(Q-x)
\virg
{\cal Q}_R:= 1 -\Pi_0 -  \L^t \gamma^{-1}\,(Q-x)
\eeq
on a:
\beq\label{QinvL}
(Q-x)^{-1}_L =  {1\over \td{g}_{d_2+1}}\,(1-{\cal Q}_L)^{-1}\,(\gamma^{-1}\L) ^{d_2}
\eeq
\beq\label{QinvR}
(Q-x)^{-1}_R = (1-{\cal Q}_R)^{-1}\,\L^t \gamma^{-1}
\eeq
la matrice $(1-{\cal Q}_L)^{-1}$ est une notation pour $\sum_{k=0}^\infty ({\cal Q}_L)^k$,
et il faut noter que pour le calcul d'un \'el\'ement de $F_n(x)$, cette somme se
r\'eduit \`a une somme finie (cf remarque  III.\ref{remassociativite}).

De m\^eme on d\'efinit les matrices respectivement triangulaires inf\'erieure et sup\'erieure:
\beq\label{tdQLRcal}
\td{\cal Q}_L:= 1 -\Pi_{d_2-1}- {1\over \td{g}_{d_2+1}}\,(\L^t\gamma^{-1}) ^{d_2}\,(Q-x)^t
\virg
\td{\cal Q}_R:= 1 -\gamma^{-1}\,\L \,(Q-x)^t
\eeq
et la matrice:
\beq\label{tdfoldinfmatrix}
\td{F}_n(x) = ({1\over \td{g}_{d_2+1}}\,(1-\td{\cal Q}_L)^{-1}\,(\L^t\gamma^{-1}) ^{d_2}
-(1-\td{\cal Q}_R)^{-1}\,\gamma^{-1}\,\L)\,  A_n^t
\eeq
qui assure le repliement des transform\'ees de Fourrier (pour tout $k=1,\dots,d_2$):
\beq\label{deftdphifolding}
\forall m=0,\dots,\infty \virg \td\phi^{(k)}_{m}(x) = \sum_{j=n-1}^{n+d_2-1} \td{F}_{mj,n}(x)\,\td\phi^{(k)}_j(x)
\eeq

Et on peut d\'efinir de fa\c con similaire les repliements des $\phi$ et $\td\psi$:
\beq\label{defphifolding}
\phi_{m}(y) = \sum_{j=n-d_1}^{n} {G_n}_{m,j}(y)\,\phi_j(y)
\virg
\td\psi^{(k)}_{m}(y) = \sum_{j=n-1}^{n+d_1-1} \td{G_n}_{m,j}(y)\,\td\psi^{(k)}_j(y)
\eeq

\section{Syst\`emes diff\'erentiels}

\bt\label{TheoremDFn}
La fen\^etre $\Psi_n(x)$ satisfait un syst\`eme d'\'equation diff\'erentielles lin\'eaires \`a coefficients polynomiaux
\beq\label{defD1}
-{T\over N} \d_x\Psi_n(x)
= {\cal D}_{1,n}(x) \Psi_n(x)
\eeq
et ${\cal D}_{1,n}(x)$ est le seul bloc non nul de la matrice:
\beq\label{defD1calc}
{\cal D}_{1,n}(x) = (\Pi_{n}-\Pi_{n-d_2-1})\,P^t\, F_n(x)
\eeq
\et
{\bf\noindent preuve:} (voir \citeBEHduality, \citeBEformulaD, mais ces syst\`emes ont \'et\'e trouv\'es dans des cas particuliers avant) on utilise le repliement sur une fen\^etre.
On \'ecrit la relation \refeq{xydxdypsiphi} ($\d_x\psi$) pour tous les polyn\^omes de la fen\^etre,
et l'on utilise \refeq{defpsifolding} pour exprimer le membre de droite sur la base des polyn\^omes de
la fen\^etre:
\bea
\forall m=n-d_2,\dots,n \virg
-{T\over N} \d_x\psi_m(x)
&=& \sum_{k=m-1}^{m+d_1} P_{km} \psi_k(x) = \sum_{k=m-1}^{m+d_1} \sum_{j=n-d_2}^{n} P_{km} F_{kj,n}(x) \psi_j(x) \cr
&=& \sum_{j=n-d_2}^n (P^t F_n(x))_{mj} \psi_j(x)
\eea
QED.

Notons que ${\cal D}_{1,n}(x)$ est une matrice de taille $d_2+1\times d_2+1$, dont
les coefficients sont des polyn\^omes en $x$ de degr\'e $\leq d_1$.

\bigskip

De m\^eme, on d\'efinit trois autres syst\`emes diff\'erentiels
(on suppose $n>{\rm max}(d_1,d_2)$, sinon il y a des termes correspondant aux petites valeurs de $n$ comme dans \refeq{eqrecphij}):
\bd Autres syst\`emes d'\'equations diff\'erentielles lin\'eaires \`a coefficients polynomiaux
\beq\label{deftdD1}
{T\over N} \d_x\td\Phi^{(k)}_n(x)
= \td{\cal D}_{1,n}(x) \td\Phi^{(k)}_n(x)
\virg
\td{\cal D}_{1,n}(x) = (\Pi_{n+d_2-1}-\Pi_{n-2})\,P\, \td{F}_n(x)
\eeq
\beq\label{defD2}
-{T\over N} \d_y\Phi_n(y)
= {\cal D}_{2,n}(y) \Phi_n(y)
\virg
{\cal D}_{2,n}(y) = (\Pi_{n}-\Pi_{n-d_1-1})\,Q^t\, G_n(y)
\eeq
\beq\label{deftdD2}
{T\over N} \d_y\td\Psi^{(k)}_n(y)
= \td{\cal D}_{2,n}(y) \td\Psi^{(k)}_n(y)
\virg
\td{\cal D}_{2,n}(y) = (\Pi_{n+d_1-1}-\Pi_{n-2})\,Q\, \td{G}_n(y)
\eeq
\ed
Notons que $\td{\cal D}_{1,n}(x)$ est une matrice de taille $d_2+1\times d_2+1$, dont
les coefficients sont des polyn\^omes en $x$ de degr\'e $\leq d_1$, comme ${\cal D}_{1,n}(x)$.
De m\^eme, ${\cal D}_{2,n}(y)$ et $\td{\cal D}_{2,n}(y)$ sont des matrices de taille $d_1+1\times d_1+1$, dont
les coefficients sont des polyn\^omes en $y$ de degr\'e $\leq d_2$.

Dans \citeBEformulaD, nous avons obtenu des expressions encore plus explicites de ces syst\`emes.
\bt\label{TheoremDexplicit} (Bertola, Eynard \citeBEformulaD)
La matrice ${\cal D}_{1,n}(x)$ s'\'ecrit:
\bea\label{D1explicit}
{\cal D}_{1,n}(x) & = &
  \pmatrix{
V'_1(Q)_{n-d_2,n-d_2} & \dots & V'_1(Q)_{n-d_2,n-1} & 0 \cr
 0 & \ddots & \vdots & \vdots \cr
0 & 0 & V'_1(Q)_{n-1,n-1} & 0 \cr
0 & \dots & 0 & V_1'(x)
}\cr
& & +  \pmatrix{\gamma(n-d_2-1) & & \cr & \ddots & \cr & & \gamma(n-1) }
\pmatrix{ - {\alpha_{d_2-1}(n-1)  \over \alpha_{d_2}(n-1)} & \dots &
  {x- \alpha_{0}(n-1)  \over \alpha_{d_2}(n-1)} & - {\gamma(n-1)
    \over \alpha_{d_2}(n-1)} \cr 1 & & &  0 \cr  & \ddots & & \vdots
  \cr 0 & \dots & 1 & 0}  \cr
& & - \pmatrix{
{V'_1(Q)-V'_1(x)\over Q-x}_{n-d_2,n-1} & \dots & {V'_1(Q)-V'_1(x)\over Q-x}_{n-d_2,n+d_2-1}\cr
\vdots & & \vdots \cr
{V'_1(Q)-V'_1(x)\over Q-x}_{n,n-1} & \dots & {V'_1(Q)-V'_1(x)\over Q-x}_{n,n+d_2-1}
}
A_N\ ,
\eea
\et
La preuve est d\'etaill\'ee dans \citeBEformulaD. Elle utilise \refeq{defD1calc} et \refeq{PVQ}.

On a bien s\^ur  des expressions similaires pour les 3 autres syst\`emes \citeBEformulaD.

\section{Dualit\'e spectrale}

Il a \'et\'e pouv\'e par Bertola-Eynard-Harnad \citeBEHduality\ que ces syst\`emes diff\'erentiels sont duaux.

\bt\label{TheoremDDconjug} (Bertola, Eynard, Harnad \citeBEHduality)
 les matrices ${\cal D}_{1,n}(x)$ et $\td{\cal D}^t_{1,n}(x)$ sont conjugu\'ees par la matrice de Christoffel--Darboux:
\beq\label{dualiteADDA}
A_n {\cal D}_{1,n}(x) = \td{\cal D}^t_{1,n}(x) A_n
\eeq
\et
{\noindent \bf preuve:}
donnons ici une id\'ee de la d\'emonstration de \citeBEHduality:

Supposons que $n>d_2+d_1$.
Consid\'erons deux solutions arbitraires des syst\`emes ${\cal D}_{1,n}(x)$ et $\td{\cal D}_{1,n}(x)$:
\beq\label{sysf}
f(x)=(f_{n-d_2}(x),\dots,f_n(x))^t
\virg
{T\over N}\partial_x f(x) = -{\cal D}_{1,n}(x) f(x)
\eeq
\beq\label{sysg}
g(x)=(g_{n-1}(x),\dots,g_{n+d_2-1}(x))^t
\virg
{T\over N}\partial_x g(x) = \td{\cal D}_{1,n}(x) g(x)
\eeq
On d\'efinit les fonctions $f_m(x)$ avec $n-d_2-1\leq m\leq n+d_1$ et  $g_m(x)$ avec $n-d_1-1\leq m\leq n+d_2$,
r\'ecursivement par:
\beq
\sum_{k=-1}^{d_2} \alpha_k(m) f_{m-k}(x) = x f_m(x)
\virg
\sum_{k=-1}^{d_2} \alpha_k(m+k) g_{m+k}(x) = x g_m(x)
\eeq
les \'equations \refeq{sysf} et \refeq{sysg} sont alors \'equivalentes \`a (voir \citeBEHduality\ pour d\'etails):
\bea
\forall m\in[n-d_2,n] \qquad -{T\over N}\partial_x f_m(x) = \sum_{k=-1}^{d_1} \beta_k(m+k) f_{m+k}(x) \cr
\forall m\in[n-1,n+d_2-1] \qquad {T\over N}\partial_x g_m(x) = \sum_{k=-1}^{d_1} \beta_k(m) g_{m-k}(x)
\eea
c'est \`a dire les relations \refeq{xydxdypsiphi} et \refeq{eqrecphij}.
On a donc:
\bea
&& {T\over N}\, g^t(x')\,\left({\td{\cal D}}^t_{1,n}(x')A_n-A_n{\cal D}_{1,n}(x)\right)\,f(x) \cr
&=& {T\over N}\, \sum_{r=n-1}^{n+d_2-1}\sum_{s=n-d_2}^{n}  g_r(x') f_s(x)\,\left({\td{\cal D}}^t_{1,n}(x')A_n-A_n{\cal D}_{1,n}(x)\right)_{r,s} \cr
&=& \sum_{r=n-d_1-1}^{n+d_2}\sum_{s=n-d_2-1}^{n+d_1} g_r(x') f_s(x) \left[P^t, A_n\right]_{r,s}   \cr
&=& \sum_{r=n-d_1-1}^{n+d_2}\sum_{s=n-d_2-1}^{n+d_1} g_r(x') f_s(x) \left[P^t, [Q,\Pi]\right]_{r,s}   \cr
&=& \sum_{r=n-d_1-1}^{n+d_2}\sum_{s=n-d_2-1}^{n+d_1} g_r(x') f_s(x) \left([[P^t,Q],\Pi] + [Q,[P^t,\Pi]]\right)_{r,s}   \cr
&=& \sum_{r=n-d_1-1}^{n+d_2}\sum_{s=n-d_2-1}^{n+d_1} g_r(x') f_s(x) \left[Q,[P^t,\Pi]\right]_{r,s}   \cr
&=& (x'-x)\,\sum_{r=n-d_1}^{n}\sum_{s=n-1}^{n+d_1-1} g_r(x') f_s(x) \left[P^t,\Pi\right]_{r,s}   \cr
&=& (x-x')\,\sum_{r=n-d_1}^{n}\sum_{s=n-1}^{n+d_1-1} g_r(x') f_s(x) \left(B^t_n\right)_{r,s}   \cr
\eea
qui s'annule pour $x'=x$.
Ceci \'etant vrai pour toute solutions $f$ et $g$, on doit avoir:
\beq
{\td{\cal D}}^t_{1,n}(x)A_n=A_n{\cal D}_{1,n}(x)
\eeq
QED.
Notons que l'on a utilis\'e la relation de Heisenberg entre $P$ et $Q$:
\beq
[P^t,Q]={T\over N}\, 1
\eeq

\bigskip

A partir de l\`a, il a \'et\'e pouv\'e par Bertola-Eynard-Harnad \citeBEHduality\ que ces quatres syst\`emes diff\'erentiels ont
la m\^eme courbe spectrale:
\bt\label{Theoremdualite} (Bertola, Eynard, Harnad \citeBEHduality)
Les 4 syst\`emes ${\cal D}_{1,n}(x)$, $\td{\cal D}_{1,n}(x)$, ${\cal D}_{2,n}(y)$, $\td{\cal D}_{2,n}(y)$ ont la m\^eme courbe spectrale:
\bea\label{defcourbespectrale}
E_n(x,y)
& = & \td{g}_{d_2+1}\,\det\left(y1-{\cal D}_{1,n}(x)\right)
= \td{g}_{d_2+1}\,\det\left(y1-\td{\cal D}_{1,n}(x)\right) \cr
& = & g_{d_1+1}\,\det\left(x1-{\cal D}_{2,n}(y)\right)
= g_{d_1+1}\,\det\left(x1-\td{\cal D}_{2,n}(y)\right)
\eea
\et
{\noindent \bf id\'ee de la preuve:}
les \'egalit\'es diagonales (par exemple entre ${\cal D}_{1,n}(x)$ et $\td{\cal D}_{2,n}(y)$)
ne font pas intervenir les propri\'et\'es d'orthogonalit\'e,
elles sont cons\'equences de la transform\'ee de Fourrier, et peuvent \^etre obtenues par simple
r\'eecriture des deux membres \citeBEHduality.

Les \'egalit\'es horizontales sont cons\'equences du th\'eor\`eme III.\ref{TheoremDDconjug}.

\section{Solution fondamentale}

Consid\'erons ici le syst\`eme $\td{\cal D}_{1,n}(x)$, et supposons $n>d_1$.
Nous savons d\'ej\`a que chacune des $d_2$ fonctions $\td\phi_n^{(k)}(x)$ avec $k=0,\dots,d_2$ est solution de ce syst\`eme.
La solution fondamentale de $\td{\cal D}_{1,n}(x)$ a \'et\'e trouv\'ee par Bertola-Eynard-Harnad dans \citeBEHRH,
et simultan\'ement par A. Kapaev \cite{Kapaev} seulement pour le cas $d_1=d_2=2$.

\bt\label{Theoremsolfondtdphi} (Bertola, Eynard, Harnad \citeBEHRH)
La matrice carr\'ee $\td{\mathbf \Phi}_n(x)$ dont les colonnes sont les vecteurs $\td\Phi_n^{(k)}(x)$
avec $k=0,\dots,d_2$, est une solution fondamentale du syst\`eme $\td{\cal D}_{1,n}(x)$:
\beq\label{defsolfondtdphi}
{T\over N}\,\d_x \td{\mathbf \Phi}_n(x) = \td{\cal D}_{1,n}(x)\,\td{\mathbf \Phi}_n(x)
\eeq
\et
{\noindent \bf preuve:}
Pour $n>d_1$, les relations de r\'ecurrences \refeq{eqrecphij} sont les m\^emes pour $j=0$ que pour $j=1,\dots,d_2$, ce
qui implique  que la fen\^etre duale $\td\Phi_n^{(0)}(x)$ (o\`u $\td\phi_n^{(0)}(x)$ d\'esigne l'une quelquonque
des fonctions enti\`eres d\'efinies par \refeq{deftrhilbertphi}) est aussi solution du syst\`eme $\td{\cal D}_{1,n}(x)$,
et donc $\td{\mathbf \Phi}_n(x)$ satisfait \refeq{defsolfondtdphi}.
Pour v\'erifier que cette matrice est bien une solution fondamentale, il suffit de calculer
son d\'eterminant, i.e. le Wronskien, et v\'erifier qu'il ne s'annule pas.
Le th\'eor\`eme III.\ref{TheoremDexplicit}, implique (voir \citeBEformulaD):
\begin{lemma} Trace de la matrice $\td{\cal D}_{1,n}(x)$:
\beq
\tr \td{\cal D}_{1,n}(x) = V'_1(x)-{\td{g}_{d_2}\over \td{g}_{d_2+1}}
\eeq
\end{lemma}
ce qui implique
\beq
{\d\over \d x}\ln{\det{\td{\mathbf \Phi}_n(x)}}
= \tr {\td{\mathbf \Phi}_n(x)}^{-1}{\d \over \d x}{\td{\mathbf \Phi}_n(x)}
= \tr \td{\cal D}_{1,n}(x)
= V'_1(x)-{\td{g}_{d_2}\over \td{g}_{d_2+1}}
\eeq
et donc
\beq
\det{\td{\mathbf \Phi}_n(x)} = C\, \ee{{N\over T}\left(V_1(x)-{\td{g}_{d_2}\over \td{g}_{d_2+1}}x\right)}
\eeq
La constante $C$ a \'et\'e calcul\'ee dans la limite $x\to\infty$ dans \citeBEHRH, \citeBEformulaD, et
elle a \'et\'e calcul\'ee directement dans \citeBEformulaD.
Elle est non nulle.
QED.

\bigskip

Avant d'\'etudier les asymptotiques \`a $x$ grand dans le paragraphe suivant, mentionons les solutions
fondamentales des trois autres syst\`emes.
La solution du syst\`eme ${\cal D}_{1,n}(x)$ a \'et\'e explicit\'ee par \citeBEHRH\  sous une forme
r\'ecursive, et explicit\'ee pr\'ecis\'ement pour certains cas particuliers: potentiels cubiques par
\cite{Kapaev}, et un cas particulier de potentiels de degr\'es $d_1=7,d_2=11$ par \citeBEHRH.
Il appara\^\i t clair que la construction de \citeBEHRH\ est g\'en\'erale, mais une expression ferm\'ee reste
\`a trouver.

On peut aussi construire une solution fondamentale de ${\cal D}_{1,n}(x)$ en utilisant la dualit\'e
\refeq{dualiteADDA}.
\bt (Bertola, Eynard, Harnad \citeBEHRH)
La matrice
\beq
{\mathbf \Psi}_n(x):= A^{-1}_n \left(\td{\mathbf \Phi}^t_n(x)\right)^{-1} \, {\mathbf C}
\eeq
o\`u ${\mathbf C}$ est n'importe quelle matrice constante inversible ind\'ependante de $x$,
est solution de $-{T\over N} \d_x{\mathbf\Psi}_n(x)= {\cal D}_{1,n}(x) {\mathbf\Psi}_n(x)$.
\et
En effet, si ${\mathbf\Psi}_n(x)$ est solution du syst\`eme ${\cal D}_{1,n}(x)$, alors on a, d'apr\`es le th\'eor\`eme III.\ref{TheoremDDconjug}:
\beq
{\d\over \d x} \left(\td{\mathbf \Phi}^t_n(x) A_n  {\mathbf \Psi}_n(x)\right)
= {T\over N} \td{\mathbf \Phi}^t_n(x) (\td{\cal D}^t_{1,n}(x) A_n-A_n {\cal D}_{1,n}(x))
{\mathbf \Psi}_n(x) = 0
\eeq
QED.

\section{Asymptotiques $x\to\infty$ et ph\'enom\`ene de Stokes}

Les asymptotiques de la solution fondamentale $\td{\mathbf \Phi}_n(x)$ lorsque $x\to\infty$ ont \'et\'e trouv\'ees par Bertola-Eynard-Harnad \citeBEHRH.

\bd On introduit les notations suivantes:

$y_1(x),\dots,y_{d_2}(x)$ sont les $d_2$ solutions de l'\'equation $V'_2(y)=x$
\beq
T_k(x):= \sum_{l=0}^{d_2} {l\over l+1}\td{g}_{l+1} y_k(x)^{l+1}
\virg
T(x)=\diag(V_1(x),T_1(x),\dots,T_{d_2}(x))
\eeq
\beq
\om:=\ee{2i\pi\over d_2}
\virg
\Omega_{00}=1
\virg \Omega_{ij}=\om^{ij} \quad {\rm si}\, i,j>0
\eeq
\beq
G=\diag(-n,{2n+1-d_2\over 2d_2},\dots,{2n-1+d_2\over 2d_2})
\eeq
\beq
Y_0= {\rm diag}(\sqrt{h_{n-1}},{\td{g}_{d_2+1}^{-{n+1\over d_2}}\over \sqrt{h_{n}}},\dots,{\td{g}_{d_2+1}^{-{n+d_2\over d_2}}\over \sqrt{h_{n+d_2-1}}})
\eeq
\ed

\bt\label{Theoremasymptdphi} (Bertola, Eynard, Harnad \citeBEHRH)
Asymptotiquement pour $x\to\infty$, on a:
\beq\label{asympxlarge}
\td{\mathbf \Phi}_n(x)
\sim
Y_0\,(1+O(x^{-1/d_2}))\,  x^{G} \Omega\, \ee{{N\over T}T(x)}
C(x)
\eeq
o\`u $C(x)$ est une matrice constante par morceaux donn\'ee dans \citeBEHRH.
$C(x)$ est constante dans des secteurs d'angle $\pi/d_2$.
\et

{\noindent \bf preuve:} Nous n'allons que survoler la preuve d\'etaill\'ee de \citeBEHRH.

Rappelons que par leur d\'efinition III.\ref{defwavefunctions}, les fonctions d'ondes $\phi_j(y)$ sont enti\`eres, et leur comportement asymptotique \`a
grand $y$ est donn\'e par:
\beq
\phi_j(y) \sim  {y^j\over \sqrt{h_j}}\, \ee{-{N\over T} V_2(y)} (1+O(1/y))
\eeq
Nous calculons alors les asymptotiques \`a grand $x$ des $\td\phi^{(k)}_j(x)$ par la m\'ethode du col.

Pour $x$ donn\'e, posons:
\beq
f_x(y):= V_2(y)-xy
\eeq
On cherche les points cols, solutions de $\d_y f_x(y)=0$, i.e.
\beq
V'_2(y)=x
\eeq
Il y a $d_2$ points cols $y_1(x),\dots,y_{d_2}(x)$.
Pour $k=1,\dots,d_2$, on d\'efinit le chemin de col ${\mathop\gamma^y}_{k}(x)$:
\beq
{\mathop\gamma^y}_{k}(x) = f_x^{-1}(f_x(y_k(x))+\R_+)
\eeq
Les $d_2$ chemins ainsi d\'efinis forment une base homologiquement \'equivalente \`a la base des
$\Gammay_{k}$. Il existe donc une matrice de changement de base $M(x)$:
\beq\label{defGcolslargex}
\Gammay_{l}=\sum_{k=1}^{d_2} M_{lk}(x) {\mathop\gamma^y}_{k}(x)
\eeq
La matrice $M(x)$ \`a coefficients entiers est constante par morceaux sur des domaines du plan complexe.
Cette matrice a \'et\'e calcul\'ee explicitement dans \citeBEHRH.
Pour chaque chemin ${\mathop\gamma^y}_{k}(x)$, on a l'approximation du col:
\beq
\int_{{\mathop\gamma^y}_{k}(x)} \phi_n(y) \ee{{N\over T} xy} \D{y}
\sim {y_k(x)^j\over \sqrt{h_j}}\,\sqrt{\pi T\over N V''_2(y_k(x))}\, \ee{-{N\over T} f_x(y_k(x))}
(1+O(x^{-1/d_2}))
\eeq
et donc:
\beq
\td\phi_j^{(l)}(x) \sim
\sum_{k} M_{lk} {y_k(x)^j\over \sqrt{h_j}}\,\sqrt{\pi T\over N V''_2(y_k(x))}\,
\ee{-{N\over T} f_x(y_k(x))} (1+O(x^{-1/d_2}))
\eeq

En ce qui concerne l'asymptotique de $\td\phi_j^{(0)}(x)$, choisissons $x\to\infty$
dans un secteur qui ne croise pas de $\Gammax_{k}$. Dans ce secteur, la relation
d'orthogonalit\'e implique:
\beq
\td\phi_j^{(0)}(x) \sim
\sqrt{h_j}\,x^{-(n+1)}\,
\ee{{N\over T} V_1(x)} (1+O(x^{-1}))
\eeq
Les asymptotiques dans les autres secteurs se calculent par la relation de discontinuit\'e
\refeq{disctdpsi0}.

La matrice $C(x)$ est en partie constitu\'ee de  la matrice $M(x)$ d\'efinie en \refeq{defGcolslargex},
et des relations de discontinuit\'e \refeq{disctdpsi0}. Elle est donn\'ee pr\'ecis\'ement dans \citeBEHRH.
QED.

\bigskip

Les discontinuit\'es de la matrice $C(x)$ repr\'esentent les matrices de Stokes:
\beq
C(x_-)^{-1}\,C(x_+)
\eeq
Celles ci sont calcul\'ees dans \citeBEHRH, leur expression prendrait trop de place ici.
Ce qui est important, c'est de remarquer que les matrices de Stokes, de m\^eme que les conditions de saut de \refeq{disctdpsi0} sont ind\'ependantes de $n$, de $x$, et des coefficients des potentiels.
Nous avons donc affaire \`a un probl\`eme {\bf isomonodromique}.

\section{Probl\`eme de Riemann--Hilbert}
\label{sectionpbRH}

Posons nous le probl\`eme suivant:

\smallskip
{\it $\bullet$ existe-t-il  une matrice  $\td{\mathbf \Phi}_n(x)$ analytique par morceaux, et inversible pour tout $x$,
qui satisfait les asymptotiques \refeq{asympxlarge}, et les conditions de saut \refeq{disctdpsi0} ?}

\smallskip

C'est le probl\`eme de Riemann--Hilbert.
La r\'eponse est oui est la solution est unique.

\smallskip

Le probl\`eme de Riemann--Hilbert pour les polyn\^omes biorthogonaux a \'et\'e trouv\'e simultan\'ement par A. Kapaev \cite{Kapaev} pour le cas o\`u $d_1=d_2=2$,
et par Bertola-Eynard-Harnad \citeBEHRH\ pour le cas g\'en\'eral.
Notons qu'un autre probl\`eme de Riemann--Hilbert a \'et\'e \'etudi\'e dans \cite{EMLbiortho}.

\bigskip
L'approche par le probl\`eme de Riemann-Hilbert est particuli\`erement utile si l'on s'int\'eresse \`a la limite $n$ grand.
En effet, les discontinuit\'es \refeq{disctdpsi0} sont ind\'ependantes de $n$, et les asymptotiques \refeq{asympxlarge} en d\'ependent d'une
fa\c con tr\`es simple.

L'id\'ee est la suivante. On introduit un ansatz $\td{\mathbf \Phi}_{n,{\rm ansatz}}(x)$, dont on conna\^\i t toutes les
discontinuit\'es et asymptotiques.
Si l'on peut montrer que pour $x\to\infty$:
\beq
\td{\mathbf \Phi}_{n,{\rm ansatz}}(x) \td{\mathbf \Phi}_{n}^{-1}(x) = 1+O(1/n)
\eeq
et que les discontinuit\'es des asymptotiques de $\td{\mathbf \Phi}_{n,{\rm ansatz}}(x) \td{\mathbf \Phi}_{n}^{-1}(x)$
sont d'ordre $O(1/n)$ partout, alors on a uniform\'ement pour tout $x$:
\beq
\td{\mathbf \Phi}_{n,{\rm ansatz}}(x) \td{\mathbf \Phi}_{n}^{-1}(x) = 1+O(1/n)
\eeq
Cette approche a \'et\'e d\'evelopp\'ee avec un immense succ\`es par Bleher et Its \cite{BlIt}, et par \cite{dkmvz, dkmvz2, EML, DeiftBook}.

Ceci reste \`a faire pour les polyn\^omes biorthogonaux, mais on peut \'esp\'erer que ce sera fait tr\`es bient\^ot \cite{BEHAMS,ansatz}.

\section{D\'eformations isomonodromiques et int\'egrabilit\'e}
\label{sectiontauiso}

Remarquons que la propri\'et\'e d'isomonodromie (i.e. le fait que les matrices de Stokes, et les matrices de saut \refeq{disctdpsi0} ne d\'ependent ni de $x$, ni de $n$, ni des potentiels \cite{FIK})
implique que la matrice $\left( {T\over N}\d_x \td{\mathbf \Phi}_n(x) \right) \,\, \td{\mathbf \Phi}_n(x)^{-1}$
n'a aucune discontinuit\'e, elle doit donc \^etre analytique enti\`ere, et du fait de son comportement \`a l'infini, ce do\^\i t \^etre une matrice polynomiale.
Elle doit \^etre \'egale \`a $\td{\cal D}_{1,n}(x)$ donn\'ee dans \refeq{D1explicit}.

De m\^eme, la propri\'et\'e d'isomonodromie implique que la matrice $\td{\cal U}_{k,n}(x)=\left( {T\over N}\d_{g_k} \td{\mathbf \Phi}_n(x) \right) \,\, \td{\mathbf \Phi}_n(x)^{-1}$
n'a aucune discontinuit\'e, elle doit donc \^etre analytique enti\`ere, et du fait de son comportement \`a l'infini, ce do\^\i t \^etre une matrice polynomiale.
De m\^eme pour les matrices
$\td{\cal V}_{k,n}(x)=\left( {T\over N}\d_{\td{g}_k} \td{\mathbf \Phi}_n(x) \right) \,\, \td{\mathbf \Phi}_n(x)^{-1}$
et $\td{R}_{n}(x)=\left(  \td{\mathbf \Phi}_{n+1}(x) \right) \,\, \td{\mathbf \Phi}_n(x)^{-1}$.
Leurs expressions explicites sont donn\'ees dans \citeBEformulaD.

\medskip
Remarquons que l'on doit avoir les relations de commutation suivantes:
\bea
\left[{T\over N}\d_x-\td{\cal D}_{1,n}(x) , {T\over N}\d_{g_k}-\td{\cal U}_{k,n}(x) \right]=0 \cr
\left[{T\over N}\d_x-\td{\cal D}_{1,n}(x) , {T\over N}\d_{\td{g}_k}-\td{\cal V}_{k,n}(x) \right]=0 \cr
\td{\cal D}_{1,n+1}(x)\td{R}_n(x) = \td{R}_n(x)\td{\cal D}_{1,n}(x) +{T\over N}\d_x\td{R}_n(x) \cr
\td{\cal U}_{k,n+1}(x)\td{R}_n(x) = \td{R}_n(x)\td{\cal U}_{k,n}(x) +{T\over N}\d_{g_k}\td{R}_n(x) \cr
\td{\cal V}_{k,n+1}(x)\td{R}_n(x) = \td{R}_n(x)\td{\cal V}_{k,n}(x) +{T\over N}\d_{\td{g}_k}\td{R}_n(x) \cr
\eea
Ces relations de compatibilit\'e de Frobenius, signifient que l'on a affaire \`a des syst\`emes int\'egrables.

Tout ceci donne lieu \`a l'existence d'une fonction $\tau$ au sens de Jimbo-Miwa-Ueno \cite{MiwaJimbo, JM},
qui est conjectur\'ee \^etre exactement la fonction de partition du mod\`ele de matrice $Z$, d\'efinie en \refeq{defZNormal},
comme cel\`a a \'et\'e prouv\'e pour le mod\`ele \`a une matrice dans \citeBEHtauiso\ par Bertola-Eynard-Harnad.

L'approche int\'egrabilit\'e des mod\`eles de matrices et polyn\^omes orthogonaux a \'et\'e abondament \'etudi\'ee
\cite{sato, uenotak, TW2, HTW, H2, dubrovin}.

\section{Traces mixtes}
\label{trmixtepol}

Nous avons mention\'e au paragraphe \ref{obstrmixtes}, qu'un probl\`eme int\'eressant est de calculer des observables
mixtes du type:
\beq\label{deftrmixtekl}
\left<\tr M_1^k M_2^l \right>
\eeq
A priori, les valeurs propres de $M_1^k M_2^l$ ne peuvent pas s'exprimer en fonction des valeurs propres de $M_1$ et de $M_2$,
et il semblait improbable de trouver une expression de \refeq{deftrmixtekl} en termes de polyn\^omes biorthogonaux.
Toutefois cel\`a est possible, en utilisant la formule de Morozov du th\'eor\`eme II.\ref{TheoremMorozov}.
Il a \'et\'e trouv\'e par Bertola-Eynard \citeBEmixed:
\bt\label{Theoremtracemixt} (Bertola, Eynard, \citeBEmixed)
La fonction g\'en\'eratrice de corr\'elations mixtes \refeq{deftrmixtekl} est donn\'ee par
\beq\label{tracemixten}
{T\over N}\left<\tr {1\over x-M_1}{1\over y-M_2}\right> = 1-\det\left(\1_{N} -\Pi_{N-1} {1\over x-Q}{1\over y-P^t}\Pi_{N-1} \right)
\eeq
\et
La g\'en\'eralisation de cette formule pour toutes les autres fonctions de corr\'elations mixtes
(i.e. produit d'un nombre quelconque de traces, chacune contenant un produit quelconque de matrices $M_1$ et $M_2$)
a \'et\'e trouv\'ee tr\`es r\'ecement par Eynard, Prats Ferrer \cite{eynprats}, en utilisant le th\'eor\`eme II.\ref{TheoremEynPrats},
mais sortirait trop du cadre de cette habilitation.

\chapter{M\'ethode des boucles et G\'eom\'etrie alg\'ebrique}
\label{chapterloop}

La m\'ethode des boucles exploite les \'equations de Schwinger--Dyson, i.e.
l'invariance de l'int\'egrale par les reparam\'etrisations infinit\'esimales \cite{DGZ, courseynard, staudacher}.
Contrairement \`a la  m\'ethode des polyn\^omes biorthogonaux,
celle ci ne fait pas appel \`a l'int\'egrale d'Itzykson--Zuber--Harish-Chandra.

La m\'ethode des boucles s'applique au mod\`ele formel comme au mod\`ele normal.
C'est toutefois pour le mod\`ele formel qu'elle est le plus utile, car elle se d\'eveloppe facilement en puissances de $1/N^2$.
En particulier, la m\'ethode des boucles donne facilement acc\`es \`a la limite $N\to\infty$.

Dans la limite $N\to \infty$, la m\'ethode des boucles fait appara\^itre une courbe alg\'ebrique, qui n'est rien d'autre que la limite
de la courbe spectrale des polyn\^omes biorthogonaux vue en \refeq{defcourbespectrale}.
Toutes les observables du mod\`ele, dans la limite $N\to\infty$, s'expriment donc \`a partir de propri\'et\'es g\'eom\'etriques de cette courbe.
C'est le lien entre mod\`eles de matrices et g\'eom\'etrie alg\'ebrique.

\section{Equations de boucles}

\subsection{Changements de variables matriciels}

On consid\`ere un changement de variable du type ($\delta$ infinit\'esimal):
\beq
M_1\to M'_1= M_1 + \delta f(M_1,M_2)+ \ovl\delta f^\dagger(M_1,M_2)
\eeq
o\`u $f$ est une fonction analytique de ses deux variables (en pratique, ce sera un polyn\^ome).

La mesure $dM_1$ est multipli\'ee par un Jacobien:
\beq
dM'_1 = |J(f(M_1,M_2))|\,dM_1
\eeq
\bd
La variation $J_1(f)$ du Jacobien \`a l'ordre 1 en $|\delta|$, est d\'efinie telle que:
\beq
|J(f(M_1,M_2))|= \left(1+2\Re\left(\delta J_1(f(M_1,M_2))\right)+O(|\delta^2|)\right) \, dM_1
\eeq
\ed

De m\^eme, l'action $\ee{-{N\over T}\tr[V_1(M_1)+V_2(M_2)-M_1M_2]}$ est chang\'ee, \`a l'ordre 1 en $\delta$ par:
\bea
&& \ee{-{N\over T}\tr[V_1(M'_1)+V_2(M_2)-M'_1M_2]} \cr
&& = \left(1-2\Re\left({N\over T}\delta \tr [f(M_1,M_2)(V'_1(M_1)-M_2)]\right) +O(|\delta^2|) \right) \ee{-{N\over T}\tr[V_1(M_1)+V_2(M_2)-M_1M_2]} \cr
\eea
L'invariance de l'int\'egrale par changement de variable  $Z=\int dM_1 dM_2 \ee{-{N\over T}\tr[V_1(M_1)+V_2(M_2)-M_1M_2]} = \int dM'_1 dM_2 \ee{-{N\over T}\tr[V_1(M'_1)+V_2(M_2)-M'_1M_2]}$
implique:
\beq
\int dM_1 dM_2 \left(J_1(f(M_1,M_2))-{N\over T}\tr [f(M_1,M_2)(V'_1(M_1)-M_2)]\right) \ee{-N\tr[V_1(M_1)+V_2(M_2)-M_1M_2]} = 0
\eeq
i.e.
\bt\label{Theoremloopeqgeneralite}
 \'equation de boucles: pour tout changement de variables $f$, on a:
\beq\label{loopeqgeneralite}
{T^2\over N^2}\left<J_1(f(M_1,M_2))\right>={T\over N}\left<\tr [f(M_1,M_2)(V'_1(M_1)-M_2)]\right>
\eeq
\et
L'\'equation \refeq{loopeqgeneralite} constitue la forme la plus g\'en\'erale possible de ce que l'on appelle les \'equations de boucles.
Il s'agit maintenant de consid\'erer des $f(M_1,M_2)$ bien choisis.

Remarquons que $J_1$ est lin\'eaire, et que les r\`egles de d\'erivations en cha\^\i ne s'appliquent si $f$ est un produit.
Il suffit donc de calculer $J_1$ pour des fonctions $f$ assez simple et ensuite les combiner.

\bl\label{Lemmesplitmerge}
Pour toutes matrices $A$ et $B$, on a:
\beq\label{splitrulediscrete}
J_1(A M_1^k B) = \sum_{l=0}^{k-1} \tr A M_1^l\,\tr M_1^{k-1-l} B
\eeq
\beq\label{mergerulediscrete}
J_1(A \tr (M_1^k B)) = \sum_{l=0}^{k-1} \tr A M_1^l B M_1^{k-1-l}
\eeq
ou, sous forme de s\'eries g\'en\'eratrices:
\beq\label{splitrule}
J_1\left(A {1\over x-M_1} B\right) =  \tr A {1\over x-M_1}\,\tr {1\over x-M_1} B
\eeq
\beq\label{mergerule}
J_1\left(A \tr \left({1\over x-M_1} B\right)\right) = \tr A {1\over x-M_1}B {1\over x-M_1}
\eeq
\el
Les relations \refeq{splitrulediscrete}, \refeq{splitrule} sont parfois appel\'ees ''split-rule'', et les relations
 \refeq{mergerulediscrete}, \refeq{mergerule} sont parfois appel\'ees ''merge-rule''.

\Remark
Les \'equations de boucles du type \refeq{loopeqgeneralite}, sont \`a prendre au sens formel, c'est \`a dire qu'elles sont vraies ordre par ordre en puissances de $T$ et
en puissances de $1/x$ lorsque l'on fait intervenir des s\'eries g\'ene\'eratrices.
Elles peuvent aussi se d\'emontrer directement par la combinatoire des surfaces discr\'etis\'ees \cite{Tutte1, Tutte2} vue au paragraphe II.\ref{pardefmodeleformel}.
Les \'equations de boucles sont un outil standard pour les physiciens de la gravitation quantique \cite{DGZ, courseynard, ACKM, Matrixsurf, BIPZ}.
Pour le mod\`ele \`a deux matrices elles ont d'abbord \'et\'e consid\'er\'ee par M. Staudacher \cite{staudacher},
puis syst\'ematis\'ees par moi m\^eme \citeeynchain, \citeeynchaint, \citeeynmultimat, \citeeynmatgzero, \citeeynmatsg.

\subsection{Equation de boucles ma\^\i tresse}

Les \'equations de boucles donnent des relations entre les moments.
Il s'agit maintenant de trouver des changements de variables $f(M_1,M_2)$ bien choisis, afin de trouver une relation ferm\'ee entre un nombre fini de moments,
et ainsi pouvoir calculer tous les moments et observables.

\bd
D\'efinissons les observables suivantes (certaines ont d\'ej\`a \'et\'e d\'efinies au paragraphe II.\ref{sectiondefobservablesseries}):
\beq
W_1(x):={1\over N}\left<\tr {1\over x-M_1}\right>
\eeq
\beq
Y(x):=V'_1(x)-TW_1(x)
\eeq
\beq
W_2(y):={1\over N}\left<\tr {1\over y-M_2}\right>
\eeq
\beq
X(y):=V'_2(y)-TW_2(y)
\eeq
\beq
U(x,y):={1\over N}\left<\tr {1\over x-M_1}{V'_2(y)-V'_2(M_2)\over y-M_2}\right>
\eeq
\beq
P(x,y):={1\over N}\left<\tr {V'_1(x)-V'_1(M_1)\over x-M_1}{V'_2(y)-V'_2(M_2)\over y-M_2}\right>
\eeq
\beq
R(x):={1\over N}\left<\tr {1\over x-M_1}V'_2(M_2)\right>
\eeq
\beq
U(x,y;x'):=\left<\tr {1\over x-M_1}{V'_2(y)-V'_2(M_2)\over y-M_2}\tr {1\over x'-M_1}\right>-{N^2} U(x,y)W_1(x')
\eeq
\ed
Remarquons que $U(x,y)$ et $U(x,y;x')$ sont des polyn\^omes de degr\'e $\leq d_2-1$ en $y$, et $P(x,y)$ est un polyn\^ome
de degr\'e $d_1-1$ en $x$ et de degr\'e $d_2-1$ en $y$.
On introduit aussi:
\bd Le polyn\^ome $E(x,y)$, de degr\'e $d_1+1$ en $x$ et de degr\'e $d_2+1$ en $y$, est d\'efini par:
\beq\label{defEloop}
E(x,y):=(V'_1(x)-y)(V'_2(y)-x)-T P(x,y)+T
\eeq
\ed
Ce polyn\^ome va jouer un r\^ole crucial dans toute la suite. La  courbe $E(x,y)=0$ sera appel\'ee ''courbe spectrale''.

\bigskip

\bt\label{thloopeqU} (Eynard \citeeynmatgzero, \citeeynmatsg, \citeeynmultimat)
Pour tout $x$ et $y$, on a (au sens des s\'eries formelles):
\beq\label{loopeqU}
 (y-Y(x)) (TU(x,y)-V'_2(y)+x) + {T^2\over N^2} U(x,y;x)  = E(x,y)
\eeq
\et
En particulier, en prenant $y=Y(x)$
on obtient
\bt\label{Cormasterloopeq} (Eynard \citeeynchain, \citeeynchaint, \citeeynmatgzero, \citeeynmatsg, \citeeynmultimat)
{\bf Equation ma\^\i tresse}:
\beq\label{masterloopeq}
\encadremath{
E(x,Y(x))=  {T^2\over N^2} U(x,Y(x);x)
}\eeq
\et

{\noindent \bf Preuve:}

L'existence d'une courbe alg\'ebrique dans la limite $N$ grand avait \'et\'e pr\'evue par M. Staudacher \cite{staudacher} et explicit\'ee pour des potentiels cubiques.
La preuve la plus g\'en\'erale est donn\'ee dans \citeeynmultimat, et elle avait d\'ej\`a \'et\'e esquiss\'ee, dans la limite $N$ grand, dans \citeeynchain\ et \citeeynchaint.

Consid\'erons le changement de variable suivant:
\beq
f(M_1,M_2) = {1\over x-M_1} {V'_2(y)-V'_2(M_2)\over y-M_2}
\eeq
Le th\'eor\`eme IV.\ref{Theoremloopeqgeneralite}, avec le Lemme IV.\ref{Lemmesplitmerge}, donne:
\bea
&& T W_1(x) U(x,y) + {T\over N^2} U(x,y;x) \cr
&& = V'_1(x)U(x,y)-P(x,y)-yU(x,y)+V'_2(y) W_1(x)-R(x)
\eea
et avec le changement de variable suivant pour la matrice $M_2$
\beq
M_2\to M_2+\delta {1\over x-M_1}
\eeq
le th\'eor\`eme IV.\ref{Theoremloopeqgeneralite} donne:
\bea
R(x) = xW_1(x)-1
\eea
On en d\'eduit:
\bea
 (y-V'_1(x)+T W_1(x)) U(x,y) + {T\over N^2} U(x,y;x)  = (V'_2(y)-x) W_1(x)-P(x,y)+1
\eea
Et, en \'ecrivant $Y(x):=V'_1(x)- T W_1(x)$, on a:
\beq
 (y-Y(x)) U(x,y) + {T\over N^2} U(x,y;x)  = (V'_2(y)-x) W_1(x)-P(x,y)+1
\eeq
En multipliant par $T$, on obtient le th\'eor\`eme IV.\ref{thloopeqU}.
En corollaire, en prenant $y=Y(x)$, on obtient le th\'eor\`eme IV.\ref{Cormasterloopeq}.

\subsection{Fonction de corr\'elation mixte}

De fa\c con tr\`es similaire, on peut obtenir une \'equation de boucle pour l'observable mixte
$W_{1,2}(x,y):={1\over N}\left<\tr {1\over x-M_1}{1\over y-M_2}\right>$.

Posons
\beq
W_{1,2;2}(x,y;y'):=\left<\tr {1\over x-M_1}{1\over y-M_2}\tr {1\over y'-M_2}\right> - {N^2}W_{1,2}(x,y)W_2(y')
\eeq

\bt\label{Theoremtracemixteloop} (Eynard \citeeynmatgzero, \citeeynmatsg, \citeeynmultimat)
 $W_{1,2}(x,y)$ satisfait l'\'equation:
\beq\label{tracemixteloop}
T W_{1,2}(x,y) = 1-{E(x,y)\over (x-X(y))(y-Y(x))}+{T^2\over N^2}\left({U(x,y,x)\over (x-X(y))(y-Y(x))} - {W_{1,2;2}(x,y;y)\over x-X(y)}\right)
\eeq
\et

{\noindent \bf preuve:}

Avec le changement de variable
\beq
M_2\to M_2+{\delta\over 2} {1\over x-M_1} {1\over y-M_2}
+{\delta\over 2} {1\over y-M_2} {1\over x-M_1}
\eeq
on obtient:
\bea
&&  T W_2(y) W_{1,2}(x,y) + {T\over N^2} W_{1,2;2}(x,y;y) \cr
&& = V'_2(y)W_{1,2}(x,y)-U(x,y)-xW_{1,2}(x,y)+W_2(y)
\eea
i.e.
\beq
 (x-V'_2(y)+T W_2(y)) W_{1,2}(x,y) + {T\over N^2} W_{1,2;2}(x,y;y)
 = W_2(y)-U(x,y)
\eeq
Posons:
\beq
X(y):=V'_2(y)- T W_2(y)
\eeq
on a:
\bea
 (x-X(y)) TW_{1,2}(x,y) + {T^2\over N^2} W_{1,2;2}(x,y;y)
 &=& V'_2(y)-X(y) -TU(x,y) \cr
& =& x-X(y)-{E(x,y)-{T^2\over N^2}U(x,y;x)\over y-Y(x)}
\eea
QED.

\medskip

\Remark
\refeq{tracemixteloop} dans la limite $N$ grand (i.e. si on n\'eglige le terme en $1/N^2$), on a:
\beq\label{tracemixtelooplargeN}
T W_{1,2}(x,y) = 1-{E(x,y)\over (x-X(y))(y-Y(x))} +O(1/N^2)
\eeq
qui n'est pas sans rappeler \refeq{tracemixten}
pour $n$ fini (il faut noter que \refeq{tracemixteloop} a \'et\'e calcul\'ee pour le mod\`ele formel, alors que \refeq{tracemixten} a \'et\'e calcul\'ee pour le mod\`ele normal).

\section{Limite $N$ grand}

Notons que pour le mod\`ele formel, par d\'efinition, les limites $N$ grand et le d\'eveloppement en puissances de $1/N^2$ existent.
Toutes les fonctions vues pr\'ec\'edement se d\'eveloppent donc:
\beq
W_1(x) = W^{(0)}_1(x) + {1\over N^2}W^{(1)}_1(x) + \dots
\eeq
\beq
Y^{(0)}(x) := V'_1(x)-T W^{(0)}_1(x)
\virg
Y^{(1)}(x) := -T W^{(1)}_1(x)
\,\, \dots
\eeq
\beq
E(x,y) = E^{(0)}(x,y) + {1\over N^2}E^{(1)}(x,y) + \dots
\eeq
etc...

\bigskip

Dans la limite $N$ grand, on enl\`eve simplement le membre de droite de \refeq{masterloopeq} et l'on obtient une \'equation alg\'ebrique
pour $Y^{(0)}$ comme fonction de $x$.
On peut \'egalement calculer la r\'esolvante $W^{(0)}_2(y)$ et $X^{(0)}(y)=V'_2(y)-T W^{(0)}_2(y)$ par la m\^eme m\'ethode.
On a donc:
\bt\label{TheoremmasterloopNlarge}
$Y^{(0)}(x)$ et $X^{(0)}(y)$ satisfont la m\^eme \'equation alg\'ebrique:
\beq\label{masterloopNlarge}
E^{(0)}(x,Y^{(0)}(x)) =  0
\virg
E^{(0)}(X^{(0)}(y),y) =  0
\eeq
\et
Comme dit plus haut, une relation alg\'ebrique \'etait connue depuis les travaux de M. Staudacher \cite{staudacher},
et l'\'equation \refeq{masterloopNlarge} a \'et\'e trouv\'ee explicitement dans \citeeynchaint\ et \citeeynchain.

\br On doit avoir:
\beq
Y^{(0)}oX^{(0)}={\rm Id}
\eeq
conform\'ement \`a ce qui a \'et\'e trouv\'e par Matytsin \cite{Matytsin} puis prouv\'e par \cite{guionet} (voir aussi \cite{ZinnZuber}).
\er

\br Rappelons que $W^{(0)}_1(x)$ a \'et\'e d\'efinie comme une s\'erie formelle en puissances de $1/x$.
Nous voyons ici que cette s\'erie a un rayon de convergence non nul. C'est une fonction analytique, et plus pr\'ecis\'ement alg\'ebrique de $x$.
Elle poss\`ede des coupures (et donc un rayon de convergence fini).
\er

\subsection{Courbe alg\'ebrique}

Nous avons donc une courbe alg\'ebrique ${\cal E}$, d'\'equation $E^{(0)}(x,y)=0$, c'est \`a dire une surface de Riemann compacte de genre $g$,
sur laquelle sont d\'efinies deux fonctions $x$ et $y$, telles que pour tout point $p$ de la courbe ${\cal E}$ on ait:
\beq
Y^{(0)}(x(p))=y(p)
\eeq
La fonction $Y^{(0)}(x)$ est solution d'une \'equation de degr\'e $d_2+1$, elle est multivalu\'ee, elle
poss\`ede $d_2+1$ branches.
Autrement dit, la fonction $x(p)$ n'est pas injective, pour chaque $x$, il existe exactement $d_2+1$ points $p\in{\cal E}$ tels que $x(p)=x$.

Commen\c cons par \'etudier ces branches pour $x$ grand.

\subsubsection{Points \`a l'infini}

\bt
Les fonctions $x$ et $y$ poss\`edent seulement deux p\^oles sur ${\cal E}$, de diviseurs $[x]=\infty_x+d_2\infty_y$, et $[y]=\infty_y+d_1\infty_x$,
et l'on a au voisinage de ces p\^oles:
\beq\label{eqresinftyxgk}
\forall k=1,\dots,d_1 \,\, , \,\,\, \Res_{\infty_x} x^{-k} y dx = -g_{k}
\eeq
\beq\label{eqresinftyytdgk}
\forall k=1,\dots,d_2 \,\, , \,\,\, \Res_{\infty_y} y^{-k} x dy = -\td{g}_{k}
\eeq
\beq\label{eqresinftyT}
\Res_{\infty_x} y dx = \Res_{\infty_y} x dy = T
\eeq
\et

{\noindent \bf preuve:}
La r\'esolvante $W_1(x)$ est d\'efinie comme une s\'erie formelle en puissances de $1/x$ telle que $W_1(x)={1\over x} + O({1\over x^2})$.
Il existe donc une solution de \refeq{masterloopNlarge}, telle que
$Y^{(0)}(x)\sim V'_1(x)-{T\over x}+O(x^{-2})$, ce qui implique qu'il existe au moins un point $\infty_x\in {\cal E}$ tel que:
\beq
x(\infty_x)=\infty
\virg
y(p)\mathop\sim_{p\to \infty_x} V'_1(x(p)) - {T\over x(p)} + O(x(p))^{-2}
\eeq
De m\^eme, pour $y$ grand, il est clair qu'il existe une solution de \refeq{masterloopNlarge}, telle que
$x\sim V'_2(Y^{(0)}(x)))-{T\over Y^{(0)}(x)}+O(Y^{(0)}(x)^{-2})$, il existe donc au moins un point $\infty_y\in {\cal E}$ tel que:
\beq
y(\infty_y)=\infty
\virg
x(p)\mathop\sim_{p\to \infty_y} V'_2(y(p)) - {T\over y(p)} + O(y(p))^{-2}
\eeq
Autrement dit, pr\`es de $\infty_y$, $y(p)\sim x(p)^{1/d_2}$, et comme il y a $d_2$ racines $d_2$-i\`emes de l'unit\'e,
il y a $d_2$ branches de la fonction $Y(x)$ pr\`es de $\infty_y$.
Comme la fonction $Y^{(0)}(x)$ poss\`ede exactement $d_2+1$ branches, cel\`a implique que les fonctions $x(p)$ et $y(p)$ n'ont
pas d'autres poles que $\infty_x$ et $\infty_y$, et que:
$\infty_x$ est un p\^ole simple de la fonction $x(p)$, et un p\^ole de degr\'e $d_1$ de la fonction $y(p)$,
et $\infty_y$ est un p\^ole simple de la fonction $y(p)$, et un p\^ole de degr\'e $d_2$ de la fonction $x(p)$.
QED.

\subsubsection{Fractions de remplissage}

La d\'efinition du mod\`ele de matrice formel II.\ref{definitionZformel} implique, que pour $T$ assez petit:
\bp
Il existe $d_1d_2$ contours disjoints ${\cal A}_i$, $i=1,\dots,d_1 d_2$ dans le plan complexe de $x$, tels que:
\beq\label{Acycleepsilonicomplex}
\forall i=1,\dots, d_1 d_2-1 \qquad \oint_{{\cal A}_i} Y(x) dx  = 2i\pi T \epsilon_i := 2i\pi \eta_i
\eeq
o\`u les $\epsilon_i$ sont les fractions de remplissages donn\'ees du mod\`ele formel (voir section II.\ref{pardefmodeleformel}),
et o\`u les $\eta_i$ sont d\'efinis par $\eta_i:=T\epsilon_i$.
\ep
{\noindent id\'ee de la preuve:}
en effet, pour $T$ tendant vers $0$, les ${\cal A}_i$ peuvent \^etre choisis comme des cercles de rayon $O(|T|)$ et centr\'es sur les $d_1 d_2$ valeurs de $x$
telles que $V'_2(V'_1(x))-x=0$ (voir section II.\ref{pardefmodeleformel}).
L'int\'egrale de la r\'esolvante $W_1(x)dx$ sur un tel contour est le ''nombre de valeurs propres'' de $M_1$ \`a l'int\'erieur du domaine d\'elimit\'e par ce contour,
c'est donc par d\'efinition $-2i\pi \epsilon_i$.
Dans la limite $N\to\infty$, la r\'esolvante et la courbe ${\cal E}$ sont des fonctions analytiques de $T$ (avec un rayon de convergence $T_c$),
et l'on obtient des contours ${\cal A}_i$ pour $T$ fini par continuit\'e.

\subsubsection{Cycles non-triviaux}

Les contours ${\cal A}_i$ peuvent \^etre relev\'es en cycles non-triviaux, not\'es aussi ${\cal A}_i$ sur la courbe alg\'ebrique ${\cal E}$.

Sur une surface de genre $g$, il y a $2g$ cylces non-triviaux ind\'ependants, et il est possible d'en choisir une base canonique
${\cal A}_i, {\cal B}_i$, $i=1,\dots, g$, tels que:
\beq
{\cal A}_i \cap {\cal A}_j  = 0
\virg
{\cal B}_i \cap {\cal B}_j  = 0
\virg
{\cal A}_i \cap {\cal B}_j  = \delta_{i,j}
\eeq
Les contours ${\cal A}_i$ satisfont:
\beq\label{Acycleepsiloni}
\forall i=1,\dots, d_1 d_2-1 \qquad \oint_{{\cal A}_i} Y(x) dx  = 2i\pi T \epsilon_i := 2i\pi \eta_i
\eeq

\subsubsection{Genre}

Une courbe alg\'ebrique de la forme \refeq{masterloopNlarge} avec $E$ de la forme \refeq{defEloop} a un genre $g\leq d_1 d_2-1$, et a g\'en\'eriquement le genre maximal.

Une courbe de genre $g$ poss\`ede exactement $g$ cycles disjoints non-triviaux ind\'ependents.
Autrement dit, la courbe ne peut \^etre d\'eg\'en\'er\'ee (i.e. $g<d_1 d_2-1$) que
si certaines combinaisons lin\'eaires enti\`eres d'int\'egrales \refeq{Acycleepsiloni} sont nulles,
autrement dit si certaines combinaisons lin\'eaires enti\`eres des fractions de remplissage sont nulles
(pour $T$ petit, si certaines fractions de remplissages sont nulles).

Nous supposerons \`a partir d'ici, que $E^{(0)}(x,y)$ est un polyn\^ome connu, de la forme \refeq{defEloop},
et que les conditions  \refeq{eqresinftyT}, \refeq{eqresinftyxgk}, \refeq{eqresinftyytdgk} et \refeq{Acycleepsiloni} sont satisfaites.

\subsubsection{Feuillets}

La fonction $x(p)$ n'est pas injective, pour chaque $x$, il existe $d_2+1$ points $p$ tels que $x(p)=x$.
On note:
\bd
\beq\label{defpix}
x(p)=x \quad \leftrightarrow \quad p\in(p_0(x),p_1(x),\dots,p_{d_2}(x))
\eeq
et l'on suppose que $p_0(x)$ est tel que $p_0(x)\to \infty_x$ lorsque $x\to\infty$, et pour $k\geq 1$ $p_k(x)\to \infty_y$.
On d\'ecoupe ainsi la courbe ${\cal E}$ en $d_2+1$ domaines disjoints appel\'es ''feuillets'', ${\cal E}_{0},\dots,{\cal E}_{d_2}$,
tels que $p_k(x)\in {\cal E}_{k}$, et la fonction $x(p)$ restreinte \`a ${\cal E}_{k}$ est une bijection.
\ed

Le feuillet num\'ero $0$, i.e. celui qui contient $\infty_x$ est appel\'e {\bf feuillet physique}.
C'est seulement dans ce feuillet que la  fonction
$W_1(x)={1\over T}(V'_1(x)-y(p_0(x)))$ se d\'eveloppe en puissances de $1/x$ pour $x$ grand, et coincide avec la fonction g\'en\'eratrice des moments \refeq{defWx}.

\medskip
De m\^eme on note:
\beq
y(p)=y \quad \leftrightarrow \quad p\in(\td{p}_0(y),\td{p}_1(y),\dots,\td{p}_{d_1}(y))
\virg
\td{p}_k(y)\in \td{\cal E}_{k}
\eeq
Le feuillet num\'ero $0$, i.e. celui qui contient $\infty_y$ est appel\'e feuillet physique.
C'est dans ce feuillet que la  fonction
$W_2(y)={1\over T}(V'_2(y)-x(\td{p}_0(y)))$ se d\'eveloppe en puissances de $1/y$ et coincide avec la fonction g\'en\'eratrice des moments \refeq{defWy}.

\medskip

Le choix des domaines ${\cal E}_{k}$ (resp. $\td{\cal E}_{k}$) et de leurs fronti\`eres est essentiellement arbitraire.
Les images des fronti\`eres par $x(p)$ (resp. $y(p)$) sont appel\'ees les coupures.

La seule contrainte est que les points multiples de $x(p)$ (resp. $y(p)$) soient sur les coupures, ce sont les points de branchements.

\subsubsection{Points de branchements et coupures}
\label{sectionptbrancht}

Les points multiples de $x(p)$ sont tels que $\exists k\neq l$ tels que $p_k(x)=p_l(x)$, ce sont les z\'eros de la diff\'erentielle $dx(p)$.
La diff\'erentielle $dx(p)$ poss\`ede deux p\^oles: un p\^ole double en $\infty_x$ et un p\^ole
de degr\'e $d_2+1$ en $\infty_y$.
cel\`a implique qu'elle doit avoir $d_2+1+2g$ z\'eros, que l'on note $e_1,\dots, e_{d_2+1+2g}$:
\beq
dx(p)=0  \quad \leftrightarrow \quad p\in(e_1,\dots, e_{d_2+1+2g})
\eeq
De m\^eme la diff\'erentielle $dy(p)$ poss\`ede deux p\^oles: un p\^ole double en $\infty_y$ et un p\^ole
de degr\'e $d_1+1$ en $\infty_x$.
cel\`a implique qu'elle doit avoir $d_1+1+2g$ z\'eros, que l'on note:
\beq
dy(p)=0  \quad \leftrightarrow \quad p\in(\td{e}_1,\dots, \td{e}_{d_1+1+2g})
\eeq

\bd
Les z\'eros $e_1,\dots, e_{d_2+1+2g}$ (resp. $\td{e}_1,\dots, \td{e}_{d_1+1+2g}$) de la diff\'erentielle $dx$ (resp. $dy$) sont appel\'es ''points de branchements'' de $x$ (resp. $y$).
\ed

Supposons ici que tous ces z\'eros sont {\bf simples et distincts}.
Pr\`es d'un z\'ero simple $e$ de $dx$, la fonction $Y(x)$ se comporte en racine carr\'ee;
\beq
Y(x)-Y(e)\sim \sqrt{x-x(e)}
\eeq

\br{Points critiques}
\label{sectionptcrit}

Les cas o\`u les z\'eros de $dx$ et $dy$ ne sont pas simples et distincts, s'appellent points critiques.
L'\'etude des points critiques est cruciale dans le cadre des applications des mod\`eles de matrices aux th\'eories conformes, mais sort largement du cadre de cette habilitation.
Voir par exemple \cite{Kazakov, KazakoVDK, DGZ, BlIt,BlEycrit}...

Notons que si l'on est a un point critique, et que $e$ est \`a la fois un z\'ero de degr\'e $q-1$ de $dx$ et un z\'ero de degr\'e $p-1$ de $dy$, on a une singularit\'e rationelle:
\beq
Y(x)-Y(e)\sim (x-x(e))^{p/q}
\eeq
Les points critiques d\'ecrivent donc des surfaces de Riemann singuli\`eres.
Il est connu que les mod\`eles de matrices avec singularit\'es rationelles d'exposant $p/q$, d\'ecrivent les mod\`eles minimaux conformes $(p,q)$, voir \cite{KazakoVDK}.

\er

\subsection{El\'ements de g\'eom\'etrie alg\'ebrique}

Voir \cite{Fay, Farkas, algeo} pour une introduction.

\subsubsection{diff\'erentielles holomorphes}
\label{sectiondiffhol}

Sur une surface de genre $g$, il existe $g$ diff\'erentielles holomorphes (i.e. sans poles) ind\'ependantes $du_i(p)$, $i=1,\dots,g$.
On peut les choisir telles que:
\beq
\oint_{{\cal A}_k} du_l(p) = \delta_{k,l}
\eeq
On d\'efinit alors la matrice des p\'eriodes par les int\'egrales sur les ${\cal B}$--cyles:
\beq
\tau_{k,l}:=\oint_{{\cal B}_k} du_l(p)
\eeq
Un r\'esultat classique de g\'eom\'etrie alg\'ebrique \cite{Farkas, Fay} montre que $\tau_{k,l}=\tau_{l,k}$ et $\Im \tau_{k,l}>0$.

\subsubsection{Fonction d'Abel}

Donnons nous un point $p_O\in{\cal E}$, arbitraire.
On d\'efinit alors la fonction d'Abel:
\bea
{\cal E}\backslash(\cup_i {\cal A}_i \cup_i {\cal B}_i) &\to& \C^g\cr
p &\to& u(p) \qquad \hbox{o\`u} \,\, u_i(p):=\int_{p_O}^{p} du_i
\eea
o\`u le chemin d'int\'egration ne do\^\i t croiser aucun des cycles ${\cal A}_i$, ${\cal B}_j$.

Cette fonction s'\'etend naturellement comme un plongement de ${\cal E}$ dans la Jacobienne $\C^g/(Z^g+\tau Z^g)$:
\bea
{\cal E} &\to& \C^g/(Z^g+\tau Z^g) \cr
p &\to& [u(p)] \,\,{\rm modulo}\,Z^g+\tau Z^g
\eea

\subsubsection{Fonctions theta}
\label{deffctiontheta}

Etant donn\'e une matrice $\tau$ de taille $g\times g$, telle que $\Im\tau>0$, on d\'efinit la fonction enti\`ere suivante:
\beq
\theta:\quad
\begin{array}{ccl}
\C^g & \to & \C \cr
u & \to & \theta(u,\tau)=\sum_{n\in Z^g}\, \ee{i\pi (n^t\tau n)}\,\ee{2i\pi (n^t u)}
\end{array}
\eeq
Par abus de notation, on ometra souvent la d\'ependance en $\tau$.

La fonction $\theta$ poss\`ede les propri\'et\'es suivantes, pour tout $n\in Z^g$:
\beq
\theta(u+n)=\theta(u)
\virg
\theta(u+\tau n)=\theta(u) \, \ee{-i\pi(2(n^t u)+(n^t \tau n))}
\virg
\theta(-u)=\theta(u)
\eeq

L'ensemble des z\'eros de la fonction $\theta$ est une sous-vari\'et\'e de codimension $1$ de $\C^g$, not\'ee $[\theta]$.
En particulier, il est clair que les points demi-entiers suivants appartiennent \`a $[\theta]$:
\beq
{n+\tau m\over 2} \in [\theta] \qquad {\rm si}\,\, (m^t n) \,\,\, {\rm est \, impair}
\eeq
Ils sont appel\'es demi--p\'eriodes impaires.

\subsubsection{Formes premi\`eres et fonctions Theta}

Soit $z$ une demi--p\'eriode impaire.
Consid\'erons la forme diff\'erentielle holomorphe suivante:
\beq
dh_z(p):= \sum_{i=1}^g \left.{\d \theta(u)\over \d u_i}\right|_{u=z} \, du_i(p)
\eeq
Alors, on d\'efinit la forme primaire sur ${\cal E}\times {\cal E}$
\beq
E(p,q):={\theta(u(p)-u(q)+z)\over \sqrt{dh_z(p) dh_z(q)}}
\eeq
dont on peut montrer qu'elle ne d\'epend pas du choix de $z$, ni du choix de l'origine $p_O$.
$E(p,q)$ s'annule si et seulement si $p=q$.

\subsubsection{Noyeau de Bergmann}

Le noyeau de Bergmann $B(p,q)$ est l'unique forme diff\'erentielle bilin\'eaire,
qui, vue comme forme diff\'erentielle en $p$ poss\`ede un unique pole double \`a $p=q$, sans r\'esidu,
tel que dans n'importequelle param\'etrisation $x(p)$ on ait:
\beq
B(p,q) \mathop\sim_{p\to q} {dx(p) dx(q)\over (x(p)-x(q))^2} + {\rm fini}
\eeq
et telle que:
\beq
\forall i=1,\dots,g \qquad \oint_{p\in{\cal A}_i} B(p,q)=0
\eeq
Cette forme est unique,  elle est sym\'etrique $B(p,q)=B(q,p)$, et elle est donn\'ee par:
\beq
B(p,q):=d_p d_q \ln{\theta(u(p)-u(q)+z)}
=d_p d_q \ln{E(p,q)}
\eeq

On a la propri\'et\'e:
\beq
\forall i=1,\dots,g \qquad \oint_{p\in{\cal B}_i} B(p,q)=2i\pi\,du_i(q)
\eeq

\subsubsection{Connexion projective}

Etant donn\'e une fonction $f(p)$ \`a valeur complexes, non constante, on d\'efinit:
\beq
{1\over 6}\,{\cal S}_f(p):=\mathop{\rm lim}_{q\to p}\, \left({B(p,q)\over df(p)df(q)}-{1\over (f(p)-f(q))^2}\right)
\eeq

\subsubsection{Diff\'erentielles de 3e esp\`eces}

Etant donn\'es deux points $p_1$ et $p_2$ distincts sur ${\cal E}$, il existe une unique forme diff\'erentielle
$dS_{p_1,p_2}(p)$ telle que:
$dS_{p_1,p_2}$ poss\`ede un pole simple \`a $p=p_1$ avec r\'esidu $+1$,
un pole simple \`a $p=p_2$ avec r\'esidu $-1$,
et pas d'autres poles, et
\beq\label{defdSpq}
\forall i=1,\dots,g \qquad \oint_{p\in{\cal A}_i} dS_{p_1,p_2}(p)=0
\eeq
Cette forme est unique, et donn\'ee par:
\beq
dS_{p_1,p_2}(p) = d \ln{E(p,p_1)\over E(p,p_2)} = \int_{q=p_2}^{p_1} B(p,q)
\eeq
o\`u cette derni\`ere int\'egrale est calcul\'ee le long d'un chemin qui ne croise aucun cycle ${\cal A}_i$, ${\cal B}_j$.

On a la propri\'et\'e:
\beq
\forall i=1,\dots,g \qquad \oint_{p\in{\cal B}_i} dS_{p_1,p_2}(p) = 2i\pi\,(u_i(p_1)-u_i(p_2))
\eeq

\subsection{Observables, \'energie libre, et leurs d\'eriv\'ees}

Les r\'esultats pr\'esent\'es dans ce paragraphe sont classiques et ont \'et\'e d\'evelopp\'es par de nombreux auteurs.
Voir \citeeynmultimat, \cite{kriechever,KazMar, WZ, marcoF}. Ils sont pr\'esent\'es pour rendre l'expos\'e complet, et car ils seront utilis\'es au chapitre V.

\bt Les fonctions \`a deux points sont donn\'ees par:
\beq\label{WxxBergmann}
W^{(0)}_{1;1}(x(p);x(q))\,\, dx(p)\,dx(q) = B(p,q) - {dx(p)dx(q)\over (x(p)-x(q))^2}
\eeq
\beq
W^{(0)}_{2;2}(y(p);y(q))\,\, dy(p)\,dy(q) = B(p,q) - {dy(p)dy(q)\over (y(p)-y(q))^2}
\eeq
\beq
W^{(0)}_{1;2}(x(p);y(q))\,\, dx(p)\,dy(q) = -B(p,q)
\eeq
\et
o\`u $B$ est le noyau de Bergmann, et
o\`u les $W^{(0)}_{k;l}$ sont les limites $N\to\infty$ des observables $W_{k;l}$ qui ont \'et\'e d\'efinies au paragraphe II.\ref{sectiondefobservablesseries}:
\beq
W_{1;1}(x;x'):=\left<\tr {1\over x-M_1}\tr {1\over x'-M_1}\right>_{\rm c}
\eeq
\beq
W_{2;2}(y;y'):=\left<\tr {1\over y-M_2}\tr {1\over y'-M_2}\right>_{\rm c}
\eeq
\beq
W_{1;2}(x;y):=\left<\tr {1\over x-M_1}\tr {1\over y-M_2}\right>_{\rm c}
\eeq

\bc
En particulier, \`a points coincidants:
\beq
W^{(0)}_{1;1}(x(p);x(p)) = {1\over 6}\,{\cal S}_x(p)
\virg
W^{(0)}_{2;2}(y(p);y(p)) = {1\over 6}\,{\cal S}_y(p)
\eeq
\ec

\bt\label{TheoremderivW} Les d\'eriv\'ees de la r\'esolvante sont donn\'ees par:
\beq
\left.{\d TW^{(0)}_1(x(p))\over \d g_k}\right|_{x(p)}\, dx(p)
=  x(p)^{k-1}\, dx(p) + {1\over k}\,\mathop\Res_{q\to\infty_x} x(q)^k\,B(p,q)
\eeq
\beq
\left.{\d TW^{(0)}_1(x(p))\over \d \td{g}_k}\right|_{x(p)}\, dx(p) = -{1\over k}\,\mathop\Res_{q\to\infty_y} y(q)^k\,B(p,q)
\eeq
\beq
\left.{\d TW^{(0)}_1(x(p))\over \d T}\right|_{x(p)}\, dx(p) = - {dS_{\infty_x,\infty_y}(p)}
\eeq
\beq
\left.{\d TW^{(0)}_1(x(p))\over \d \eta_i}\right|_{x(p)}\, dx(p) = -2i\pi {du_i(p)}
\eeq
\et

\subsection{Energie libre et ses d\'eriv\'ees premi\`eres et secondes}

De m\^eme, les formules suivantes sont classiques (voir \citeeynmultimat, \cite{marcoF, kriechever,KazMar, WZ}.

\bt\label{energielibreleading}
La limite $N$ grand de l'\'energie libre est donn\'ee par:
\beq\label{energielibre}
F^{(0)}={1\over 2}\left( \sum_{k=0}^{d_1} g_k {\d F^{(0)}\over \d g_k} + \sum_{k=0}^{d_2} \td{g}_k {\d F^{(0)}\over \d \td{g}_k}
+ \sum_{i=1}^{g} \eta_i {\d F^{(0)}\over \d \eta_i} + T  {\d F^{(0)}\over \d T} - {1\over2}\mathop\Res_{\infty_x} xy^2 dx \right)
\eeq
avec:
\beq
{\d F^{(0)}\over \d g_k} = {1\over k}\,\mathop\Res_{\infty_x} x^k y dx
\virg
{\d F^{(0)}\over \d \td{g}_k} = {1\over k}\,\mathop\Res_{\infty_y} y^k  x dy
\virg
{\d F^{(0)}\over \d \eta_i} = \oint_{{\cal B}_i} y dx
\virg
\eeq
\bea
{\d F^{(0)}\over \d T} &=& \int_{\infty_x}^p (y-V'_1(x)+{T\over x}) dx + \int_{\infty_y}^p (x-V'_2(y)+{T\over y}) dy \cr
&& +V_1(x(p))+ V_2(y(p))-x(p)y(p)-T\ln{x(p)}-T\ln{y(p)}
\eea
cette derni\`ere expression \'etant ind\'ependante du point $p$.
On peut la voir comme une r\'egularisation de l'int\'egrale $\int_{\infty_x}^{\infty_y} ydx$.
\et

\br
si l'on introduit un coefficent $h$ devant le $\Tr M_1 M_2$:
\beq
Z = \int dM_1 dM_2 \ee{-{N\over T}\Tr[V_1(M_1)+V_2(M_2)-h M_1 M_2]}
\eeq
alors on a:
\beq
-{1\over2}\mathop\Res_{\infty_x} xy^2 dx = \left.{\d F^{(0)}\over \d h}\right|_{h=1}
\eeq
ce  qui rend l'expression eq.\ref{energielibre} homog\`ene.
\er

\bt\label{thenergielibrederiv2}
Les deriv\'ees secondes de $F^{(0)}$ sont donn\'ees par:
\beq
{\d^2 F^{(0)}\over \d g_k\d g_j} = \Res_{\infty_x}\Res_{\infty_x} x(p)^k x(q)^j B(p,q)
\eeq
\beq
{\d^2 F^{(0)}\over \d \td{g}_k\d \td{g}_j} = \Res_{\infty_y}\Res_{\infty_y} y(p)^k y(q)^j B(p,q)
\eeq
\beq
{\d^2 F^{(0)}\over \d g_k\d \td{g}_j} = \Res_{\infty_x}\Res_{\infty_y} x(p)^k y(q)^j B(p,q)
\eeq
\beq\label{ddFdetadeta}
{\d^2 F^{(0)}\over \d \eta_i\eta_j} = -2i\pi\, \tau_{ij}
\virg
{\d^2 F^{(0)}\over \d T\d\eta_i} = -2i\pi (u_i(\infty_x)-u_i(\infty_y))
\eeq
\beq
{\d^2 F^{(0)}\over \d T^2} = -\ln{\gamma\td\gamma}
\virg
\gamma:=\mathop{\rm lim}_{p\to\infty_x} x(p){E(p,\infty_x)\over E(p,\infty_y)}
\virg
\td\gamma:=\mathop{\rm lim}_{p\to\infty_y} y(p){E(p,\infty_y)\over E(p,\infty_x)}
\eeq
...etc.
\et

\subsection{Autres d\'eriv\'ees}

Les d\'eriv\'ees troisi\`emes de l'\'energie libre ont \'et\'e trouv\'ees par plusieurs auteurs \cite{kriechever, marcoF},
elles se r\'eduisent esentiellement \`a la formulle variationelle de Rauch \cite{Farkas, Fay}.
Pour calculer les d\'eriv\'ees suivantes, il suffit en principe d'appliquer r\'ecursivement la formule variationelle de Rauch,
mais en pratique les expressions obtenues ainsi sont assez fastidieuses, et leur structure ne saute pas aux yeux.

Une m\'ethode diagrammatique a r\'ecement \'et\'e invent\'ee, par moi m\^eme \cite{eynloop1mat} pour le mod\`ele
\`a une matrice, et a \'et\'e g\'en\'eralis\'ee au mod\`ele \`a deux matrices avec mon \'etudiant N. Orantin \cite{eoloop2mat}.
Elle permet d'\'ecrire directement n'importe quelle d\'eriv\'ee, comme une somme de diagrammes de Feynman en arbres.
Cette m\'ethode est non r\'ecursive, et met en \'evidence une structure tr\`es riche.
H\'elas, elle sort trop du cadre de cette habilitation...

\section{D\'eveloppement topologique}

On peut d\'evelopper l'\'equation \refeq{masterloopeq} au premier ordre en $1/N^2$:
\beq
W_1(x) = W_1^{(0)}(x) + {1\over N^2} W_1^{(1)}(x) + \dots
\eeq
Nous venons de voir que $W^{(0)}_1(x)$ est solution d'une \'equation alg\'ebrique.
Nous voulons maintenant calculer $W^{(1)}_1(x)$.
Ceci a \'et\'e fait dans \citeeynmatgzero (pour $g=0$) ou \citeeynmatsg (pour $g$ quelconque).
Je vais pr\'esenter ici, une m\'ethode diff\'erente de \citeeynmatgzero ou \citeeynmatsg, et inspir\'ee de mon r\'ecent travail \cite{eynloop1mat, eoloop2mat} pour obtenir le th\'eor\`eme suivant:

\bt\label{Theoremresolvantesubleading} (Eynard, Kokotov, Korotkin \citeeynmatgzero, \citeeynmatsg), (Eynard, Orantin \cite{eoloop2mat}),
 La r\'esolvante \`a l'ordre $1/N^2$ est donn\'ee par:
\beq\label{Wdevtopores}
TW_1^{(1)}(x(p))\, dx(p)
=  {1\over 2}\, \sum_{i=1}^{d_2+1+2g} \mathop\Res_{q\to e_i}  \mathop\Res_{p'\to q} {dS_{q,p'}(p)\,B(q,p')\over (y(q)-y(p'))(x(q)-x(p'))}
\eeq
\et

{\noindent \bf Id\'ee de la preuve:}

L'\'equation ma\^\i tresse \refeq{masterloopeq} d\'evelopp\'ee au premier ordre en $1/N^2$ donne:
\beq\label{masterloopnextorder}
T W_1^{(1)}(x) = - {T^2U^{(0)}(x,Y(x);x) + T P^{(1)}(x,Y(x))\over E_y(x,Y(x))}
\eeq
De plus, on a (nous ne d\'etaillerons pas ici le calcul (voir \citeeynmatgzero, \citeeynmatsg), qui permet de trouver $U^{(0)}(x,y;x')$  \`a partir de \refeq{loopeqU} et \refeq{WxxBergmann}):
\beq\label{Uxyx}
{T^2 U^{(0)}(x(q),y(q);x(q)) \over E_y(x(q),y(q))}\, dx(q)= - \sum_{k=1}^{d_2} {B(q,p_k(q))\over (y(q)-y(p_k(q)))\,dx(p_k(q))}
\eeq
o\`u les $p_k(q)$ sont les solutions de $x(p_k(q))=x(q)$ dans les autres feuillets (on suppose $p_0(q)=q$).

En manipulant \refeq{masterloopnextorder}, on peut montrer (cf \citeeynmatsg) que $W_1^{(1)}(x(p))\,dx(p)$ est une diff\'erentielle sur ${\cal E}$.
L'hypoth\`ese \refeq{Acycleepsiloni} du mod\`ele formel, implique que:
\beq\label{Acycleepsiloninextorder}
\forall i=1,\dots, g \qquad \oint_{{\cal A}_i} W_1^{(1)}(x)\, dx  = 0
\eeq
et implique aussi que $W_1^{(1)}(x)\, dx$ ne poss\`ede pas d'autres r\'esidus sur ${\cal E}$.
En particulier $W_1^{(1)}(x)\, dx$ ne poss\`ede pas de poles aux z\'eros de $E_y(x,Y(x))$ autres que les points de branchements.
Tout ceci implique que $W_1^{(1)}(x)\, dx$ est une diff\'erentielle sur ${\cal E}$ avec des poles (de degr\'e $4$) seulement aux points de branchements $e_i$, et
qui satisfait \refeq{Acycleepsiloninextorder}.

La formule de Cauchy permet d'\'ecrire (pour tout $\alpha\in{\cal E}$):
\beq
W_1^{(1)}(x(p))\, dx(p)= - \mathop\Res_{q\to p} dS_{q,\alpha}(p)\, W_1^{(1)}(x(q))\, dx(q)
\eeq
o\`u le r\'esidu est calcul\'e comme une int\'egrale le long d'un petit cercle entourant le point $p$.
On peut d\'eformer ce contour d'int\'egration, de fa\c con \`a transformer le r\'esidu en $p$, en r\'esidus aux points de branchements.
En vertu de l'identit\'e bilin\'eaire de Riemann \cite{Fay, Farkas}, et gr\^ace aux  relations \refeq{defdSpq} et \refeq{Acycleepsiloninextorder},
il n'y a pas de contribution des cycles ${\cal A}_i$, ${\cal B}_j$. Il reste donc:
\beq\label{CauchyWnextorder}
W_1^{(1)}(x(p))\, dx(p)= \sum_{i=1}^{d_2+1+2g} \mathop\Res_{q\to e_i} dS_{q,\alpha}(p)\, W_1^{(1)}(x(q))\, dx(q)
\eeq

On ins\`ere donc \refeq{masterloopnextorder} dans \refeq{CauchyWnextorder}, et l'on remarque que ${P^{(1)}(x,Y(x))\over E_y(x,Y(x))}\,dx$
n'a pas de r\'esidu aux points de branchements car $P^{(1)}(x,y)$ est un polyn\^ome.
D'o\`u:
\beq\label{CauchyWnextorderU}
W_1^{(1)}(x(p))\, dx(p)= - T \sum_{i=1}^{d_2+1+2g} \mathop\Res_{q\to e_i} dS_{q,\alpha}(p)\, {U^{(0)}(x(q),y(q);x(q)) \over E_y(x(q),y(q))}\, dx(q)
\eeq
Et donc, en utilisant \refeq{Uxyx}:
\bea
T\, W_1^{(1)}(x(p))\, dx(p)
&=&  \sum_{i=1}^{d_2+1+2g} \mathop\Res_{q\to e_i} dS_{q,\alpha}(p)\, \sum_{k=1}^{d_2} {B(q,p_k(q))\over (y(q)-y(p_k(q)))\,dx(p_k(q))}  \cr
&=&  \sum_{i=1}^{d_2+1+2g} \mathop\Res_{q\to e_i} dS_{q,\alpha}(p)\, \sum_{k=1}^{d_2} \mathop\Res_{p'\to p_k(q)} {B(q,p')\over (y(q)-y(p'))(x(p')-x(q))}  \cr
&=& - \sum_{i=1}^{d_2+1+2g} \mathop\Res_{q\to e_i} dS_{q,\alpha}(p)\, \mathop\Res_{p'\to q} {B(q,p')\over (y(q)-y(p'))(x(p')-x(q))}  \cr
&& - \sum_{i=1}^{d_2+1+2g} \mathop\Res_{q\to e_i} dS_{q,\alpha}(p)\, \sum_{k=1}^{d_1} \mathop\Res_{p'\to \td{p}_k(q)} {B(q,p')\over (y(q)-y(p'))(x(p')-x(q))}  \cr
&=&  \sum_{i=1}^{d_2+1+2g} \mathop\Res_{q\to e_i} dS_{q,\alpha}(p)\, \mathop\Res_{p'\to q} {B(q,p')\over (y(q)-y(p'))(x(q)-x(p'))}  \cr
&& + \sum_{i=1}^{d_2+1+2g} \mathop\Res_{q\to e_i} dS_{q,\alpha}(p)\, \sum_{k=1}^{d_1} {B(q,\td{p}_k(q))\over dy(\td{p}_k(q))(x(\td{p}_k(q))-x(q))}  \cr
&=&  \sum_{i=1}^{d_2+1+2g} \mathop\Res_{q\to e_i}  \mathop\Res_{p'\to q} {dS_{q,\alpha}(p)\,B(q,p')\over (y(q)-y(p'))(x(q)-x(p'))}  \cr
&=&  {1\over 2}\, \sum_{i=1}^{d_2+1+2g} \mathop\Res_{q\to e_i}  \mathop\Res_{p'\to q} {dS_{q,p'}(p)\,B(q,p')\over (y(q)-y(p'))(x(q)-x(p'))}  \cr
\eea
L'\'egalit\'e entre la ligne 1 et 2 est une simple r\'eecriture de $x(p_k(q))=x(q)$ sous forme de r\'esidus.
L'\'egalit\'e entre la ligne 2 et 3 est obtenue en d\'eplacant le contour, pour prendre tous les autres r\'esidus,
i.e. $p'=q$ et les $d_1$ autres solutions de $y(p')=y(q)$, que l'on appelle $\td{p}_k(q)$, $k=1,\dots,d_1$.
Dans la quatri\`eme ligne, on calcule explicitement les r\'esidus en $p'=\td{p}_k(q)$, et l'on voit qu'ils n'ont pas de poles lorsque $q=e_i$.
Le passage de la cinqui\`eme \`a la sixi\`eme ligne, est obtenu en \'echangeant l'ordre des r\'esidus, il montre que cette expression est bien ind\'ependante du choix du point $\alpha$.
On trouve donc \refeq{Wdevtopores}, QED.

\medskip
\`a partir de l\`a, on obtient:
\bt\label{TheoremFsubleading} (Eynard, Kokotov, Korotkin \citeeynmatgzero, \citeeynmatsg)
L'\'energie libre \`a l'ordre $1/N^2$ est donn\'ee par:
\beq\label{Fsubltaufprodyprime}
F^{(1)}=-{1\over 24}\ln{\left(\td{g}_{d_2+1}^{1-1/d_2}\,\,\tau_x^{12} \,\,\prod_{i=1}^{d_2+1+2g} y'(e_i) \right)}
\eeq
o\`u
\beq
\zeta_i(p):=\sqrt{x(p)-x(e_i)}
\eeq
\beq
y'(e_i):={dy(e_i)\over d\zeta_i(e_i)}
\eeq
et $\tau_x$ est d\'etermin\'ee par \citeeynmatsg, \cite{kk}:
\beq\label{deftaux}
{\d \ln\tau_x\over \d x(e_i)} = -{1\over 12}\,{\cal S}_{{\zeta_i}}(e_i)
\eeq
\et

La formule variationnelle de Rauch implique que \refeq{deftaux} d\'efinit bien une fonction $\tau_x$, parfois appel\'ee fonction $\tau$ de Bergmann \cite{kk}.
On peut d\'efinir une fonction $\tau_y$ de fa\c con similaire, et on peut v\'erifier que \refeq{TheoremFsubleading} est bien symm\'etrique dans l'\'echange $x\leftrightarrow y$.

{\noindent \bf Id\'ee de la preuve:}

Il s'agit de v\'erifier que l'expression \refeq{Fsubltaufprodyprime} satisfait bien toutes les relations du type \refeq{defobstrsimplesconn}, i.e. par exemple:
\beq\label{subldFdgkResWxk}
{\d F^{(1)}\over \d g_k} = -{T\over k}\, \mathop\Res_{\infty_x} x^k\,W_1^{(1)}(x) dx
\eeq
On calcule le membre de gauche de \refeq{subldFdgkResWxk}, i.e. les d\'eriv\'ees par rapport \`a $g_k$, en utilisant les relations du th\'eor\`eme IV.\ref{TheoremderivW},
et on calcule le membre de droite de de \refeq{subldFdgkResWxk} en utilisant le th\'or\`eme IV.\ref{Theoremresolvantesubleading}.
La preuve, fastidieuse. est expos\'ee dans \citeeynmatsg, et elle est trop longue pour apara\^\i tre ici.

Par ailleurs, la fonction $\tau_x$ est connue pour \^etre proportionelle au determinant d'un Laplacien sur ${\cal E}$ (voir \citeeynmatsg, \cite{kk, Zeldich}).

\bigskip
Le d\'eveloppement syst\'ematique de la r\'esolvante et de ses d\'eriv\'ees a \'et\'e obtenu tr\`es r\'ecement par (Eynard, Orantin) \cite{eoloop2mat},
par une m\'ethode diagramatique, que je ne pr\'esenterai pas ici.
Toutefois, cette m\'ethode diagrammatique n'a pas encore pu s'appliquer au calcul du d\'eveloppement de l'\'energie libre en $1/N^2$.
Ceci reste \`a comprendre.
Pour l'instant, il n'existe pas d'expression connue pour $F^{(2)}$ par exemple.


\chapter{Asymptotiques des polyn\^omes biorthogonaux}
\label{chapterasymp}

Dans ce chapitre, nous allons voir comment on peut deviner les asymptotiques \`a $n$ grand, pour les polyn\^omes biorthogonaux.
Les chapitres pr\'ec\'edents n'\'etaient qu'une introduction \`a ce calcul, pour poser les notations, et mettre en place les \'el\'ements n\'ec\'essaires.

L'id\'ee est la suivante:
la formule de Heine \refeq{Heinepsi} permet de voir les polyn\^omes orthogonaux comme des fonctions de partition d'un mod\`ele normal.
Le mod\`ele normal est lui m\^eme la transform\'ee de Fourrier d'un mod\`ele normal \`a symm\'etrie bris\'ee.
Le mod\`ele \`a symm\'etrie bris\'ee, a, sous certaines hypoth\`eses, le m\^eme d\'eveloppement asymptotique qu'un mod\`ele formel.
Par la m\'ethode des \'equations de boucles, nous pouvons trouver les asymptotiques du mod\`ele formel en termes de g\'eom\'etrie alg\'ebrique.
Il suffit alors de refaire la transform\'ee Fourrier inverse pour obtenir les asymptotiques des polyn\^omes.
Cett m\'ethode a \'et\'e introduite dans \citeeynchain et \citeeynchaint, et a \'et\'e pr\'esent\'ee aux Houches en 2004 \cite{eynHouches}.

Les asymptotiques ainsi obtenus s'expriment naturellement en termes de g\'eom\'etrie alg\'ebrique d'une courbe complexe, et l'on voit ais\'ement que cette courbe
complexe n'est rien d'autre que la limite $n$ grand de la courbe spectrale vue en \refeq{defcourbespectrale}.

Il faut noter que les asymptotiques pr\'esent\'es ci-dessous ne sont que des {\bf conjectures} (\`a ce jour),
soutenues par des arguments physiques tr\`es raisonables,
mais ils ne sont pas d\'emontr\'es math\'ematiquement.
La m\'ethode la plus prometeuse pour les d\'emontrer est la m\'ethode de Riemann--Hilbert introduite par \cite{BlIt, BlIt1} et \cite{dkmvz, dkmvz2, DeiftBook, EML}.
Il apara\^\i t \'egalement probable que l'on puisse un jour justifier rigoureusement les hypoth\`eses pr\'esent\'ees ci-dessous, sur lesquelles se base l'intuition physique,
sans passer par Riemann--Hilbert.

Les asymptotiques pr\'esent\'es ci-dessous ont \'et\'e introduits pour la premi\`ere fois en 1997 dans \citeeynchain et \citeeynchaint pour le cas o\`u la courbe spectrale est de genre $0$,
et ils ont \'et\'e pr\'esent\'es \`a la conf\'erence de l'AMS \`a Montr\'eal en 2002 \cite{BEHAMS}.

\section{La courbe spectrale}

Supposons qu'il existe une courbe alg\'ebrique ${\cal E}$ d'\'equation $E(x,y)=0$ avec:
\beq
E(x,y)=(V'_1(x)-y)(V'_2(y)-x)-P(x,y)+1
\eeq
o\`u $\deg P=(d_1-1,d_2-1)$,
et telle que pour tout cycle sur ${\cal E}$, on ait:
\beq
\Re\oint y dx =0
\eeq
Supposons que la courbe est de genre  $g\leq d_1 d_2-1$,
et que l'on ait pu choisir des cycles ${\cal A}_I$, entourant des coupures $[a_{2I-1},a_{2I}]$, tels que les:
\beq
\epsilon_I := {1\over 2i\pi}\oint_{{\cal A}_I} y dx
\virg
\epsilon_{g+1} := 1-\sum_{I=1}^g \epsilon_I
\eeq
soient tous positifs.

\br
L'existence d'une telle courbe est obtenue \`a partir de la courbe alg\'ebrique du mod\`ele formel, vue au chapitre pr\'ec\'edent.
Pour chaque $\epsilon_I\geq 0$, il existe une courbe alg\'ebrique, et la partie r\'eelle de l'\'energie libre associ\'ee est une fonction convexe
des $\epsilon$ (en effet, sa d\'eriv\'ee seconde est $-2i\pi\tau$, dont la partie r\'eelle est toujours positive).
Elle admet donc un unique minimum, et c'est ce minimum que l'on consid\`ere ici.
Le minimum peut \^etre atteint aux bords, dans ce cas certains $\epsilon_I$ sont nuls, et le genre est donn\'e par le nombre de fractions de remplissage non nulles.
Au minimum d'un $\epsilon_I\neq 0$, on a ${\d \Re F\over \d \epsilon_I}=0$, i.e. $\Re \oint_{{\cal B}_I} y dx=0$.
\er

\subsection{Feuillets}

Comme nous l'avons vu au chapitre pr\'ec\'edent, cette courbe poss\`ede  deux points \`a l'infini, not\'es $\infty_x$ et $\infty_y$,
et poss\`ede $d_2+1$ (resp. $d_1+1$) feuillets en $x$ (resp. en $y$).
C'est \`a dire que pour tout $x\in \C$ (resp. $y\in\C$), il existe $d_2+1$ (resp. $d_1+1$) points de ${\cal E}$ tels que:
\beq
\forall i=0,\dots, d_2, \qquad x(p_i(x))=x
\virg
({\rm resp.}\,\,
\forall i=0,\dots, d_1, \qquad y(\td{p}_i(y))=y\,\,)
\eeq
On note $p_0(x)$ (resp. $\td{p}_0(y)$) le point qui est dans le feuillet physique, i.e. qui s'approche de $\infty_x$ (resp. $\infty_y$) lorsque $x\to\infty$ (resp. $y\to\infty$).
Les autres points $p_i(x)$ (resp. $\td{p}_i(y)$) avec $i>0$ s'approchent de $\infty_y$ (resp. $\infty_x$) lorsque $x\to\infty$ (resp. $y\to\infty$).

\subsection{coupures, points de branchelents et bases d'homologies}
\label{defcheminsalgasymp}

Les points de branchement en $x$ (resp. en $y$) sont les z\'eros de $dx$ (resp. $dy$), on les note $e_i$ (resp. $\td{e}_i$).
On supposera ici qu'ils sont simples et tous distincts.
On notera:
\beq
a_i:=x(e_i)
\virg
({\rm resp.}\,\, b_i:=y(\td{e}_i)\,)
\eeq
Chaque point de branchement du feuillet physique $a_i$ (resp. $b_i$), est tel que $\exists j>0$ tel que:
\beq
p_j(a_i)=p_0(a_i)
\virg
({\rm resp.\,\,}
\td{p}_j(b_i)=\td{p}_0(b_i)
\,)
\eeq
D\'efinissons le potentiel effectif:
\beq
V_{1,{\rm eff}(i)}(x):=\int_{p_j(x)}^{p_0(x)} y dx
\virg
({\rm resp.}\,\,\,
V_{2,{\rm eff}(i)}(y):=\int_{\td{p}_j(y)}^{\td{p}_0(y)} x dy
\,)
\eeq
On a:
\beq
V_{1,{\rm eff}(i)}(a_i)=0
\virg
({\rm resp.}\,\,\,
V_{2,{\rm eff}(i)}(b_i)=0
\,)
\eeq
et $V_{1,{\rm eff}(i)}(x)$ (resp. $V_{2,{\rm eff}(i)}(y)$) se comporte en $(x-a_i)^{3/2}$ (resp. $(y-b_i)^{3/2}$) au voisinage de $a_i$ (resp. $b_i$).
Il existe des directions o\`u $V_{1,{\rm eff}(i)}(x)$ (resp. $V_{2,{\rm eff}(i)}(y)$) est r\'eel et strictement croissant.
Ceci d\'efinit (au moins) un chemin allant de $a_i$ (resp. $b_i$) \`a l'infini.
Pour chaque coupure $I$, i.e. reliant deuxpoints de branchements $[a_{2I-1},a_{2I}]$ (resp. $[b_{2I-1},b_{2I}]$), on a donc deux chemins, partant chacun d'une extr\'emit\'e de la coupure.
Notons $\Gammax_I$ (resp. $\Gammay_I$) le chemin allant de $\infty$ \`a $\infty$, constitu\'e de la r\'eunion de la coupure et des deux chemins \`a potentiel effectifs croissant \`a partir des deux bords.
Les chemins $\Gammax_I$ ainsi obtenus sont n\'ec\'essairement sans intersections, et vont n\'ec\'essairement \`a l'infini dans des directions o\`u $\Re V_1(x)>0$ (resp. $\Re V_2(y)>0$).
On peut aussi facilement montrer qu'ils sont homologiquement non nuls, et que les $\Gammax_I\times \Gammay_I$ forment bien une base d'homologie
(cf chemins similaires dans \cite{moore}).

Nous utiliserons cette base pour d\'efinir le produit scalaire des polyn\^omes biorthogonaux, i.e. il faut
multiplier la matrice $\kappa$ vue en \ref{defGamma} qui a servi \`a d\'efinir le produit scalaire des  polyn\^omes biorthogonaux.
Nous supposerons d\'esromais que $\kappa$ est la matrice des coefficients qui d\'ecrit le chemin $\Gamma$ des polyn\^omes biorthogonaux dans cette base:
\beq
\Gamma = \sum_{I} \kappa_I \Gammax_I\times \Gammay_I
\eeq
\beq
\int_\Gamma \pi_n(x)\sigma_m(y)\,\ee{-(V_1(x)+V_2(y)-xy)}\,dx\,dy = h_n \delta_{nm}
\eeq

Nous noterons:
\beq
\kappa_I=\ee{2i\pi \nu_I}
\eeq

Remarquons que dans les applications les plus fr\'equentes, les $\kappa_I$ valent $\pm 1$,
et donc les $\nu_I$ sont des entiers ou demi-entiers.

Si un coefficient $\kappa_I$ est nul, il faut choisir $\epsilon_I=0$.
Le genre de la courbe est donn\'e par le nombre de coefficients non nuls.
Remarquons que si le chemin est l'axe r\'eel, celui ci est somme d'au plus la moiti\'e des chemins, et donc le genre est au plus:
\beq
g\leq {d_1+1\over 2}\,{d_2+1\over 2}
\eeq

\subsection{Notations}

Rappelons que
\beq\label{courbealgeta}
\displaystyle
\left\{
{\begin{array}{lcl}
\displaystyle
\mathop\Res_{\infty_x} y dx = \mathop\Res_{\infty_y} x dy = 1
&,&
\displaystyle
\forall I\,\, , \,\,\, {1\over 2i\pi}\oint_{{\cal A}_I} y dx = \epsilon_I \geq 0
\virg
{1\over 2i\pi}\oint_{{\cal B}_I} y dx =\zeta_I \,\, (\in \R) \cr
\displaystyle
\forall X\,\, , \,\,\, \mathop\Res_{\infty_x} {1\over x-X}y  dx = V'_1(X)
&,&
\displaystyle
\forall Y\,\, , \,\,\, \mathop\Res_{\infty_y} {1\over y-Y} x dy = V'_2(Y)
\end{array}}\right.
\eeq

On choisit une base canonique de diff\'erentielles holomorphes:
\beq
\oint_{{\cal A}_I} du_J :=\delta_{IJ}
\eeq
et la matrice des p\'eriodes associ\'ee $\tau$ (on a $\tau^t=\tau$ et $\Im\tau>0$):
\beq
\oint_{{\cal B}_I} du_J :=\tau_{IJ}
\eeq
On choisit une demi p\'eriode impaire $z={n+\tau m\over 2}$ avec $\sum_I n_I m_I \in 2Z+1$,
et on consid\`ere $\theta_z(u,\tau):=\theta(u+z,\tau)$.
On d\'efinit la diff\'erentielle holomorphe:
\beq
dh_z(p):= \sum_I du_I(p)\,\left.{\d\theta_z(v,\tau)\over \d v_I}\right|_{v=0}
\eeq

On d\'efinit (sur un domaine fondamental du recouvrement universel, et avec une coupure le long d'une ligne $[\infty_x,\infty_y]$):
\beq
T(p):=V_1(x(p))-\ln{(x(p))} + \int_{q=\infty_x}^p \left(y(q)-V'_1(x(q))+{1\over x(q)}\right) dx(q)
\eeq
\beq
\td{T}(p):=V_2(y(p))-\ln{(y(p))} + \int_{q=\infty_y}^p \left(x(q)-V'_2(y(q))+{1\over y(q)}\right) dy(q)
\eeq
Notons que $\ee{T(p)}$ n'a pas de coupure le long de $[\infty_x,\infty_y]$.

On d\'efinit:
\beq
\mu:=T(p)+\td{T}(p)-x(p)y(p)
\eeq
qui est ind\'ependant du point $p$ (en effet $d\mu=ydx+xdy-d(xy)=0$).

On d\'efinit aussi:
\beq
\L(p):={\theta_z(u(p)-u(\infty_y),\tau)\over \theta_z(u(p)-u(\infty_x),\tau)}
\virg
\gamma:=\mathop{\rm lim}_{p\to\infty_x} {x(p)\over \L(p)}
\virg
\td\gamma:=\mathop{\rm lim}_{p\to\infty_y} {y(p) \L(p)}
\eeq

\beq
H(p):= \gamma\,{dh_z(p)\, \theta_z(u(\infty_y)-u(\infty_x))\over dx(p)\,\theta_z(u(p)-u(\infty_x))^2}
\virg
\td{H}(p):= \td\gamma\,{dh_z(p)\, \theta_z(u(\infty_x)-u(\infty_y))\over dy(p)\,\theta_z(u(p)-u(\infty_y))^2}
\eeq
Notons que $H(\infty_x)=1$, resp. $\td{H}(\infty_y)=1$, et
\beq
{H(p)\over \td{H}(p)}=-{\gamma\over \td\gamma}\,\L(p)^2\,{dy(p)\over dx(p)}
\eeq

Posons:
\beq
\eta_k:= N(\tau\epsilon-\zeta) -\nu + k(u(\infty_x)-u(\infty_y))
\eeq
et
\beq
\td{h}_k:= {2\pi\over (\gamma\td\gamma)^k}\,\sqrt{2\pi\gamma\td\gamma\over N}\, \ee{-N\mu} \,\ee{-2i\pi N\epsilon(u(\infty_x)-u(\infty_y))}\, {\theta(\eta_{k-1},\tau)\over \theta(\eta_{k},\tau)}
\eeq

\bd
Nous d\'efinissons les fonction suivantes:
\beq\label{deffkpasymp}
f_k(p)
:= \sqrt{H(p)\over \td{h}_{k}}\,(\gamma\L(p))^{-k}  \,\ee{2i\pi N\epsilon (u(p)-u(\infty_x))} {\theta(\eta_{k} +u(p)-u(\infty_x),\tau)\over \theta(\eta_{k},\tau)}
\eeq
\beq
\td{f}_k(p)
:= \sqrt{\td{H}(p)\over \td{h}_{k}}\,\left({\td\gamma\over \L(p)}\right)^{-k}  \,\ee{-2i\pi N\epsilon (u(p)-u(\infty_y))} {\theta(\eta_{k} -u(p)+u(\infty_y),\tau)\over \theta(\eta_{k},\tau)}
\eeq
\beq
g_k(p)
:= {1\over 2i\pi}\,\sqrt{\td{h}_{1-k}\,\td{H}(p)}\,\left({\L(p)\over \td\gamma}\right)^{k}  \,\ee{2i\pi N\epsilon (u(p)-u(\infty_y))} {\theta(\eta_{1-k} +u(p)-u(\infty_x),\tau)\over \theta(\eta_{-k},\tau)}
\eeq
\beq
\td{g}_k(p)
:= {1\over 2i\pi}\,\sqrt{\td{h}_{1-k}\,H(p)}\,\left({\gamma\L(p)}\right)^{-k}  \,\ee{-2i\pi N\epsilon (u(p)-u(\infty_x))} {\theta(\eta_{1-k} -u(p)+u(\infty_y),\tau)\over \theta(\eta_{-k},\tau)}
\eeq
\ed

Notons que $f_k(p)\ee{-NT(p)}$ n'a pas de discontinuit\'e le long des cycles ${\cal B}_I$, et est multipli\'ee par une phase $\ee{2i\pi(\nu_I+z_I)}=\ee{2i\pi z_I}\,\kappa_I$ lorsqu'on traverse un cycle ${\cal A}_{I}$.

$f_k(p)$ se comporte asymptotiquement \`a l'infini comme:
\beq
f_k(p) \ee{-NT(p)} \mathop\sim_{p\to\infty_x} x(p)^{N-k}\, \ee{-NV_1(x(p))}\,(1+O(1/x))
\eeq

\bd
D\'efinissons les matrices carr\'ees suivantes de taille $d_2+1$
\beq
{\mathbf T}(x):=\diag(T(p_0(x)),T(p_1(x)),\dots,T(p_{d_2}(x)))
\eeq
\beq
F_{ij}(x):=f_i(p_j(x)) \virg i,j=0,\dots,d_2
\eeq
\beq
\td{G}_{ij}(x):=\td{g}_i(p_j(x)) \virg i,j=0,\dots,d_2
\eeq
et les matrices carr\'ees suivantes de taille $d_1+1$
\beq
{\mathbf {\td{T}}}(y):=\diag(\td{T}(\td{p}_0(y)),\td{T}(\td{p}_1(y)),\dots,\td{T}(\td{p}_{d_1}(y)))
\eeq
\beq
\td{F}_{ij}(y):=\td{f}_i(\td{p}_j(y)) \virg i,j=0,\dots,d_1
\eeq
\beq
G_{ij}(y):=g_i(\td{p}_j(y)) \virg i,j=0,\dots,d_1
\eeq

\ed

\section{Conjecture}

\bconj\label{conjasymp} (Eynard \citeeynchain, \citeeynchaint)
les asymptotiques des syst\`emes fondamentaux des polyn\^omes biorthogonaux (d\'efinis par le theoreme \ref{Theoremsolfondtdphi}), dans le r\'egime
\beq
n\to\infty, N\to\infty, |n-N|=O(1)
\eeq
sont donn\'es par:
\bea
{\mathbf\Psi}_{N}(x) \sim  F(x)\, \ee{-N{\mathbf T}(x)}\,C_x (1+O(1/N))
\eea
\bea
{\mathbf\Phi}_{N}(y) \sim \td{F}(y)\,\ee{-N\td{\mathbf{T}}(y)}\,C_y (1+O(1/N))
\eea
\bea
{\mathbf{\td\Psi}}_{N}(y) \sim G(y)\,\ee{N\td{\mathbf T}(y)}\,\td{C}_y (1+O(1/N))
\eea
\bea
{\mathbf{\td\Phi}}_{N}(x) \sim \td{G}(x)\,\ee{N {\mathbf T}(x)}\,\td{C}_x (1+O(1/N))
\eea

o\`u $C_x$, $\td{C}_x$, $C_y$, $\td{C}_y$ sont des matrices carr\'ees constantes par morceaux.
\econj

\br
Ces asymptotiques ont \'et\'e introduits pour la premi\`ere fois en 1997 dans \citeeynchain\ et \citeeynchaint\ pour le cas o\`u la courbe spectrale est de genre $0$,
et ils ont \'et\'e pr\'esent\'es \`a la conf\'erence de l'AMS \`a Montr\'eal en 2002 \cite{BEHAMS}.
Ces asymptotiques sont des ansatz, qu'il faudrait maintenant prouver, par exemple par la m\'ethode de Riemann--Hilbert \cite{BlIt} et \cite{dkmvz}.
\er

\br
Les polyn\^omes orthogonaux $\psi_{N-i}(x)$ sont obtenus dans la premi\`ere colonne de la matrice ${\mathbf\Psi}_{N}(x)$:
\beq
\psi_{N-i}(x) = \left({\mathbf\Psi}_{N}(x)\right)_{i0}
\eeq
\er

\br
Ces asymptotiques sont uniforme pour tout $x$, hors du voisinage des points de branchements.
Les lignes de discontinuit\'e des $C_x$, sont donn\'ees par:
\beq
\Re \int_{p_i(x)}^{p_j(x)} y dx =0
\eeq
\er

\br
La m\'ethode euristique pour obtenir ces asymptotiques est pr\'esent\'ee dans \citeeynchain\ et \citeeynchaint, de m\^eme que dans \cite{eynbetapol} pour le genre z\'ero.
Lorsque la courbe a un genre plus \'elev\'e, la sommation sur les fractions de remplissages (cf eq.\ref{ZNormNormb}), produit une fonction $\theta$, comme dans \cite{BDE}.
De plus, cette m\'ethode euristique fonctionne \`a $x$ fix\'e, i.e. on peut calculer la matrice $C_x$ tr\`es explicitement pour un $x$ donn\'e, sans avoir \`a se pr\'eocupper des discontinuit\'es.
La matrice $C_x$ provient de la matrice de changement de base d'homologie pour calculer les transform\'ees de Fourrier et Hilbert, par la m\'ethode du col (m\'ethode du col habituelle en dimension $1$).
Les discontinuit\'es sont obtenues \`a posteriori.
Les lignes de discontinuit\'es apparaissent d'abord car les $p_j(x)$ sont par d\'efinition discontinus le long des coupures, puis, par un ph\'enom\`ene de Stokes
lorsque l'on calcule les transform\'ees de Fourrier et transform\'ees de Hilbert par la m\'ethode du col.
Les matrices de Stokes et de saut le long des coupures pourraient \^etre calcul\'ees tr\`es explicitement par cette m\'ethode euristique.
\er

\section{V\'erifications}

V\'erifions que cet ansatz satisfait certaines propri\'et\'es attendues:

\subsection{Asymptotiques}

On a:
\beq
f_k(p) \mathop\sim_{p\to\infty_x} {1\over \sqrt{\td{h}_k}}\,x(p)^{-k}\, (1+O(1/x))
\eeq
\beq
f_k(p) \mathop\sim_{p\to\infty_y} O\left(y(p)^{k-{d_2+1\over 2}}\right)
\eeq
ce qui garantit bien que:
\beq
p_{N-k}(x)\mathop\sim_{x\to\infty} {C_x}_{0,0}\,x^{N-k}\, (1+O(1/x))
\eeq
Il faut donc choisir ${C_x}_{0,0}=1$ dans les secteurs o\`u $x\to\infty$.

De fa\c con g\'en\'erale, on trouve que les syst\`emes fondamentaux donn\'es dans la conjecture V.\ref{conjasymp}, satisfont les m\^emes asymptotiques \`a $x$ grand que ceux
donn\'es au th\'eor\`eme III.\ref{Theoremasymptdphi}.

\subsection{Transform\'ees de Fourrier et Hilbert}

On peut v\'erifier que pour $j>0$,
\beq
\td\psi^{(j)}_{N-i}(y)=\int \psi_{N-i}(x) \,\ee{Nxy}\,dx \,\,\, (1+O(1/N))
\eeq
l'int\'egration se fait par la m\'ethode du col habituelle. Le chemin d'int\'egration non pr\'ecis\'e ici, d\'epend de $x$ de fa\c con constante par morceaux, car il d\'epend de la matrice $C_x$,  et de la classe d'homologie des chemins passant par les points cols.

On peut aussi v\'erifier par la m\'ethode du col et la formule de Cauchy, que
\beq
\td\psi^{(0)}_{N-i}(y)=\int\int {1\over y-y'} \psi_{N-i}(x) \,\ee{Nxy'}\,dx\, dy' \,\,\, (1+O(1/N))
\eeq

\subsection{Orthogonalit\'e}

On peut v\'erifier que:
\beq
\int\int \psi_{N-i}(x)\, \phi_{N-j}(y) \,\,\ee{Nxy}\,\, dx\,dy = \delta_{ij} +O(1/N)
\eeq
en effet, l'int\'egration se r\'eduit \`a un r\'esidu au point $\infty_x$ (ou au point $\infty_y$) sur la courbe alg\'ebrique, de la forme:
\bea
\sqrt{\td{h}_i\over\td{h}_j}\,\mathop\Res_{p\to\infty_x} {dh_z(p)\,\theta_z(u(\infty_x)-u(\infty_y))\,(\gamma\L(p))^{j-i}\over \theta_z(u(p)-u(\infty_x)) \theta_z(u(p)-u(\infty_y))}\,{\theta(\eta_{i}+u(p)-u(\infty_x))\,\theta(\eta_{j}-u(p)+u(\infty_y))\over \theta(\eta_i)\theta(\eta_{j-1})} = \delta_{ij} \cr
\eea
si $i>j$, il n'y a pas de pole en $\infty_x$ et le r\'esidue s'annule, de m\^eme, si $j>i$, il n'y a pas de p\^ole en $\infty_y$ et le r\'esidue s'annule,
et si $i=j$, le r\'esidue vaut bien $1$.
Ceci montre que le coefficient $h_n$ (def.III.\ref{defpolbiorth}) vaut:
\beq
h_n \sim \td{h}_{N-n}\,(1+O(1/N))
\eeq

de m\^eme, le coefficient $\gamma_n$ (def.\ref{defgammanalphabeta}) vaut:
\beq
\gamma_n  \sim \sqrt{\gamma\td\gamma}\,\sqrt{\theta(\eta_{N-n})\theta(\eta_{N-n-2})\over \theta(\eta_{N-n-1})^2}\,\, (1+O(1/N))
\eeq

\subsection{Syst\`emes diff\'erentiels, courbe spectrale et dualit\'e spectrale}

Les 4 syst\`emes de la conjecture V.\ref{conjasymp} satisfont 4 syst\`emes d'\'equations diff\'erentiels:
\beq
{\cal D}_1(x) \sim F(x)\,{\mathbf T}'(x)\,F(x)^{-1} +O(1/N)
\eeq
\beq
\td{\cal D}_1(x) \sim \td{G}(x)\,{\mathbf T}'(x)\,\td{G}(x)^{-1} +O(1/N)
\eeq
\beq
{\cal D}_2(y) \sim \td{F}(y)\,\td{\mathbf T}'(y)\,\td{F}(y)^{-1} +O(1/N)
\eeq
\beq
\td{\cal D}_2(y) \sim G(y)\,\td{\mathbf T}'(y)\,G(y)^{-1} +O(1/N)
\eeq
Les valeurs propres de ${\cal D}_1(x)$ sont donc celles de ${\mathbf T}'(x)$, i.e. ce sont les $y(p_i(x))$, autrement dit la courbe alg\'ebrique que nous avons introduite
est bien la limite de la courbe spectrale, et ce pour les 4 syst\`emes.
Nous retrouvons le th\'or\`eme de dualit\'e spectrale:
\bt
Les 4 syst\`emes diff\'erentiels ci-dessus ont la m\^eme courbe spectrale, qui est la courbe alg\'ebrique ${\cal E}$.
\et

\br
La matrice ${\cal D}_1(x)$ n'a pas n\'ec\'essairement de limite lorsque $N\to\infty$, car $F(x)$ d\'epend explicitement de $N$.
\er

\subsection{Dualit\'e et Christoffel--Darboux}

\bt
La matrice
\beq
A= (F(x)\td{G}(x)^t)^{-1}
\eeq
ne d\'epend pas de $x$, et a la forme de la matrice de Christoffel Darboux (cf th\'eor\`eme III.\ref{ChrisDarb}, \'ecrite dans une autre base):
$A_{ij} = 0$ si ($i=0$ et $j\neq 0$) ou si ($j=0$ et $i\neq 0$) ou si ($i>0$, $j>0$,  et $i+j>d_2$).
On a $A_{00}=-\gamma_{N-1}$.

\et

id\'ee de la preuve: Chacun des $A_{ij}$ peut \^etre \'ecrit comme des combinaisons  de fonctions sur la courbe ${\cal E}$.
On v\'erifie que cette fonction est bien une fonction m\'eromorphe sur ${\cal E}$, i.e. qu'elle reprend bien sa valeur apr\`es un tour autour de chaque ${\cal A}_I$
et chaque ${\cal B}_I$, et qu'elle n'a pas de poles. Toute fonction m\'eromorphe sans p\^ole \'etant constante, on trouve que $A$ ne d\'epend pas de $x$.
La valeur des $A_{ij}$ peut donc \^etre calcul\'ee pour $x$ grand, et l'on trouve le r\'esultat annonc\'e.

\section{Id\'ee euristique sous-tendant la conjecture}

Nous allons exposer rapidement les id\'ees qui conduisent \`a cette conjecture.
Elles se basent sur les articles \citeeynchain et \citeeynchaint, et sur \cite{BDE} (voir aussi \cite{eynHouches}).

\subsection{Formule de Heine et mod\`ele normal avec potentiel rationel}

La formule de Heine \refeq{Heinepsi} permet de voir les polyn\^omes orthogonaux comme des fonctions de partition d'un mod\`ele normal:
\beq\label{pinHeineasymp}
\pi_n(\xi) = {Z_{{\rm Norm}}(N,T,r,\kappa)\over Z_{{\rm Norm}}(N,T,0,\kappa)}
\eeq
o\`u $T={n\over N}$ et $r=-{1\over N}$ et:
\beq
Z_{{\rm Norm}}(N,T,r,\kappa):=\int_{(H_{n}\times H_{n})(\Gamma)} dM_1\, dM_2 \,
\ee{-{n\over T}\tr \left[V_1(M_1)+r\ln{(\xi-M_1)}+V_2(M_2)-M_1 M_2\right]}
= \ee{-{n^2\over T^2}\,F_{{\rm Norm}}(N,T,r,\kappa)}
\eeq
autrement dit, on a simplement ajout\'e un terme logarithmique au potentiel $V_1$, et une temp\'erature $T={n\over N}$.
Jusqu'ici, nous n'avons consid\'er\'e que des potentiels polynomiaux, mais il est connu \cite{BERatlFree}
que tout les raisonements pr\'ec\'edents se g\'en\'eralisent bien au cas o\`u $V'_1$ est une fraction rationelle.
De plus, ici, le terme logarithmique est multipli\'e par $r=-1/N$ qui est petit, et nous allons effectuer un d\'eveloppement de Taylor
au voisinage d'un potentiel polynomial.

\subsection{Transform\'ee de Fourrier d'un mod\`ele normal bris\'e}

Pour un chemin d'int\'egration $\Gamma=\sum_{I} \kappa_{I} \Gammax_{I}\times \Gammay_{I}$, on a, d'apr\`es \refeq{ZNormNormb}:
\beq\label{pinormnormb}
Z_{{\rm Norm}}(N,T,r,\kappa) = \sum_{\sum_{I} m_{I}=n} \kappa_{I}^{m_{I}}\,\,Z_{{\rm Normb}}(N,T,r,m_I)
\eeq
avec:
\bea
Z_{{\rm Normb}}(N,T,r,m_I)
&:=& \ee{-{n^2over T^2}F_{{\rm Normb}}(N,T,r,m_I)} \cr
&:=& \prod_{I} \int_{H_{m_{I}}(\Gammax_{I})\times H_{m_I}(\Gammay_{I})} dM_{x,I} dM_{y,I} \cr
&& \qquad\ee{-{n\over T}\tr \left[V_1(M_{x,I})+r\ln{(\xi-M_{x,I})}+V_2(M_{y,I})-M_{x,I} M_{y,I}\right]} \cr
&& \prod_{(I)<(k,l)} \det{\left(M_{x,I}\otimes \1_{n_{k,l}}-\1_{n_{I}}\otimes M_{x,k,l}\right)} \cr
&& \prod_{(I)<(k,l)} \det{\left(M_{y,I}\otimes \1_{n_{k,l}}-\1_{n_{I}}\otimes M_{y,k,l}\right)}
\eea

On fait l'hypoth\`ese, que si les chemins $\Gamma_I$ sont ''bien choisis'' (i.e. section \ref{defcheminsalgasymp}),
alors l'\'energie libre du mod\`ele normal bris\'e poss\`ede un d\'eveloppement en puissances de $1/N^2$ pour $N$ grand,
i.e. on identifie le mod\`ele normal bris\'e \`a un mod\`ele formel (\`a des termes exponentiellement petits pr\`es):
\beq\label{eqZnormformnstar}
Z_{{\rm Normb}}(N,T,r,m_I) \equiv Z_{{\rm Form}}(N,T,r,{1\over N}m_I) = \ee{-{n^2\over T^2}\,F_{{\rm Form}}(N,T,r,{1\over N}m_I)}
\eeq

\subsection{D\'eveloppement \`a N grand du mod\`ele formel}

Nous savons que le mod\`ele formel peut se d\'evelopper en puissances de $1/N^2$:
\beq
-{1\over N^2}\ln{Z_{{\rm Form}}(N,T,r,\eta_{I})} = F^{(0)}(T,r,\eta_{I}) + {1\over N^2} F^{(1)}(T,r,\eta_{I}) + O({1\over N^4})
\eeq
o\`u $F^{(0)}(T,r,\eta_{I})$, $F^{(1)}(T,r,\eta_{I})$, ..., etc, sont donn\'ees par Th.\ref{energielibreleading}, Th.\ref{TheoremFsubleading},
et sont analytiques dans tous leurs param\`etres.

Nous poserons $F=F^{(0)}$ car nous n'aurons pas besoin de l'ordre suivant.

En utilisant les th\'eor\`emes IV.\ref{TheoremderivW}, IV.\ref{energielibreleading} et  IV.\ref{thenergielibrederiv2},
on a (en \'ecrivant $\xi=x(q)$):
\beq
F_\eta:={\d F\over \d \eta_I} = -\oint_{{\cal B}_I} ydx = -2i\pi \zeta_I
\eeq
\beq
F_r:={\d F\over \d r} = -(T(q)-V_1(x(q)))
\eeq
\beq
F_T:={\d F\over \d T} = \mu
\eeq

\beq
F_{\eta\eta}:={\d^2 F\over \d \eta_I\d \eta_J} = -2i\pi \tau_{IJ}
\eeq
\beq
F_{rr}:={\d^2 F\over \d r^2} = -\ln{H(q)}
\eeq
\beq
F_{TT}:={\d^2 F\over \d T2} = -\ln{\gamma\td\gamma}
\eeq
\beq
F_{\eta r}:={\d^2 F\over \d \eta_I\d r} = -2i\pi (u(q)-u(\infty_x))
\eeq
\beq
F_{\eta T}:={\d^2 F\over \d \eta_I\d T} = -2i\pi (u(\infty_x)-u(\infty_y))
\eeq
\beq
F_{rt}:={\d^2 F\over \d r\d T} = \ln{\gamma\L(q)}
\eeq

Au voisinage de $T=1$, $r=0$ et $\eta=\epsilon$,  en posant $m=N\eta$, la formule de Taylor \`a l'ordre $2$ donne:
\bea
\ee{-{n^2\over T^2}F(T,r,\eta)}
&\sim&
\ee{-N^2 F(1,0,\epsilon)}\,
\ee{-N (m-N\epsilon)F_\eta}\,
\ee{N F_r}\,
\ee{-N (n-N)F_T}\, \cr
&& \ee{-{1\over 2}(m-N\epsilon)^t F_{\eta\eta}(m-N\epsilon)}\,
\ee{-{1\over 2}F_{rr}}\,
\ee{-{1\over 2}(n-N)^2 F_{TT}}\, \cr
&& \ee{ F_{\eta r} (m-N\epsilon)}\,
\ee{-(n-N)F_{\eta T}(m-N\epsilon)}\,
\ee{(n-N) F_{rT}}\, \cr
&\sim&
(\gamma\td\gamma)^{{1\over 2}(n-N)^2}\,
\ee{-N^2 F(1,0,\epsilon)}\,
\ee{-N (n-N)\mu}\, \cr
&& \sqrt{H(q)}\, \L(q)^{n-N}\, \ee{-N (T(q)-V_1(x(q)))}\, \cr
&&  \ee{i\pi (m-N\epsilon)^t \tau (m-N\epsilon)}\,\,\,
\ee{2i\pi N \zeta^t(m-N\epsilon)}\, \cr
&&  \ee{ -2i\pi (u(q)-u(\infty_x))^t (m-N\epsilon)}\,
\ee{2i\pi (n-N)(u(\infty_x)-u(\infty_y))^t(m-N\epsilon)}\, \cr
\eea
o\`u toutes les d\'eriv\'ees sont calcul\'ees en $T=1$, $r=0$ et $\eta=\epsilon$.

\subsection{Sommation sur les fractions de remplissage}

Comme la matrice des p\'eriodes $\tau_{IJ}$ a toujours une partie imaginaire positive (voir paragraphe \ref{sectiondiffhol}), $\Re F(1,0,\eta)$ est une fonction convexe des $\eta_I$.
Notons $\epsilon$ son minimum. La condition de minimalit\'e s'\'ecrit:
\beq
{\d \Re F\over \d \eta_I} = 0 = 2\pi \Im \zeta_I
\eeq
ce qui implique que les $\zeta_I$ sont r\'eels.
La convexit\'e de $\Re F$, implique que dans la somme sur les $m_I$, seuls les $m_I$ tels que $|m_I-N\epsilon_I|\sim O(1)$ contribuent
notablement \`a la somme.
ceci entraine, d'une part, on peut se contenter de l'approximation de Taylor de l'\'energie libre au 2e ordre, et d'autre part, on peut sommer sure les $m_I$ allant de $-\infty$ \`a $+\infty$,
ce qui produit une fonction $\theta$:
\bea
 Z_{{\rm Norm}}(N,T,r,\kappa)
&\sim&
(\gamma\td\gamma)^{{1\over 2}(n-N)^2}\,
\ee{-N^2 F(1,0,\epsilon)}\,
\ee{-N (n-N)\mu}\, \cr
&& \sqrt{H(q)}\, \L(q)^{n-N}\, \ee{-N (T(q)-V_1(x(q)))}\, \cr
&& \sum_{m_I} \ee{2i\pi \nu^t m }
 \ee{i\pi (m-N\epsilon)^t \tau (m-N\epsilon)}\,\,\,
\ee{2i\pi N \zeta^t(m-N\epsilon)}\, \cr
&& \ee{ -2i\pi (u(q)-u(\infty_x))^t (m-N\epsilon)}\,
\ee{2i\pi (n-N)(u(\infty_x)-u(\infty_y))^t(m-N\epsilon)}\, \cr
&\sim&
(\gamma\td\gamma)^{{1\over 2}(n-N)^2}\,
\ee{-N^2 F(1,0,\epsilon)}\,
\ee{-N (n-N)\mu}\, \cr
&& \sqrt{H(q)}\, \L(q)^{n-N}\, \ee{-N (T(q)-V_1(x(q)))}\, \cr
&& \ee{i\pi N \epsilon^t \tau N\epsilon}\, \ee{-2i\pi N \zeta^t N\epsilon}\,
\ee{ 2i\pi (u(q)-u(\infty_x))^t N\epsilon}\,
\ee{-2i\pi (n-N)(u(\infty_x)-u(\infty_y))^t N\epsilon}\, \cr
&& \theta( \eta_{N-n}+ u(q)-u(\infty_x)  ,\tau)
\cr
\eea
et le rapport $ Z_{{\rm Norm}}(N,T,r,\kappa)/ Z_{{\rm Norm}}(N,T,0,\kappa)$ vaut dans cette approximation:
\bea
\pi_n(x(q)) = {Z_{{\rm Norm}}(N,T,r,\kappa)\over Z_{{\rm Norm}}(N,T,0,\kappa)}
&\sim&
 \sqrt{H(q)}\, \L(q)^{n-N}\, \ee{-N (T(q)-V_1(x(q)))}\,  \cr
&& \ee{ 2i\pi N\epsilon^t (u(q)-u(\infty_x)) }\, {\theta( \eta_{N-n}+ u(q)-u(\infty_x)  ,\tau)\over \theta( \eta_{N-n},\tau)}
\cr
\eea

i.e. on retrouve la fonction $f_{N-n}(p)$ d\'efinie \`a l'\'equation \ref{deffkpasymp}.

\bigskip

Les arguments pr\'esent\'es ici sont euristiques et indicatifs.
Ils servent juste \`a deviner la bonne forme \`a donner \`a la conjecture.

Je pense qu'il do\^\i t \^etre possible de justifier rigoureusement toute cette approche, et donc
de prouver la conjecture, sans passer par une m\'ethode de Riemann--Hilbert.
Mais ceci reste \`a faire...

 \vfill\eject

\chapter{Conclusion}

Ce m\'emoire fait un r\'esum\'e de plus de 10 ann\'ees de mes travaux sur le mod\`ele \`a deux matrices.
Les travaux les plus anciens ne concernaient que les courbe spectrale de genre $0$, et c'est petit \`a petit que
la g\'eom\'etrie alg\'ebrique un peu plus \'evolu\'ee, a fait son apparition, et a permis d'\'etudier les courbes de
genre quelconque.

Mes travaux avaient d\'ebut\'e en temps que physicien, le but \'etant de comprendre un peu mieux la gravitation quantique
et les th\'eories conformes. Dans ce cadre, les physiciens \'etaient tr\`es int\'eress\'es par le d\'eveloppement topologique du
mod\`ele formel, et seul le cas de genre z\'ero correspondait vraiement \'a la ''gravit\'e quantique'' (en effet, la fonction de partition est une
somme sur des surfaces discr\'etis\'ees col\'ees seulement par leur bords, pas par leurs centres).

L'id\'ee de regarder les courbes de genre plus \'elev\'e, c'est impos\'ee naturellement, d'une part par souci de compl\'etude,
d'autre part pour r\'esoudre le ''puzzle de Br\'ezin et Deo'' (depuis r\'esolu par Deift \& co \cite{dkmvz}), et enfin car la th\'eorie de Dijkgraaf et Vafa a mis en \'evidence
le r\^ole jou\'e par les courbes de genre plus \'elev\'e en th\'eorie des cordes.

\bigskip

Dans ce m\'emoire, j'ai commenc\'e par d\'efinir les mod\`eles de matrices, de la fa\c con dont je les comprends,
puis j'ai expliqu\'e ce que l'on sait faire avec la m\'ethode des polyn\^omes biorthogonaux,
et ensuite par la m\'ethode des \'equations de boucles.
Le dernier chapitre montre que lorsque l'on combine les deux m\'ethodes, on obtient une compr\'ehension
beaucoup plus grande, et ceci permet par exemple de deviner la forme des asymptotiques
des polyn\^omes biorthogonaux.
Bref, on gagne toujours \`a attaquer un probl\`eme par plusieurs cot\'es, et \`a conn\^\i tre aussi bien
les travaux des math\'ematiciens que des physiciens.

\bigskip

Les perspectives de ces travaux sont nombreuses, tant en math\'ematique qu'en physique.

Cot\'e math\'ematique, il reste bien \'evidement \`a d\'emontrer les asymptotiques,
mais il reste aussi de nombreuses questions en suspens, comme la fonction tau isomonodromique.

Cot\'e physique, les matrices al\'eatoires ne sont qu'un outil. Le mod\`ele \`a deux matrices semble
riche de perspectives pour mieux comprendre, voire pour donner un sens rigoureux, aux
th\'eories de champs conformes, en particulier sur les surfaces avec bords \cite{Kostov}.
Les op\'erateurs de bords dans les th\'eories conformes sont encore assez mal connus, et le mod\`ele
\`a deux matrices semble fournir un tr\`es bon outil pour avancer sur ce sujet, surtout depuis que l'on
sait calculer des observables contenant des traces mixtes.

D'une fa\c con plus lointaine et ambitieuse, il semble que de nombreux concepts physiques et math\'ematiques
sont en train de confluer: il s'agit de comprendre en profondeur la relation entre int\'egrabilit\'e(s) (au sens
isomonodromique, au sens Yang--Baxter, au sens Ansatz de Behte), g\'eom\'etrie alg\'ebrique,
combinatoire,...
Mais ceci est plus un espoir qu'une r\'ealit\'e \`a l'heure actuelle...

\section*{Remerciements}

Je tiens avant tout \`a remercier tous mes collaborateurs durant ces ann\'ees de recherhes, par ordre alphab\'etique:
Marco Bertola, Pavel Bleher, Gabrielle Bonnet, Fran\c cois David, Philippe Di Francesco, Emmanuel Guitter, John Harnad, Jacques Hurtubise,
Alexey Kokotov, Dmitri Korotkin, Charlotte Kristjansen, Madan Lal Mehta, Nicolas Orantin, Aleix Prats Ferrer, Jean Zinn-Justin.
Je tiens \`a remercier \'egalement tous ceux avec qui j'ai eu des discussions fructueuses sur ce sujet, en particulier dans mon laboratoire Ivan Kostov,
Jean-Bernard Zuber, Michel Berg\`ere, Jean-Marie Normand.
Je dois remercier mon laboratoire le Service de Physique Th\'eorique du CEA \`a Saclay, ainsi que tous les laboratoires qui m'ont accueilli lorsque je travaillais sur ces sujets:
L'Universit\'e de Durham (UK), Le Niels Bohr Institutet (Copenhague, Danemark), l'Universit\'e de Colombie Britanique (Vancouver, Canada, BC), le CRM (Montr\'eal, canada QC).
Je dois remercier \'egalement les r\'eseaux Europ\'eens Eurogrid et Euclide dont le support mat\'eriel ainsi que les \'echanges qu'ils m'ont permis de d\'evelopper ont jou\'e
un r\^ole essentiel dans ma recherche.
Je remercie \'egalement Anne Boutet de Monvel et Jean-Jacques Sansuc, qui m'ont encourag\'e et permis de soutenir cette habilitation \`a l'universit\'e Paris VII.
Pour finir, je remercie ma femme Sandra et mes filles Salom\'e et Ma\`eve, pour m'avoir accompagn\'e tout ce temps.

\vfill\eject


\section*{Publications de l'auteur contenues dans le m\'emoire}

\vspace {0.6cm}

\bibeynmatsg
\vspace {0.3cm}

\noindent\bibBEformulaD
\vspace {0.3cm}

\noindent\bibeynmultimat
\vspace {0.3cm}

\noindent\bibBEmixed
\vspace {0.3cm}

\noindent\bibeynmatgzero
\vspace {0.3cm}

\noindent\bibBEHRH
\vspace {0.3cm}

\noindent\bibBEHtauiso
\vspace {0.3cm}

\noindent\bibBEHduality
\vspace {0.3cm}

\noindent\bibeynchaint
\vspace {0.3cm}

\noindent\bibEMchain
\vspace {0.3cm}

\noindent\bibeynchain

\end{document}